\documentclass[11pt]{article}
\usepackage{array}
\usepackage{float}%
\usepackage{graphicx}%
\usepackage{multirow}%
\usepackage{amsmath,amssymb,amsfonts}%
\usepackage{amsthm}%支持公式环境
\usepackage{mathrsfs}%
\usepackage[title]{appendix}%
\usepackage{xcolor}%
\usepackage{textcomp}%
\usepackage{manyfoot}%
\usepackage{booktabs}%
\usepackage{algorithm}%
\usepackage{epstopdf}
\usepackage{algorithmicx}%
\usepackage{algpseudocode}%
\usepackage{makecell}%
\usepackage{graphicx,epstopdf} % <- Preamble
\usepackage{pifont}
\usepackage[caption=false]{subfig}
\usepackage{listings}%
\usepackage{tikz}%
\usepackage{booktabs}%
\usepackage{enumerate}%
\usepackage{chngcntr} %修改编号的包
\usepackage[title]{appendix} % 附录环境
\usepackage{etoolbox}
\usepackage{epsfig}
\usepackage[numbers,sort&compress]{natbib}

\theoremstyle{definition}

\makeatletter
\def\@biblabel#1{[#1]}
\makeatother

\makeatletter \@addtoreset{equation}{section}

\begin{document}
%\begin{CJK*}{GBK}{song}

\begin{titlepage}
\title{\bf{Resonance interaction of vector breathers and exotic dynamics in the coupled Hirota equation
\footnote{%This work is supported by the National Natural Science Foundation of China under Grant No.11975306.\protect\\
Corresponding authors.\protect\\
\hspace*{3ex} E-mail addresses: xbwang@cumt.edu.cn, xiubinwang@163.com}
}}
\author{Li-Juan Bo, Xiu-Bin Wang$^{*}$\\
%%%%%%%%%%%%%%%%%%%%%%%%%%%%%%%%%%%%%%%%%%%%%%%%%%%%%%%%%%%%%%%%%%%%%%%%%%%%%%%%%%%%%%%%%
%%%%%              以下两行为作者单位
%%%%%%%%%%%%%%%%%%%%%%%%%%%%%%%%%%%%%%%%%%%%%%%%%%%%%%%%%%%%%%%%%%%%%%%%%%%%%%%%%%%%%%%%%
%\small $^{\textrm{a}}$\emph{School of Mathematics and Statistics, Anyang Normal University, Anyang 455000,} \\
%\small \emph{People's Republic of China}\\
\small \emph{School of Mathematics, China University of Mining and Technology, Xuzhou 221116,} \\
\small \emph{People's Republic of China}\\
\date{}}
\thispagestyle{empty}
\end{titlepage}
\maketitle

\vspace{-0.5cm}
\begin{center}
\rule{15cm}{1pt}\vspace{0.3cm}

\parbox{15cm}{\small
{\bf Abstract}\\
\hspace{0.5cm}
This work presents a systematic theoretical investigation of vector breathers and their nonlinear interactions in the coupled Hirota equation. This model describes the simultaneous propagation of ultrashort optical pulses in optical fibers and contains a key parameter \(\epsilon\) that characterizes the strength of higher-order linear and nonlinear effects, such as third-order dispersion and self-steepening. For the model, we focus on the focusing-focusing case. On a nonzero plane-wave background, we derive explicit expressions for multi-breather solutions of the coupled Hirota equation by the dressing method. For the single-breather solution, by choosing special constant vectors, the vector breathers can be classified into three fundamental types: Type I, Type II, and Type III. We then analyze their characteristic parameters, including spatial localization and group velocity, as well as important special cases such as the Kuznetsov-Ma breathers, Peregrine soliton, and Akhmediev type breathers etc. Through the above analysis, we find that the higher-order effect parameter \(\epsilon\) only modifies the propagation velocity and oscillation period of the breathers, while the resonance conditions remain satisfied: when the wave numbers and frequencies match, inelastic resonant interactions can occur between Type I and Type II breathers, manifesting either as a fusion process (\(\mathrm{I} + \mathrm{II} \rightarrow \mathrm{III}\)) or the reverse decay process (\(\mathrm{III} \rightarrow \mathrm{I} + \mathrm{II}\)). Furthermore, we present the explicit expression for the two breather solutions of the equation and analyze its rich interaction scenarios, including two breather resonances and elastic collisions, and we point out that the special case of resonance corresponds to the limiting case of merging eigenvalues. Finally, we carry out numerical simulations to verify the accuracy of the obtained solutions.
}

\vspace{0.5cm}
\parbox{15cm}{\small{

\vspace{0.3cm} \emph{Key words:} The coupled Hirota equation; dressing method; resonant interactions; vector breathers.

 } }
\end{center}
\vspace{0.3cm} \rule{15cm}{1pt} \vspace{0.2cm}

\section{Introduction}\label{sec1}

Breathers, as nonlinear wave structures that coherently pulsate on an unstable background, represent a core research direction in nonlinear science and are highly challenging. They can be regarded as a natural generalization of soliton theory to an unstable plane wave background \cite{Akhmediev1997}.
The study of breather waves possesses both profound theoretical value and broad practical application prospects.
From a theoretical perspective, many important physical models, such as the nonlinear Schr\"{o}dinger equation (NLSE) \cite{Frisquet2016,VEEE2013},
the derivative NLSEs \cite{LIUCC2019,HJJS2011,GLLL2013}, the Davey-Stewartson equations \cite{YJKKK2012},
the vector NLSEs \cite{TTBBB2021}, the Ablowitz-Ladik equation \cite{ALKKAA2010} etc,
are exactly solvable, and breather wave solutions constitute a vital component of the nonlinear dynamics of such integrable systems.
From an applied perspective, these integrable models have been successfully employed to describe the propagation of optical solitons in optical fibers, rogue waves in fluid mechanics, localized structures in plasmas, and coherent matter waves \cite{A2012,B2004,C2018} in Bose-Einstein condensates (BECs) etc. The theory of breather waves for numerous scalar equations has been deeply developed over the past decades, revealing a series of fundamental building blocks such as the Kuznetsov \cite{Kuznetsov1977,Ma1979}, Akhmediev \cite{Akhmediev1986}, Peregrine \cite{Peregrine1983}, and Tajiri-Watanabe solutions, as well as novel phenomena like superregular interactions and ``ghost'' breather waves. These nonlinear wave field evolution scenarios have been confirmed experimentally in platforms including optics, fluid mechanics, and plasma physics. Breather waves play a crucial role in the formation of rogue waves \cite{Zhao2013,Lu2012,Ling2015,Liu-2022,Zhang-2017,WXXB1,WXXB2,WXXB3,WXXB4,WXXB5}, the development of modulation instability \cite{Chen2021,Chen2022,Liu2022,Chen2023}, and the dynamics and statistical properties of complex nonlinear random wave fields \cite{JMMM1,JMMM2}.

In recent years, the research focus has expanded from scalar systems to multicomponent vector systems to address more complex physical phenomena, such as polarization effects in optical fibers \cite{Kaup1993,Solli2007,Erkintalo2009}. The nonlinear interactions among multiple coupled wave components frequently arise in a wide range of more complex physical systems, including optical fibers and Bose-Einstein condensates \cite{Kevrekidis2009} confined in either parabolic traps or optical lattices \cite{Bludov2009}. To provide an adequate description of such phenomena, it is essential to extend beyond the framework of the scalar NLSE. A key aspect is that, instead of a single amplitude, multiple amplitudes must be taken into account due to the presence of various interacting wave components in coupled-wave systems. By extending the description from an isotropic medium \cite{Zakharov19721} to an anisotropic one, a pair of coupled equations were given by Manakov as the following forms
\begin{align}\label{NLSE1}
 \begin{aligned}
&\textrm{i}q_{1,t} + q_{1,xx} + 2(|q_1|^2 + |q_2|^2)q_1  = 0,\\
&\textrm{i}q_{2,t} + q_{2,xx} + 2(|q_1|^2 + |q_2|^2)q_2 = 0,
                     \end{aligned}
\end{align}
where \(q_{1,2}\) is a two-component complex wave field, and \(t\) is time, \(x\) is spatial coordinate. The Manakov system, as a two-component generalization of the NLSE, is a core model describing light propagation in birefringent optical fibers and has attracted widespread attention. In the presence of a constant background, soliton solutions of the Manakov system transform into vector breather wave solutions. Kraus et al., based on the Riemann-Hilbert approach of the inverse scattering transform \cite{Manakov1974}, systematically classified vector breather waves into three fundamental types: Type I, Type II, and Type III. Among them, Type I is a direct vector generalization of scalar breather waves, while Types II and III exhibit distinct nonlinear dynamical behaviors unique to the vector case. Our recent work has revealed that resonance interactions can occur among these vector breather waves. For instance, Type I and Type II breather waves can merge into a Type III breather wave (denoted as (\( \text{I} + \text{II} \rightarrow \text{III} \))).
However, with in-depth research on ultrashort pulses, higher-order effects \cite{Malomed2013,Pan-2024,PanL-2024} such as third-order dispersion, self-steepening \cite{Morris1979,Wang-2024}, and stimulated Raman scattering become non-negligible. In such cases, the standard  NLSE is no longer sufficient to accurately describe the system dynamics. Consequently, the coupled Hirota equation, which incorporates higher-order terms, has emerged as a more accurate model for describing the simultaneous propagation of multicomponent optical fields in optical fibers \cite{Agrawal2012}.
		
In this paper, we focus on an important integrable vector model, which is the coupled Hirota equation, which introduces higher-order linear and nonlinear terms (including self-steepening and third-order dispersion, among others), enables a more precise characterization of the propagation dynamics of ultrashort optical pulses in optical fibers. The standard form of this model is as follows		
\begin{align}\label{HT}
 \textrm{i}q_{1,t} + q_{1,xx} + &2(\sigma_1|q_1|^2 + \sigma_2|q_2|^2)q_1 +\notag\\
 &\textrm{i}\epsilon \left[ q_{1,xxx}+(6\sigma_1|q_1|^2 + 3\sigma_2|q_2|^2)q_{1,x} + 3\sigma_2 q_1 q_2^* q_{2,x} \right] = 0, \notag\\
 \textrm{i}q_{2,t} + q_{2,xx} + &2(\sigma_1|q_1|^2 + \sigma_2|q_2|^2)q_2 + \notag\\
 &\textrm{i}\epsilon \left[ q_{2,xxx}+ (6\sigma_1|q_2|^2 + 3\sigma_2|q_1|^2)q_{2,x} + 3\sigma_1 q_2 q_1^* q_{1,x} \right] = 0,
\end{align}
where \(q_1\) and \(q_2\) are the two-component electric fields, subscripts \(x\) and \(t\) are the scaled spatial and temporal partial derivatives, \(\sigma_1\) and \(\sigma_2\) refer to the self-phase modulation (SPM) and cross-phase modulation, respectively. Here, \(\epsilon\) is a real parameter characterizing the strength of higher-order effects. Although the integrability of this model has been confirmed, and solutions such as bright-bright solitons and dark-dark solitons have been constructed in different parameter regimes (focusing-focusing, defocusing-defocusing, and mixed cases), existing studies have mainly focused on localized waves in the mixed case (\(\sigma_1=-\sigma_2\)) \cite{Chai-2019}. A systematic and in-depth investigation of vector breathers and their complex interactions, particularly resonance phenomena, in the focusing-focusing case (\(\sigma_1=\sigma_2=1\)) remains lacking.
Furthermore, it conducted a detailed study of the initial value problem  based on the Riemann-Hilbert method within the inverse scattering transform framework and proposed a universal classification system for vector breathers based on the analytic properties of the Jost functions of the wave field. According to this classification \cite{Raskovalov-2022,Gelash-2023}, vector breathers in the Hirota equation can be divided into three fundamental types: Type I, Type II, and Type III. Type I is a direct analog of the scalar Hirota equation breather, whereas Type II and Type III exhibit fundamentally different dynamics unique to the vector case. Recent studies have revealed that vector breathers can participate in resonant interactions \cite{Raskovalov-2022,Gelash-2023}. This resonance manifests as a three-breather coupling mechanism: two breathers of Type I and Type II merge into a Type III breather (symbolically represented as (\( \text{I} + \text{II} \rightarrow \text{III} \)), or conversely, undergo a decay process (\( \text{III} \rightarrow \text{I} + \text{II}\)).

In this work, we study interactions of the vector breathers, including the resonance situations. By means of the dressing method, we derive multi-breather solutions to the Hirota equation. First, we analyze single breathers and their physical properties. Then, we study the resonanse breather interactions. Compared to the Manakov system \cite{Raskovalov-2022,Gelash-2023}, the coupled Hirota equation remains to satisfy the resonance condition. The high-order effect \(\epsilon\) can change the propagation velocities, oscillation periods, and the spatiotemporal positions of the merging points of the breathers. In addition, we mainly consider the two breather solutions of the equation and analyze its rich interaction scenarios.

The paper is organized as follows. In the next Section 2, we construct the general scheme of the vector dressing method and find the \(N\)-breather solution of the coupled Hirota equation in the plane-wave background. In Section 3, we analyze the single-breather solutions of the coupled Hirota equation and their mathematical structures and physical characteristics. In Section 4, we consider the case of resonant interactions. In Section 5, we analyze different types of interactions between two-pole solutions by selecting various eigenvalues and integration constants. Furthermore, we also investigate important particular cases of vector two-breather solution. In Section 6, we present numerical simulations. The last Section 7 summarizes our conclusions.

\section{Derivation of the multibreather solutions }\label{sec2}

In this section, we build a dressing method \cite{Gelash2023,Liu2024,Yang2018,Gelash-2014,DARVOU-1991,NNANVV-1991} scheme for constructing multibreather solutions to the Eq.\eqref{HT}.
%Note, that the dressing method \cite{Gelash-2014}, also known as the Darboux dressing scheme, being a powerful tool in mathematical physics for constructing accurate solutions to nonlinear partial differential equations.
In order to construct the vector breather solutions of Eq.\eqref{HT}, we first represent Eq.\eqref{HT} in the form of two linear equations with \(3 \times 3\) matrix operators
\begin{equation}\label{eq2}
\Phi_{x}=\textbf{U}\Phi,~~
\Phi_{t}=\textbf{V}\Phi,
\end{equation}
where \(\Phi = (\Phi_1, \Phi_2, \Phi_3)^T\) is the vector eigenfunction, \(\Phi_1, \Phi_2\) and \(\Phi_3\) are all the complex functions of \(x\) and \(t\), the superscript \(T\) denotes the transpose for a vector/matrix, and the matrices \(U\) and \(V\) are written as
\begin{align*}
 \begin{aligned}
&\textbf{U}=-\textrm{i}\lambda \sigma + \textrm{i}Q, \\
&\textbf{V}=-4\textrm{i}\epsilon \sigma \lambda^3 + (4\textrm{i}\epsilon Q - 2\textrm{i}\sigma)\lambda^2 + (2\textrm{i}\epsilon \sigma Q^2 + 2\textrm{i}Q + 2\epsilon Q_x \sigma)\lambda \nonumber \\
			&\quad -2\textrm{i}\epsilon Q^3 + \textrm{i}\sigma Q^2 - \textrm{i}\epsilon Q_{xx} + Q_x \sigma - \epsilon Q_x Q + \epsilon Q Q_x,
                     \end{aligned}
\end{align*}
with
\begin{equation*}
Q=\left[
    \begin{array}{ccc}
      0 & q_1^* &  q_2^*  \\
      q_1 & 0 & 0   \\
      q_2 & 0 & 0
    \end{array}
  \right],~\sigma = \left[\begin{array}{ccc}
				-1 & 0 & 0 \\
				0 & 1 & 0 \\
				0 & 0 & 1
			\end{array}
  \right].
\end{equation*}
Here, \(\lambda\) is the complex spectral parameter and \(*\) stands for the complex conjugate.

In order to construct the fundamental (first-order) vector breather solutions, we start with the vector plane wave \(q_j^{[0]}, j =1 , 2\) as the seed solution. The vector plane wave solution of Eq.\eqref{HT} can be written as

\begin{equation}\label{BJ}
q_j^{[0]}=a_j\exp(\textrm{i} b_j t),
\end{equation}
where
\begin{equation*}
b_j = 2(a_1^2  + a_2^2 ).
\end{equation*}		
Here, \(a_j\)'s represent the background amplitude.
		
Then using the following matrix transformation
\begin{equation}\label{eq3}
\Phi=\textbf{G}\Phi_0,~~
\textbf{G}=\left[
    \begin{array}{ccc}
      1 & 0 & 0 \\
	0 & \exp(\textrm{i} b_1 t) & 0 \\
	0 & 0 & \exp(\textrm{i} b_2 t)
    \end{array}
  \right],
\end{equation}
and substituting the above transformation \eqref{eq3} into Lax pair \eqref{eq2}, we have
\begin{align}\label{LP}
&\Phi_{0x}=\textbf{U}_0\Phi_0,\notag\\
&\Phi_{0t}=\textbf{V}_0\Phi_0=\textrm{i}\left[ \epsilon  \textbf{U}_1^3 + (3\lambda \epsilon + 1)\textbf{U}_1^2 + (3\epsilon \lambda^2 + 2\lambda - 3a_1^2  \epsilon - 3a_2^2  \epsilon)\textbf{U}_1 + wI \right] \Phi_0,
\end{align}
with
\begin{align*}
 \begin{aligned}
&\textbf{U}_0=\textrm{i}\textbf{U}_1=\left[
    \begin{array}{ccc}
      \textrm{i}\lambda & \textrm{i}a_1  & \textrm{i}a_2  \\
			\textrm{i}a_1 & -\textrm{i}\lambda  & 0 \\
			\textrm{i}a_2 & 0 & -\textrm{i}\lambda
    \end{array}
  \right],\\
&w = -3\epsilon \lambda^3 - \lambda^2 - (3a_1^2  + 3a_2^2 ) \epsilon \lambda - 2a_1^2  - 2a_2^2  ,
                     \end{aligned}
\end{align*}
where \(U_1\) is a \(3 \times 3\) matrix independent of \(x\) and \(t\), \(G\) is a \(3 \times 3\) nonsingular matrix and \(\Phi_0\) is a \(3 \times 3\) matrix function, \(\lambda = \text{Re}(\lambda) + \textrm{i}\text{Im}(\lambda)\), \(\text{Re}(\cdot)\) and \(\text{Im}(\cdot)\) denote the real and image parts of the complex number, respectively.

From Eq.\eqref{eq2}, we find the following auxiliary equations
\begin{align}\label{YH}
&\Phi_{x t} = \Phi_{t x},\notag \\
&\Phi_{x}^{- 1} = -\Phi^{-1}\textbf{U},~~\Phi_{t}^{-1} = -\Phi^{-1}\textbf{V},\notag \\
&\Phi_{x}^{\dagger} = \Phi^{\dagger}\textbf{U}^{\dagger},~~ \Phi_{t}^{\dagger} = \Phi^{\dagger}\textbf{V}^{\dagger}.
\end{align}
Here, the sign \(\dagger\) means Hermitian conjugation. It follows from \eqref{YH} and the symmetry properties
\begin{align*}
 \begin{aligned}
			&U^\dagger(\lambda^*) = -	U(\lambda),\\
			&V^\dagger(\lambda^*) = -	V(\lambda)
\end{aligned}
\end{align*}
that
\begin{equation}\label{DC}
\Phi^{-1}(\lambda) =\Phi^\dagger(\lambda^*).
\end{equation}		
At the first procedure, we can get the ``seed'' solution		
\begin{equation*}\label{ZZJ}
\Phi_{\mathrm{0}}(x,t,\lambda)=\left[
    \begin{array}{ccc}
     0 & \exp(-\phi) & -\textrm{i}r\exp(\phi)\\ -\frac{a_{2}}{A}\exp(\phi_{0}) & \frac{a_{1}}{A}r\exp(-\phi) & \textrm{i}\frac{a_{1}}{A}\exp(\phi)\\ \frac{a_{1}}{A}\exp(\phi_{0}) & \frac{a_{2}}{A}r\exp(-\phi) & \textrm{i}\frac{a_{2}}{A}\exp(\phi)
    \end{array}
  \right],
\end{equation*}	
where
\begin{equation*}\label{DC1}
r = A / (\lambda +\zeta),~~ \zeta = \sqrt{\lambda^{2} + A^{2}},~~ A = \sqrt{a_{1}^{2} +a _{2}^{2}} ,
\end{equation*}		
and the functions \(\phi_{0}\) and \(\phi\) read
\begin{equation}\label{DC2}		
\phi_{0} = -\textrm{i}\lambda x - {\textrm{i}}\left(\lambda^{2} + \zeta^{2}\right)t-4\textrm{i}\epsilon\lambda^{3} t,~~
\phi = -\textrm{i}\zeta x - 2\textrm{i}\lambda \zeta t+2\textrm{i}\epsilon(\zeta^{3}-3\lambda^{2}\zeta) t.
\end{equation}		
At the second procedure, we introduce the dressing function \cite{Gelash-2014} \(\chi\) as
\begin{equation*}\label{DC3}			
\chi = \Phi \Phi_0^{-1},
\end{equation*}			
which admits the asymptotic condition
\begin{equation*}	
\chi (\lambda)\to I + \frac{{R}}{\lambda} +O(\lambda^{-2}),~~ |\lambda |\to \infty ,
\end{equation*}	
we then find	
\begin{equation*}	
\chi^{-1} (\lambda)\to I - \frac{{R}}{\lambda} +O(\lambda^{-2}),~~|\lambda |\to \infty ,
\end{equation*}	
in which \({R}\) denotes a constant matrix and \(I\) denotes the $3\times3$ unit matrix.
It follows from \eqref{DC} that the function \(\chi (x,t,\lambda)\) satisfies
\begin{equation*}\label{DC6}			
\chi^{\dagger}(\lambda^{*}) =\chi^{-1}(\lambda).
\end{equation*}	
Then, using \eqref{LP}  and \eqref{YH}, we obtain
\begin{align*}	
\chi_{x}^{-1} = -\chi^{-1}{U} + {U}_{0}\chi^{-1},\notag\\
 \chi_{t}^{-1} = -\chi^{-1}{V} + {V}_{0}\chi^{-1}.
\end{align*}		
From the first equation, we can get
\begin{equation}\label{YH2}
{U} = \chi{U}_{0}\chi^{-1}- \chi\chi_{x}^{-1}.
\end{equation}		
Substitute expansions of \(\chi\) and \(\chi^{-1}\) into Eq.\eqref{YH2}, we have
\begin{equation*}		
{U} = {U}_{0}+\lambda^{-1}({R}{U}_{0}-{U}_{0}{R}+{R}_{x})+O(\lambda^{-2} ).
\end{equation*}			

Equate \(\lambda^{0}\)-coefficients on both sides, we find out the final formulas about the two components \(q_{1,2}\)
\begin{equation}\label{YJJ1}		
q_{1} = e^{2\textrm{i}A^{2}t}(a_{1} + 2R_{21}) ,~~ q_{2} = e^{2\textrm{i}A^{2}t}(a_{2} + 2R_{31}).
\end{equation}		
Next step, we assume that \(\chi(\lambda)\) has a simple pole at \(\lambda = \lambda_1\) (\(\operatorname{Im}\lambda_1>0\)) and is analytic elsewhere in the complex plane. Based on this assumption, \(\chi(\lambda)\) can be written in the following form \cite{Zakharov19721}
\begin{equation}\label{TZ1}
			\chi(\lambda) = I + \frac{R}{\lambda - \lambda_1}, ~~ R(x,t) \in \mathbb{C}^{3\times 3}.
\end{equation}
The constant \(\lambda_{1}\) represents discrete eigenvalue of the system \eqref{LP}.
From the reduction condition \(\chi^\dagger(\lambda^*) = \chi^{-1}(\lambda)\), we can find \(\chi^{-1}(\lambda)\) also has a similar pole structure
\begin{equation}\label{TZ2}
			\chi^{-1}(\lambda) = I + \frac{R^\dagger}{\lambda - \lambda_1^*}.
\end{equation}
Substituting \eqref{TZ1} and \eqref{TZ2} into the identity \(\chi \chi^{-1} = I\), we obtain
\begin{equation*}
			\left(I + \frac{R}{\lambda - \lambda_1}\right)\left(I + \frac{R^\dagger}{\lambda - \lambda_1^*}\right) = I.
\end{equation*}		
Expanding the left-hand side yields
\begin{equation*}
I + \frac{R}{\lambda - \lambda_1} + \frac{R^\dagger}{\lambda - \lambda_1^*} + \frac{R R^\dagger}{(\lambda - \lambda_1)(\lambda - \lambda_1^*)} = I.
\end{equation*}
Multiplying both sides by \((\lambda - \lambda_1)(\lambda - \lambda_1^*)\), we get
\begin{equation}\label{XS}
			R(\lambda - \lambda_1^*) + R^\dagger(\lambda - \lambda_1) + R R^\dagger = 0.
\end{equation}
Eq.\eqref{XS} holds for arbitrary \(\lambda\), so we can compare coefficients of powers of \(\lambda\).
\begin{itemize}
		\item Comparing the coefficient of \(\lambda\) (linear term):
			\begin{equation*}
				R + R^\dagger = 0 \quad\Longrightarrow\quad R^\dagger = -R.
			\end{equation*}
			
			\item Comparing the constant term (zero-order term):
			\begin{equation}\label{XS2}
				-\lambda_1^* R - \lambda_1 R^\dagger + R R^\dagger = 0.
				\end{equation}
		\end{itemize}
Substituting \(R^\dagger = -R\) into \eqref{XS2}, we obtain
\begin{equation}\label{XS3}
			-\lambda_1^* R + \lambda_1 R - R^2 = 0 \quad\Longrightarrow\quad R ^2 = (\lambda_1 - \lambda_1^*) R.
\end{equation}
Denote \(\delta = \lambda_1 - \lambda_1^* = 2\textrm{i}\operatorname{Im}\lambda_1 \neq 0\). Then \eqref{XS3} can be written as
\begin{equation}\label{XS4}
			R^2 = \delta R.
\end{equation}
Eq.\eqref{XS4} indicates that \(R\) is a nilpotent matrix (since \(R^2\) is proportional to \(R\)). Furthermore, it can be proved that \(R\) has rank \(1\). Let
\begin{equation}\label{XS5}
			R = p q^T, ~~ p,q \in \mathbb{C}^3,
\end{equation}
then \(R^\dagger = q^* p^\dagger\). Substituting it into \(R^\dagger = -R\) gives
\begin{equation*}
			q^* p^\dagger = -p q^T.
\end{equation*}

This shows that \(p\) is proportional to \(q^*\). Set \(p = \alpha q^*\) and substitute into \eqref{XS5} to determine the coefficient \(\alpha\)
\begin{equation*}
			R^2 = (p q^T)(p q^T) = p (q^T p) q^T.
\end{equation*}
Since \(q^T p\) is a scalar, from \eqref{XS3} we have \(q^T p = \lambda_{1} - \lambda_{1}^{*}\). Moreover,
\begin{equation*}
			 q^T p = q^T (\alpha q^*) = \alpha q^T q^* = \alpha |q|^2.
\end{equation*}		
Solving for \(\alpha\) yields
\begin{equation*}
			\alpha = \frac{\lambda_1 - \lambda_1^*}{|q|^2}.
\end{equation*}
Hence
\begin{equation*}
			p = \frac{\lambda_1 - \lambda_1^*}{|q|^2} q^*.
\end{equation*}

In order to cancel the additional pole at \(\lambda = \lambda_{1}^{*}\) in \eqref{YH2}, we subject the vector \(\mathbf{q}(x,t)\) to the condition
\begin{equation*}
\partial_{x}\mathbf{q}^{*} - \mathbf{U}_{0}(x,t,\lambda_{1}^{*})\mathbf{q}^{*} = 0.
\end{equation*}		
Then we know that the vector \(\mathbf{q}\) satisfies the following form
\begin{equation}\label{ZZ}		
\mathbf{q}(x,t) = \Phi_{0}^{*}(x,t,\lambda_{1}^{*})\mathbf{C},
\end{equation}		
where
\begin{equation*}
\mathbf{C} = (C_0,C_1,C_2)^T
\end{equation*}			
is an arbitrary three-component complex vector with the superscript $T$ meaning transposing.

%Thereby, the one-pole function \(\chi (x,t,\lambda)\) from Eq.\eqref{TZ1} is completely defined.
By using \eqref{YJJ1}, we obtain the components \(q_{1,2}\) of the single-eigenvalue solution of Eq.\eqref{HT} as follows	
\begin{align}\label{YJJ2}
q_{1} = e^{2\textrm{i}A^{2}t}\left[a_{1} + \frac{2(\lambda_{1} - \lambda_{1}^{*})q_{2}^{*}q_{1}}{|q_1|^2+|q_2|^2+|q_3|^2}\right],\notag\\
q_{2} = e^{2\textrm{i}A^{2}t}\left[a_{2} + \frac{2(\lambda_{1} - \lambda_{1}^{*})q_{3}^{*}q_{1}}{|q_1|^2+|q_2|^2+|q_3|^2}\right].
\end{align}

Similar to \cite{Gelash-2014}, if the matrix \(\chi (x,t,\lambda)\) have \(N\) poles, \(\lambda = \lambda_{j}, j = 1, \ldots , N\),
then by making use of the Cramer's rule, the corresponding \(N\)-eigenvalue solution of Eq.\eqref{HT} can be obtained as
\begin{align}\label{YJ2}		
q_{1} =e^{2\textrm{i}A^2 t}\left[ a_{1} + 2\frac{\widetilde{\mathbf{M}}_{12}}{\mathbf{M}} \right],\notag\\
q_{2} =e^{2\textrm{i}A^2 t}\left[ a_{2} + 2\frac{\widetilde{\mathbf{M}}_{13}}{\mathbf{M}}  \right].
\end{align}
Here, \(\widetilde{\mathbf{M}}_{\alpha ,\beta}\) and \(\mathbf{M}\) are the following determinants
\begin{equation*}\label{YJJ3}		
\widetilde{\mathbf{M}}_{\alpha\beta} =
\left|
\begin{array}{c|c}
0 &
\begin{array}{ccc} q_{1,\beta}^* & \dots & q_{N,\beta}^* \end{array} \\
\hline
\begin{array}{c} q_{1,\alpha} \\ \vdots \\ q_{N,\alpha} \end{array} &
M^{T}
\end{array}
\right|,
~~ \mathbf{M} = \operatorname *{det}(\emph{M}),~~\emph{M}_{nm} = \frac{(\mathbf{q}_{n}^{*}\cdot \mathbf{q}_{m})}{\lambda_{n} - \lambda_{m}^{*}},
\end{equation*}
with \(\alpha , \beta = 1, 2, 3\) and \(n, m = 1, 2, \ldots ,N\).
From Eq.\eqref{ZZ}, we find the vectors \(\mathbf{q}_n\) can be represented as
\begin{equation*}\label{YJJ4}
\mathbf{q}_n = \left[ \begin{array}{ccc}0 & \exp(\phi_n) & \textrm{i}r_n \exp(-\phi_n)\\ -\frac{a_2}{A}\exp(-\phi_{0n}) & \frac{a_1}{A}r_n \exp(\phi_n) & -\textrm{i}\frac{a_1}{A} \exp(-\phi_n)\\ \frac{a_1}{A} \exp(-\phi_{0n}) & \frac{a_2}{A}r_n \exp(\phi_n) & -\textrm{i}\frac{a_2}{A} \exp(-\phi_n) \end{array} \right]\left[ \begin{array}{c}C_{n0}\\ C_{n1}\\ C_{n2} \end{array} \right],
\end{equation*}		
which can be rewritten in the following form
\begin{align}\label{YJJ5}
&{q_{n1} = \exp(\phi_n)C_{n1} + \textrm{i}r_n \exp(-\phi_n)C_{n2}},\notag\\
&{q_{n2} = \frac{1}{A} [-a_2\exp(-\phi_{0n})C_{n0} + a_1(r_n \exp(\phi_n)C_{n1} -\textrm{i} \exp(-\phi_n)C_{n2})],}\notag\\
 & {q_{n3} = \frac{1}{A} [a_1\exp(-\phi_{0n})C_{n0} + a_2(r_n \exp(\phi_n)C_{n1} -\textrm{i} \exp(-\phi_n)C_{n2})].}
\end{align}	
In addition, from Eq.\eqref{DC2}, we know that the functions \(\phi_{0n}\) and \(\phi_n\) can be expressed by	
\begin{align}\label{YJJ6}	
&{\phi_{0n} = -\textrm{i}\lambda_nx - \textrm{i}\left(\lambda_n^2 +\zeta_n^2\right)t-4\textrm{i}\epsilon \lambda_n^3  t = u_{0n} - \textrm{i}v_{0n},}\notag\\
 &{\phi_n = -\textrm{i}\zeta_nx - 2\textrm{i}\lambda_n\zeta_nt+2\textrm{i}\epsilon(\zeta_n^3-3 \lambda_n^2 \zeta_n )  t = u_n - \textrm{i}v_n.}
\end{align}		
By distinguishing real and imaginary parts of the functions \(\phi_{0n}\) and \(\phi_n\),
 the functions \(u_n, v_n, u_{0n}, v_{0n}\) can be represented as
\begin{align}\label{YJJ7}	
&{u_{0n} = \mathrm{Im}[\lambda_n]x + \mathrm{Im}[\lambda_n^2 +\zeta_n^2 ]t+4\epsilon\mathrm{Im} [\lambda_n^3]t ,}\notag\\
&{v_{0n} = \mathrm{Re}[\lambda_n]x + \mathrm{Re}[\lambda_n^2 +\zeta_n^2 ]t+4\epsilon\mathrm{Re} [\lambda_n^3]t,}\notag\\
&{u_n = \mathrm{Im}[\zeta_n]x +2 \mathrm{Im}[\lambda_n\zeta_n ]t-2\epsilon\mathrm{Im}[\zeta_n^3-3 \lambda_n^2 \zeta_n ]  t,}\notag\\
&{v_n = \mathrm{Re}[\zeta_n]x +2 \mathrm{Re}[\lambda_n\zeta_n ]t-2\epsilon\mathrm{Re}[\zeta_n^3-3 \lambda_n^2 \zeta_n ]  t.}
\end{align}			
	
For the sake of convenience, we next utilize the square of the modulus \(\mathbf{q}_n\), which can be expressed as
\begin{equation*}
|\mathbf{q}_n|^2 = |\exp(\phi_n)C_{n1} + \textrm{i}r_n \exp(-\phi_n)C_{n2}|^2 +|r_n \exp(\phi_n)C_{n1}
-\textrm{i} \exp(-\phi_n)C_{n2}|^2 +|\exp(-\phi_{0n})C_{n0}|^2.
\end{equation*}		

\noindent
\textbf{Remark 1:}
Since the transformation
\begin{equation*}	
\mathbf{q}_n\rightarrow \kappa\mathbf{q}_n
\end{equation*}
by the complex constant \(\kappa\) does not change the form of the solution \eqref{YJJ2}, an arbitrary choice of the vector \(\mathbf{C}\) actually corresponds to only four real degrees of freedom. To obtain nontrivial solutions of the coupled Hirota equation, at least two components of \(\mathbf{C}\) must be nonzero. The physical meanings of these parameters and the different types of solutions resulting from various choices of \(\mathbf{C}\) will be discussed in the following sections.

\section{Characteristics of types I, II and III vector breathers}\label{sec3}

In this section,
 %based on the previously proposed dressing method and the classification of the analytic properties of Jost functions, see Section 2, the expressions for three types of breather solutions are given by setting different components of the vector \(\mathbf C\) to zero.
the mathematical structures and physical characteristics of the three fundamental types (type I, II, and III) of vector breathers for Eq.\eqref{HT} are systematically constructed and analyzed.
%, including their spatial localization, group velocity, asymptotic phase, and important special cases (such as the Kuznetsov, Peregrine, and Akhmediev types). In particular, type I breathers can be obtained as a direct generalization of scalar breather solutions, while types II and III exhibit polarization dynamics specific to the vector system.
		
To simplify the algebra, we adopt the uniformization variables
\begin{align}\label{CSH}
&\lambda = A\sinh (\xi +\textrm{i}\alpha),\notag\\
&\zeta = A\cosh (\xi +\textrm{i}\alpha),\notag\\
& r = e^{-\xi -\textrm{i}\alpha}.
\end{align}

The single-eigenvalue solution \eqref{YJJ2} with \( N = 1 \) serves as our starting point, different choices of the nonzero components of \( C \) lead to the three types.

First of all, for the type-I breather, we set the components of the vector \( C \) as follows
\begin{equation*}\label{LX1}		
C_0 = 0,~~ C_1 =1,~~ C_2 = 1,
\end{equation*}
and from the general solution \eqref{YJJ2}, the explicit real and imaginary parts for type I breather can be written as
\begin{align}\label{LX2}		
 & {\mathrm{Re}q_{1,2}^{I} = e^{2\mathrm{i}A^{2}t}\left[a_{1,2} -  \frac{2a_{1,2}\sin\alpha\cosh\xi[\cos(2v_{I})\cosh\xi + \cosh(2u_{I})\sin\alpha]}{\cosh\xi\cosh(2u_{I}) + \sin\alpha\cos(2v_{I})}\right],}\notag\\
 & {\mathrm{Im}q_{1,2}^{I} = e^{2\mathrm{i}A^{2}t}\left[ \frac{2a_{1,2}\sin\alpha\cosh\xi[\sin(2v_{I})\sinh\xi + \sinh(2u_{I})\cos\alpha]}{\cosh\xi\cosh(2u_{I}) + \sin\alpha\cos(2v_{I})}\right] ,}
\end{align}
where
\begin{equation*}\label{LX3}		
2u_{I} = l_{I}^{-1}(x + V_{I}t ),~~ 2v_{I} = k_{I}x + \omega_{I}t,
\end{equation*}		
with the breather characteristic length \(l_{I}\), group velocity \(V_{I}\), wave number \(k_{I}\), and characteristic frequency \(\omega_{I}\)
\begin{align}\label{LX4}
&{l_{I} = (2\mathrm{Im}[\zeta ])^{-1} = (2A\sin \alpha \sinh \xi)^{-1},}\notag\\
&{V_{I} = \frac{2\mathrm{Im}[\lambda \zeta ]-2\epsilon\mathrm{Im}[\zeta^3-3 \lambda^2 \zeta ]}{\mathrm{Im}[\zeta ]} ,}\notag\\
&{k_{I} = 2\mathrm{Re}[\zeta ] = 2A\cos \alpha \cosh \xi ,}\notag\\
&{\omega_{I} = 4\mathrm{Re}[\lambda \zeta ]-4\epsilon\mathrm{Re}[\zeta^3-3 \lambda^2 \zeta ]  .}
\end{align}	
In the case, the resulting breather represents a simple vector generalization of the solution of the scalar Hirota equation, and we can find the two wavefield components are proportional, it satifies
\begin{equation}\label{BL}		
q_2 = (a_2 / a_1) q_1,
\end{equation}
so that each component individually satisfies the scalar Hirota equation
\begin{equation}\label{DFL}		
\textrm{i} u_t + u_{xx} + 2|u|^2 u+ \textrm{i}\epsilon[u_{xxx}+6|u|^2u_x]= 0.
\end{equation}
From this, we  also can get that the vector breathers of type I can be obtained from the known breather solutions of Eq.\eqref{DFL} through the following transformation
\begin{align}\label{NJ1}
&q_1(x, t) = \frac{a_1}{A} u \left( x, t \right),\notag \\
&q_2(x, t) = \frac{a_2}{A} u \left( x, t \right).
\end{align}

Notably, \(\epsilon\) appears only in the expressions for \(V_{I}\) and \(\omega_{I}\),  but does not affect \(l_{I}\) and \(k_{I}\). Thus, the higher-order effect alters the propagation dynamics and the temporal pulsation period, whereas the spatial localisation scale remains unchanged.

Figure 1 shows a general case, we choose the following parameters
\begin{equation}\label{W}		
a_1=1,~~a_2=1,~~\alpha=\pi /6,~~\xi=1/5 ,
\end{equation}	
in the picture, type I breather behaves as a moving wave packet with proportional amplitudes in both components, and the breather is localized in space and moves with a nonzero group velocity.

$~~~~~~~~~~~~~~~~~~~~~~~~~~$
{\rotatebox{0}{\includegraphics[width=6.2cm,height=4.2cm,angle=0]{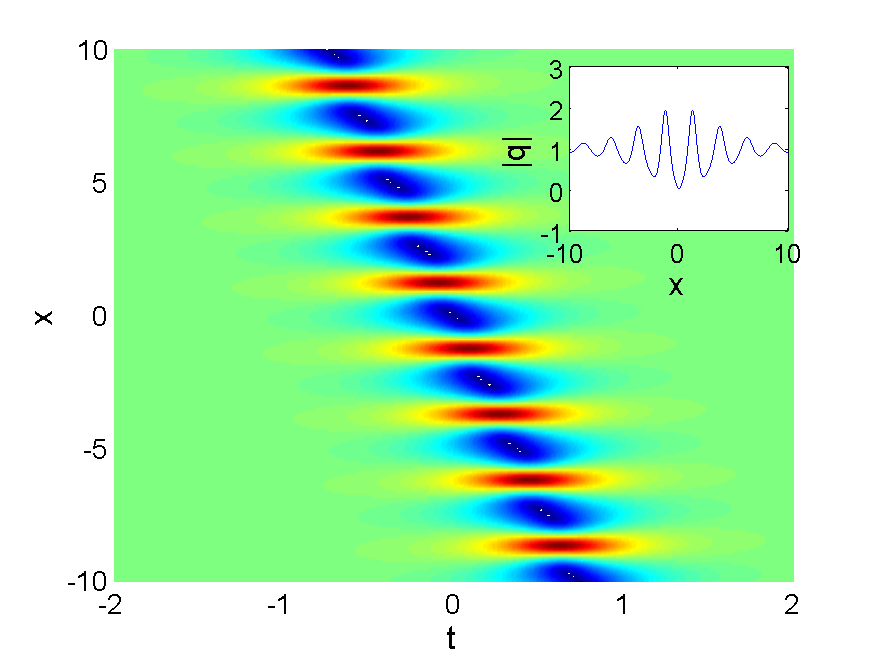}}}
\\

\noindent { \small \textbf{Figure 1.} Vector breather of type I, see Eq.\eqref{LX2}, which can be obtained from scalar breather solution
using the transformation \eqref{NJ1}. The solution parameters are defined in \eqref{W}.  Spatio-temporal evolution of \(q_1\). The wavefield component \(q_2\) is not shown because it coincides with the first one after rescaling the amplitude, see Eq.\eqref{BL}.\\}

\noindent

Secondly, for the type-II breather, we choose the components of the vector \( C \) as follows
\begin{equation*}\label{LX21}
C_0 =1,~~C_1 = 0,~~C_2 = 1.
\end{equation*}
In this case, we obtain
\begin{align}\label{LX22}
q_1^{\mathrm{II}} =e^{2\textrm{i}A^2 t}\left[a_1 - \frac{4\textrm{i}\mathrm{e}^{-\textrm{i}\alpha}\sin\alpha\cosh\xi(a_1 + a_2\mathrm{e}^{-u_{\mathrm{II}} - \textrm{i}v_{\mathrm{II}}}\textrm{i})}{e^{-2u_{\mathrm{II}} + \xi} + 2\cosh\xi}\right],\notag\\
 q_2^{\mathrm{II}} =e^{2\textrm{i}A^2 t}\left[ a_2 - \frac{4\textrm{i}\mathrm{e}^{-\textrm{i}\alpha}\sin\alpha\cosh\xi(a_2 - a_1\mathrm{e}^{-u_{\mathrm{II}} - \textrm{i}v_{\mathrm{II}}}\textrm{i})}{e^{-2u_{\mathrm{II}} + \xi} + 2\cosh\xi}\right],
\end{align}			
where
\begin{equation*}\label{LX23}
u_{\mathrm{II}} =u_0-u= l_{\mathrm{II}}^{-1}(x + V_{\mathrm{II}}t ),~~ v_{\mathrm{II}} =v_0-v= k_{\mathrm{II}}x + \omega_{\mathrm{II}}t,
\end{equation*}		
and the physical characteristics of type II breather are writed as
\begin{align}\label{LX24}
&{l_{\mathrm{II}} = (\mathrm{Im}[\lambda ] - \mathrm{Im}[\zeta ])^{-1} = (Ae^{-\xi}\sin \alpha)^{-1},}\notag\\
&{V_{\mathrm{II}} = \frac{\mathrm{Im}[\lambda^2 +\zeta^2]  -2 \mathrm{Im}[\lambda\zeta ]+4\epsilon\mathrm{Im}[\lambda^3]+2\epsilon\mathrm{Im}[\zeta^3-3 \lambda^2 \zeta ]}{\mathrm{Im}[\lambda] - \mathrm{Im}[\zeta]},}\notag\\
&{k_{\mathrm{II}} = \mathrm{Re}[\lambda ] - \mathrm{Re}[\zeta ] = -A\cos \alpha e^{-\xi},}\notag\\
& {\omega_{\mathrm{II}} = \mathrm{Re}[\lambda^2 +\zeta^2 ] - 2\mathrm{Re}[\lambda \zeta ] +4\epsilon\mathrm{Re}[\lambda^3]+2\epsilon\mathrm{Re}[\zeta^3-3 \lambda^2 \zeta ]\ .}
\end{align}

Unlike type I, the two solutions do not satisfy a simple proportionality relation between the components.

Figure 2 shows an example of the type II localized breather, we choose the parameters \eqref{W}, which moves with a nonzero group velocity.

\noindent
{\rotatebox{0}{\includegraphics[width=4.8cm,height=3.2cm,angle=0]{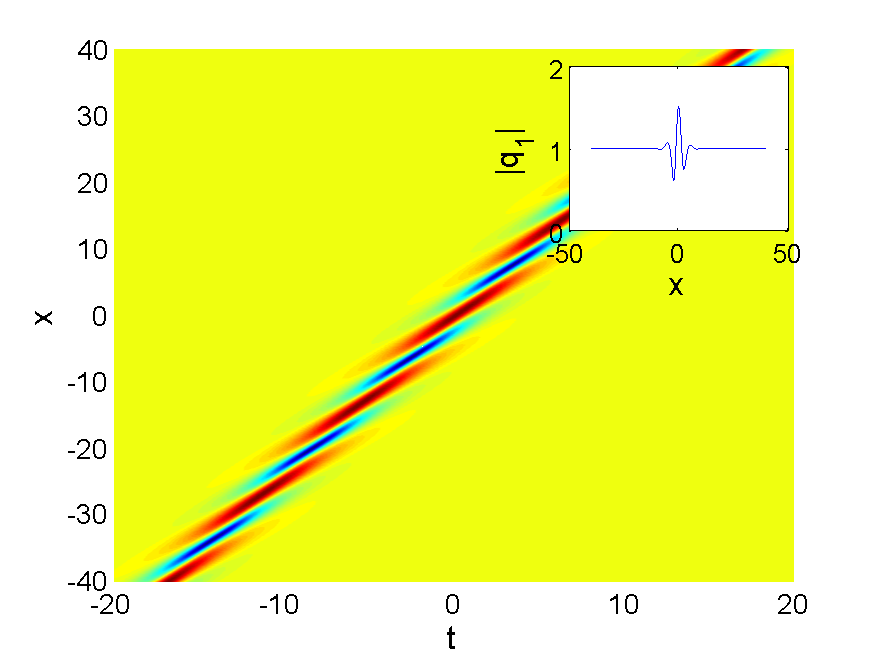}}}
{\rotatebox{0}{\includegraphics[width=4.8cm,height=3.2cm,angle=0]{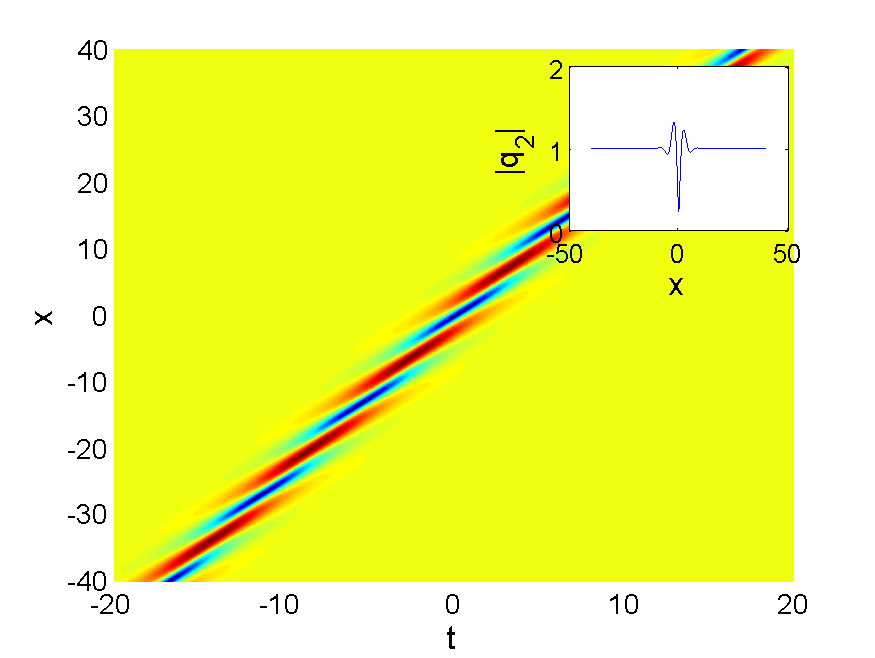}}}
{\rotatebox{0}{\includegraphics[width=4.8cm,height=3.2cm,angle=0]{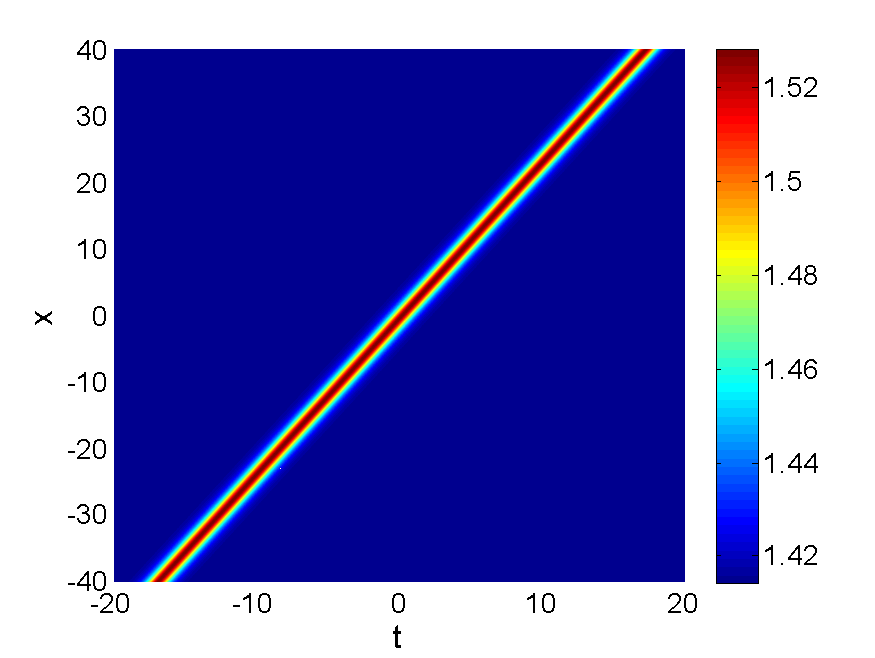}}}\\
$~~~~~~~~~~~~~~~~~~(\textbf{a})
~~~~~~~~~~~~~~~~~~~~~~~~~~~~~~~~~~~(\textbf{b})
~~~~~~~~~~~~~~~~~~~~~~~~~~~~~~~~(\textbf{c})$\\

\noindent { \small \textbf{Figure 2.} Amplitude distributions of vector breather of type II, with integration constants ( $C_{0} = 1 , ~C_{1} = 0, ~C_{2} = 1$). From left to right, \(|q_1|\), \(|q_2|\), and \(\sqrt{|q_1|^2+|q_2|^2}\). Other parameters are defined as \(a_1 = 1, ~a_2 = 1, ~\xi = 0.2, ~\alpha = \pi /6 \).\\}
\\

\noindent

Finally, for the type-III breather, we choose the components of the vector \( C \) in the form
\begin{equation*}\label{LX31}
C_0 = 1,~~\textrm{i} r C_1 = 1,~~
C_2 = 0.
\end{equation*}		
In this case, the solution is writed as
\begin{align}\label{LX32}
q^{\mathrm{III}}_1 =e^{2\textrm{i}A^2 t}\left[a_1 + \frac{4\textrm{i} \sin\alpha e^{\textrm{i}\alpha}\cosh\xi \left(a_1 + a_2 e^{-u_{\mathrm{III}} - \textrm{i} v_{\mathrm{III}}}\textrm{i}\right)}{e^{-2u_{\mathrm{III}} - \xi} + 2\cosh\xi}\right], \notag\\
q^{\mathrm{III}}_2 =e^{2\textrm{i}A^2 t}\left[a_2 + \frac{4\textrm{i} \sin\alpha e^{\textrm{i}\alpha}\cosh\xi \left(a_2 - a_1 e^{-u_{\mathrm{III}} - \textrm{i} v_{\mathrm{III}}}\textrm{i}\right)}{e^{-2u_{\mathrm{III}} - \xi} + 2\cosh\xi}\right],
\end{align}		
where
\begin{equation}\label{LX33}
u_{\mathrm{III}} =u_0+u= l_{\mathrm{III}}^{-1}(x + V_{\mathrm{III}}t ),~~ v_{\mathrm{III}} =v_0+v= k_{\mathrm{III}}x + \omega_{\mathrm{III}}t,
\end{equation}		
and the physical characteristics of type III breather have the form
\begin{align}\label{LX34}
&{l_{\mathrm{III}} = (\mathrm{Im}[\lambda ] + \mathrm{Im}[\zeta ])^{-1} = (Ae^{\xi}\sin \alpha)^{-1},}\notag\\
 &{V_{\mathrm{III}} = \frac{\mathrm{Im}[\lambda^2 +\zeta^2]  +2 \mathrm{Im}[\lambda\zeta ]+4\epsilon\mathrm{Im}[\lambda^3]-2\epsilon\mathrm{Im}[\zeta^3-3 \lambda^2 \zeta ]}{\mathrm{Im}[\lambda] + \mathrm{Im}[\zeta]} ,}\notag\\
 &{k_{\mathrm{III}} = \mathrm{Re}[\lambda ] + \mathrm{Re}[\zeta ] = A\cos \alpha e^{\xi},}\notag\\
 &{\omega_{\mathrm{III}} = \mathrm{Re}[\lambda^2 +\zeta^2 ] + 2\mathrm{Re}[\lambda \zeta ]+4\epsilon\mathrm{Re}[\lambda^3]-2\epsilon\mathrm{Re}[\zeta^3-3 \lambda^2 \zeta ].}
\end{align}

Figure 3 shows an example of type III localized breather through the parameters \eqref{W}, which moves with a nonzero group velocity.

\noindent
{\rotatebox{0}{\includegraphics[width=4.8cm,height=3.2cm,angle=0]{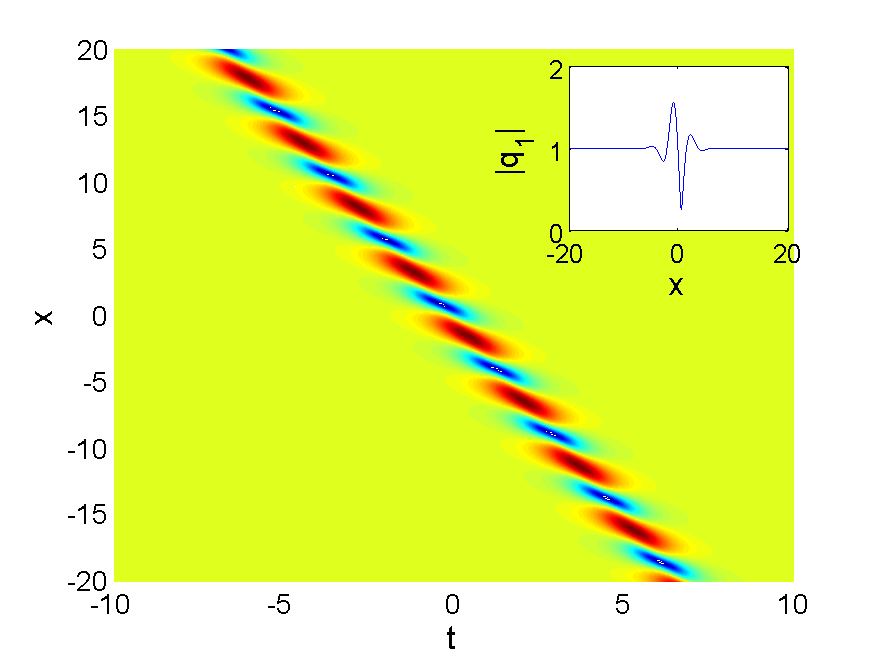}}}
{\rotatebox{0}{\includegraphics[width=4.8cm,height=3.2cm,angle=0]{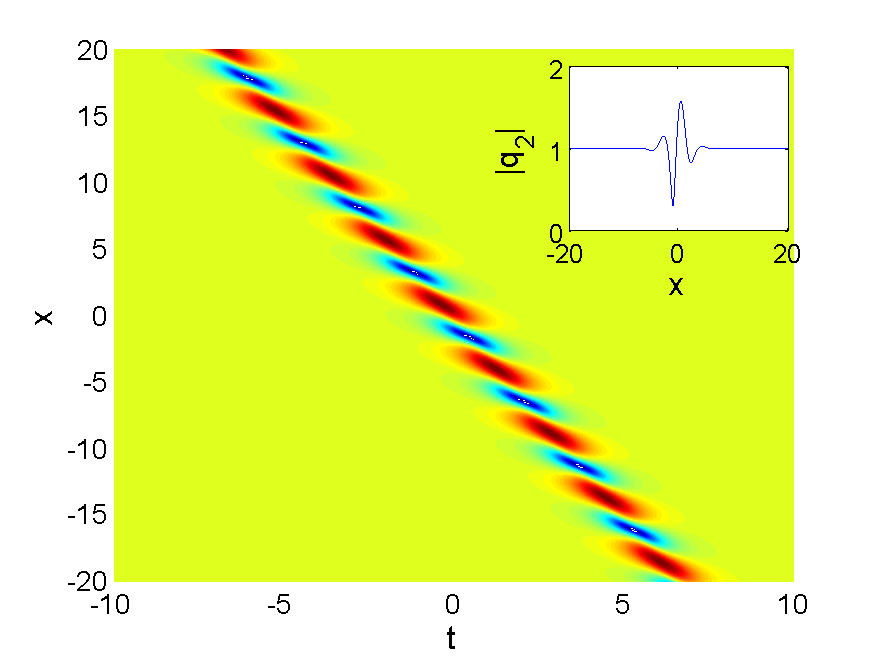}}}
{\rotatebox{0}{\includegraphics[width=4.8cm,height=3.2cm,angle=0]{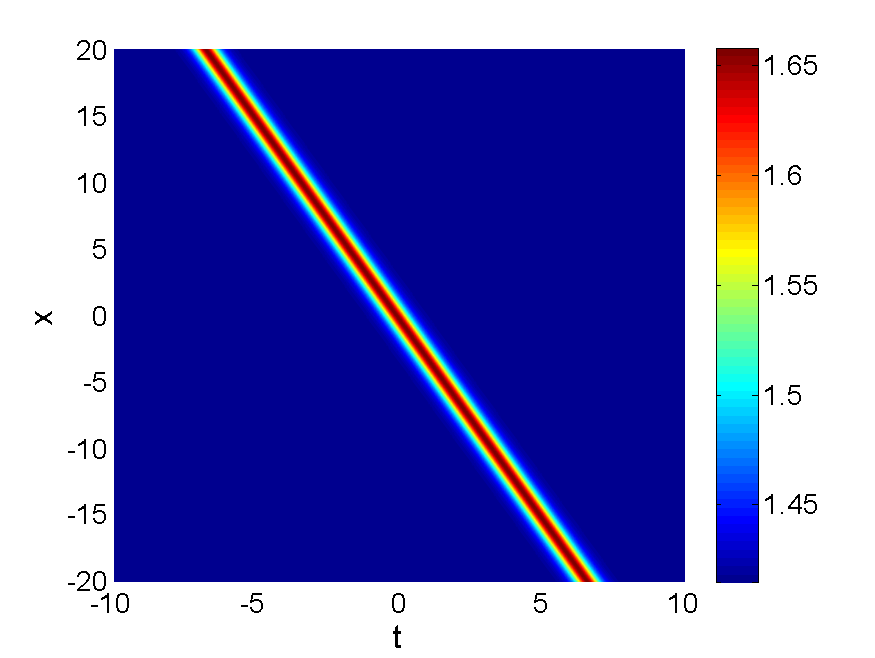}}}\\
$~~~~~~~~~~~~~~~~~~(\textbf{a})
~~~~~~~~~~~~~~~~~~~~~~~~~~~~~~~~~~~(\textbf{b})
~~~~~~~~~~~~~~~~~~~~~~~~~~~~~~~~(\textbf{c})$\\

\noindent { \small \textbf{Figure 3.} Amplitude distributions of vector breather of type III. From left to right, \(|q_1|\), \(|q_2|\), and \(\sqrt{|q_1|^2+|q_2|^2}\). Other parameters are defined as \(a_1 = 1, ~a_2 = 1, ~\xi = 0.2, ~\alpha = \pi /6 \).\\}
\\

\noindent

Through the above content, we can find the solutions of type II and III have a similar structure. From the corresponding physical parameters, see Equations \eqref{LX24} and \eqref{LX34}. A direct comparison shows that \(\mathrm{II}\) and \(\mathrm{III}\) have opposite signs for the same eigenvalue, and the inequalities \(|k_{\mathrm{II}}|< |k_{\mathrm{III}}|\) and \(|l_{\mathrm{II}}| > |l_{\mathrm{III}}|\) always hold. Moreover, the transformation
\begin{equation*}		
\xi \to -\xi ,~~ \alpha \to \pi -\alpha
\end{equation*}
maps type II into type III; applied to type I, it merely produces a phase shift. These properties are inherited from the Manakov case, but all velocity and frequency formulas now carry explicit \(\epsilon\) dependent corrections, reflecting the role of higher-order terms in the Hirota equation.

From the Figures 1-3 through choose the parameters, type I behaves as a moving wave packet with proportional amplitudes in both components. Types II and III, on the other hand, display asymmetric polarisation patterns and distinct amplitude distributions in each component. These figures confirm that while \(\epsilon\) quantitatively changes the speed and the internal oscillation rhythm, the localised nature and the fundamental classification remain intact.

\textbf{Special limiting cases:} The scalar Hirota equation admits three famous particular breather regimes: Kuznetsov, Peregrine, and Akhmediev. Their vector counterparts are obtained by restricting the spectral parameters to the regions \eqref{A}, \eqref{B}, and \eqref{C}, respectively. We summarise the key features below, with emphasis on how they differ from the Manakov analogues.

The first region is writed as
\begin{equation}\label{A}
\mathrm{(a)}~~ \{\mathrm{Re}[\lambda ] = 0,~~ \mathrm{Im}[\lambda ] > A\} , ~~ \{\alpha = \pi /2,~~\xi >0\}.
\end{equation}		
In this case, we obtain a vector analogue of the excitation known as the Kuznetsov breather
\begin{align*}
 \begin{aligned}
&\mathrm{Re}\,q_{1,2}^{I} = e^{2\mathrm{i}A^{2}t}\left[a_{1,2} - \frac{2a_{1,2}\cosh\xi\bigl[\cosh(2u_{I}) + \cosh\xi\cos(2v_{I})\bigr]}{\cosh\xi\cosh(2u_{I}) + \cos(2v_{I})}\right],\\
&\mathrm{Im}\,q_{1,2}^{I} = e^{2\mathrm{i}A^{2}t}\left[\frac{2a_{1,2}\sinh\xi\cosh\xi\sin(2v_{I})}{\cosh\xi\cosh(2u_{I}) + \cos(2v_{I})}\right].
\end{aligned}
\end{align*}
The Kuznetsov breather is like an isolated wave packet it contracts and expands periodically in time, but its spatial location remains unchanged. It resembles a static, periodically oscillating local structure. It oscillates periodically over time, with a period of
\begin{equation*}
T = \frac{2\pi}{\omega_{I}},
\end{equation*}
where
\begin{equation*}
{\omega_{I} = 4\mathrm{Re}[\lambda \zeta ]-4\epsilon\mathrm{Re}[\zeta^3-3 \lambda^2 \zeta ].}
\end{equation*}
Note that
\begin{align*}
 \begin{aligned}
		T \to \infty, ~~ \xi \to 0 , \\
		T \to 0, ~~\xi \to \infty.
\end{aligned}
\end{align*}
It was reported that recently Kuznetsov breather was observed experimentally in optical fibres.
\begin{equation}\label{B}		
 \mathrm{(b)}~~ \{\mathrm{Re}[\lambda ] = 0,~~\mathrm{Im}[\lambda ] = A\} , ~~ \{\alpha = \pi /2,\quad \xi = 0\}.
\end{equation}		
In this case, expression becomes a vector called Peregrine breather. The Peregrine breather is localized in both space and time, is the limiting case of the breather.  The Peregrine breather describes a wave event: a huge wave crest (with an amplitude three times that of the background) suddenly appears at a specific time (t = 0) and a specific position (x = 0), and then quickly disappears. This characteristic of ``coming from nothing and then returning to nothing'' makes it a standard model for studying extreme events.
In the limit \( \xi \to 0 \), \( T \to \infty \) the Kuznetsov solution turns into the Peregrine breather.
\begin{equation}\label{C}			
 \mathrm{(c)}~~ \{\mathrm{Re}[\lambda ] = 0,~~ \mathrm{Im}[\lambda ]< A\} , ~~ \{\alpha \neq \pi /2,~~ \xi = 0\}.
\end{equation}
In this case, expression becomes a vector analogue of the Akhmediev excitation
\begin{align*}
		\begin{aligned}
			\mathrm{Re}\,q_{1,2}^{I} &= e^{2\mathrm{i}A^{2}t}\left[a_{1,2} - \frac{2a_{1,2}\sin\alpha\bigl[\cos(2v_{I}) + \cosh(2u_{I})\sin\alpha\bigr]}{\cosh(2u_{I}) + \sin\alpha\cos(2v_{I})}\right],\\
			\mathrm{Im}\,q_{1,2}^{I} &= e^{2\mathrm{i}A^{2}t}\left[\frac{2a_{1,2}\sin\alpha\sinh(2u_{I})\cos\alpha}{\cosh(2u_{I}) + \sin\alpha\cos(2v_{I})}\right].
		\end{aligned}
\end{align*}
The Akhmediev breather describes a physical process: a spatially periodically modulated wave experiences a large fluctuation in time (reaching its maximum around t = 0) and then returns to its original state. This is an exact solution for the nonlinear phase modulation instability.
The three types of breathers do not exist in isolation; rather, they are interconnected through continuous variations of parameters.
When  $\xi \rightarrow 0$, the time period of the Kuznetsov breather tends to infinity, and in the limiting case, the Peregrine breather is obtained.
When $\alpha \rightarrow \pi/2$, the spatial period of the Akhmediev breather tends to infinity, and in the limiting case, the Peregrine breather is also obtained.
The Peregrine breather serves as the ``confluence point'' of the Kuznetsov and Akhmediev breathers, making it the most special member of the breather family.

The continuous transitions among the above three limits are preserved, but the detailed evolution of types II and III across these limits is qualitatively different from the scalar or Manakov pictures. Importantly, \(\epsilon\) influences only the temporal periods, velocities, and phase accumulations; it does not alter the existence or the qualitative classification of the limiting regimes.

\noindent
{\rotatebox{0}{\includegraphics[width=4.8cm,height=3.2cm,angle=0]{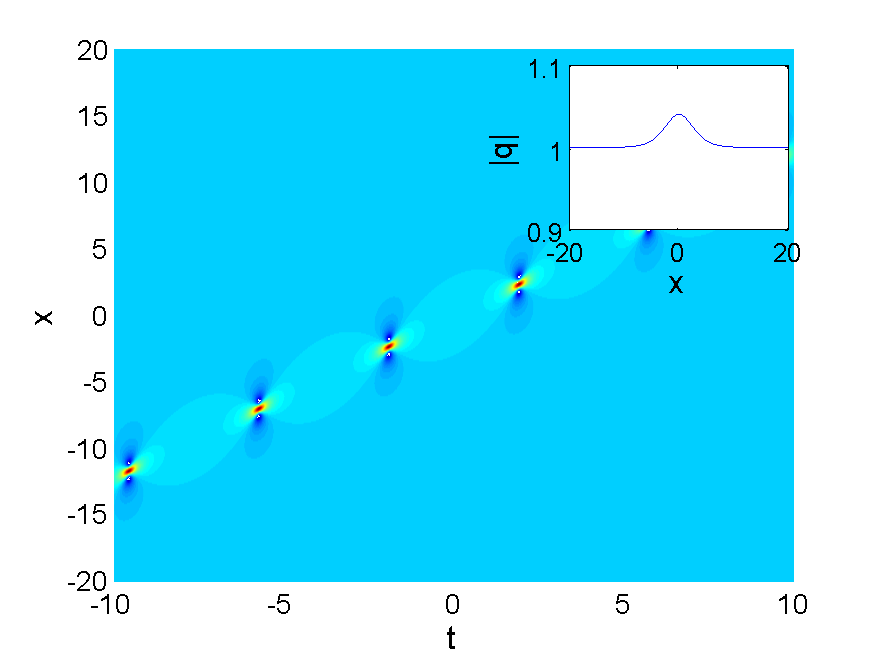}}}
{\rotatebox{0}{\includegraphics[width=4.8cm,height=3.2cm,angle=0]{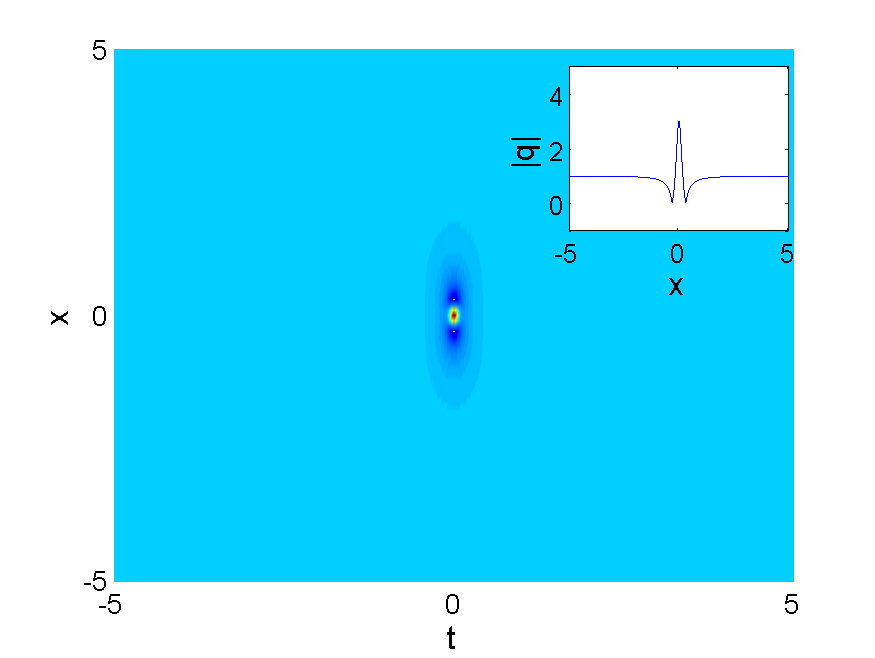}}}
{\rotatebox{0}{\includegraphics[width=4.8cm,height=3.2cm,angle=0]{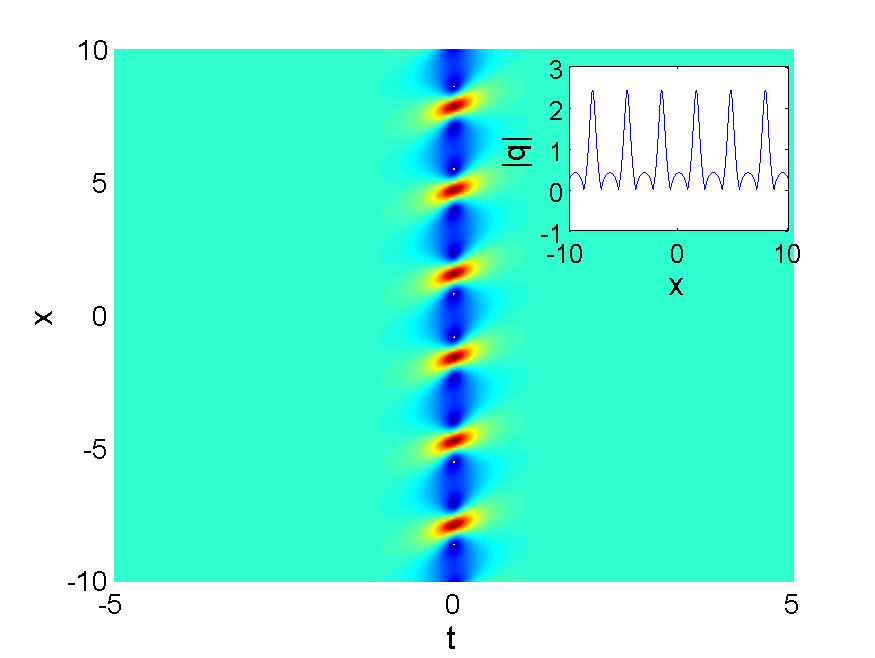}}}\\
$~~~~~~~~~~~~~~~~~~(\textbf{a})
~~~~~~~~~~~~~~~~~~~~~~~~~~~~~~~~~~~(\textbf{b})
~~~~~~~~~~~~~~~~~~~~~~~~~~~~~~~~~~~(\textbf{c})$\\

\noindent { \small \textbf{Figure 4.} Amplitude distributions of vector breather of type I, see Eq.\eqref{LX4}, which can be obtained from solutions of scalar Hiorta equation, with different parameter regions for \eqref{A}, \eqref{B}, \eqref{C}. (a) Kuznetsov solution, (b) Peregrine solution, and (c) Akhmediev solution.\\}
\\

\noindent
{\rotatebox{0}{\includegraphics[width=4.8cm,height=3.2cm,angle=0]{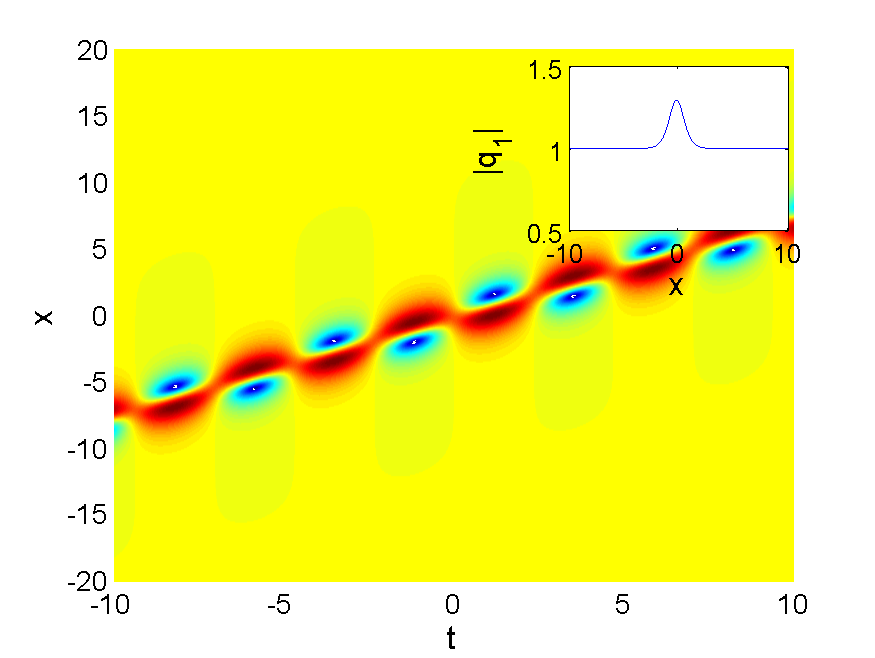}}}
{\rotatebox{0}{\includegraphics[width=4.8cm,height=3.2cm,angle=0]{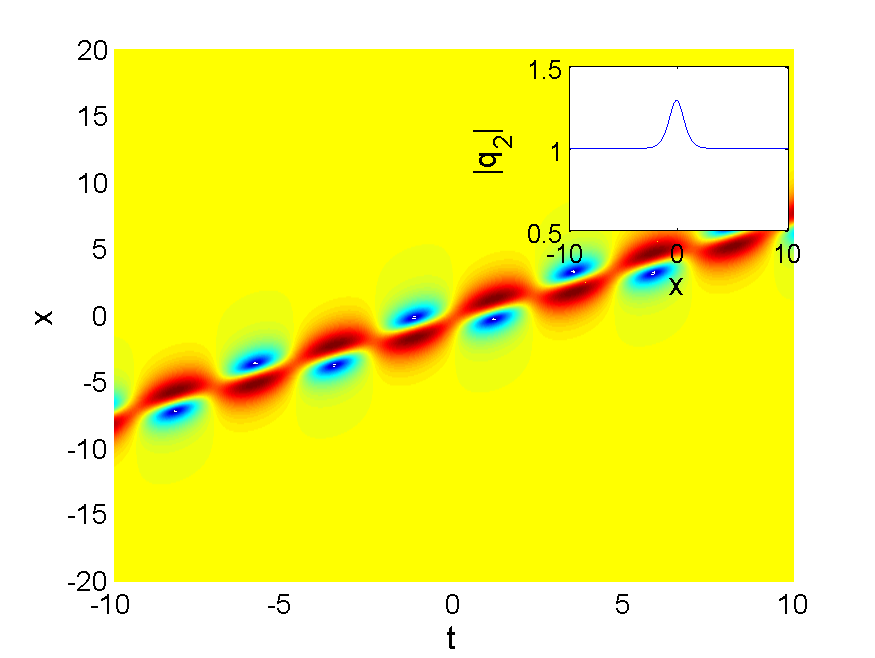}}}
{\rotatebox{0}{\includegraphics[width=4.8cm,height=3.2cm,angle=0]{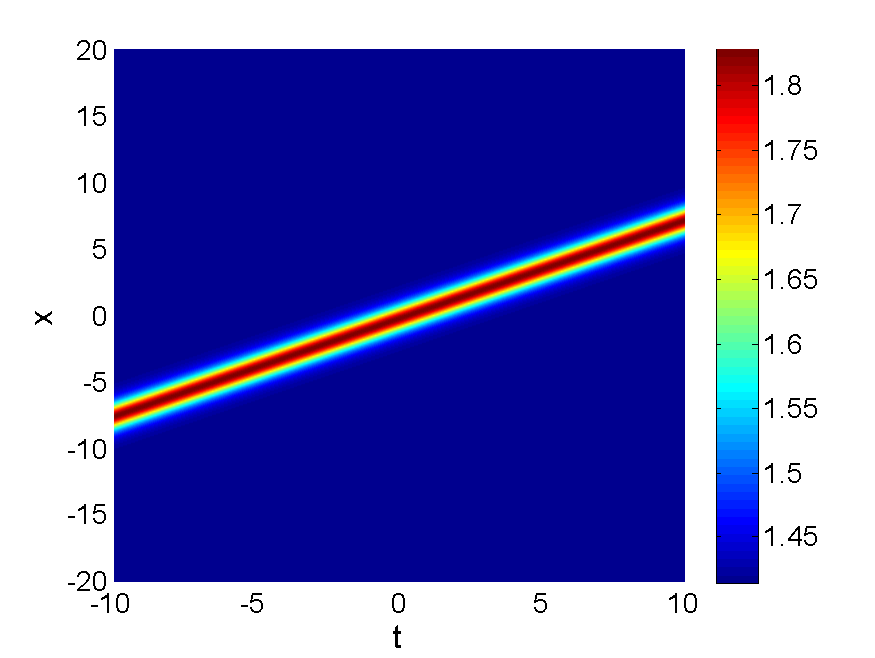}}}\\
$~~~~~~~~~~~~~~~~~~(\textbf{a1})
~~~~~~~~~~~~~~~~~~~~~~~~~~~~~~~~~(\textbf{a2})
~~~~~~~~~~~~~~~~~~~~~~~~~~~~~~(\textbf{a3})$\\

\noindent
{\rotatebox{0}{\includegraphics[width=4.8cm,height=3.2cm,angle=0]{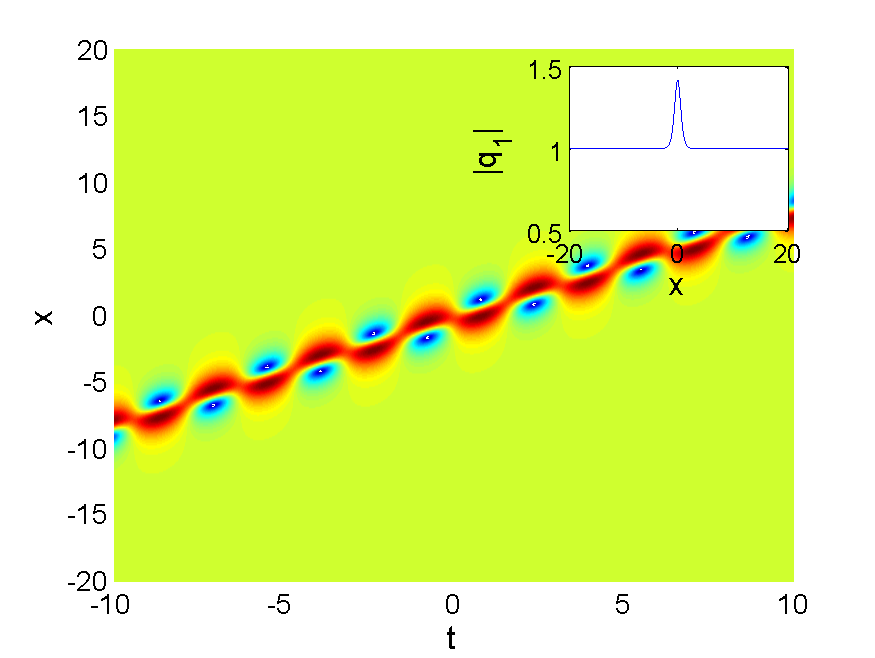}}}
{\rotatebox{0}{\includegraphics[width=4.8cm,height=3.2cm,angle=0]{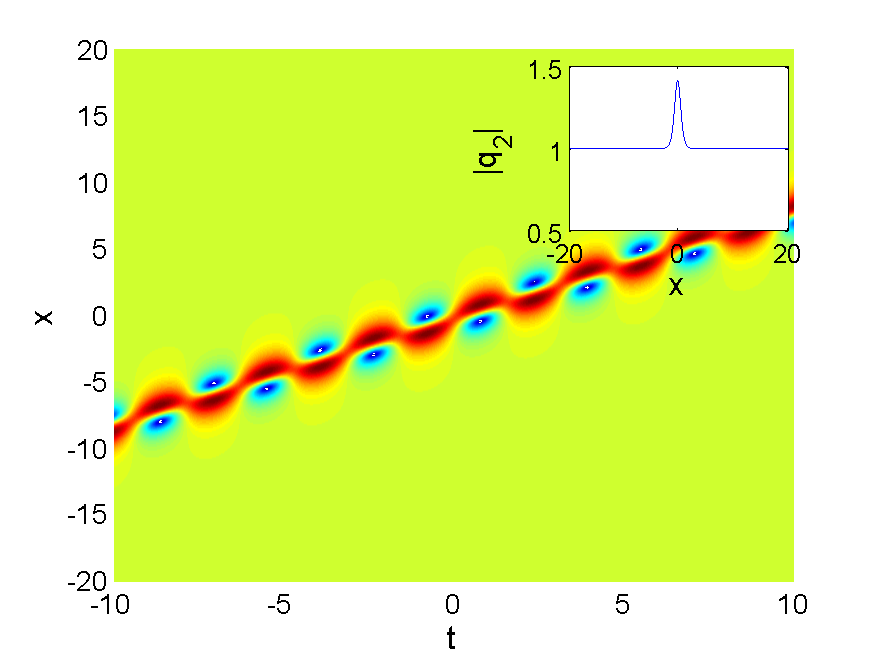}}}
{\rotatebox{0}{\includegraphics[width=4.8cm,height=3.2cm,angle=0]{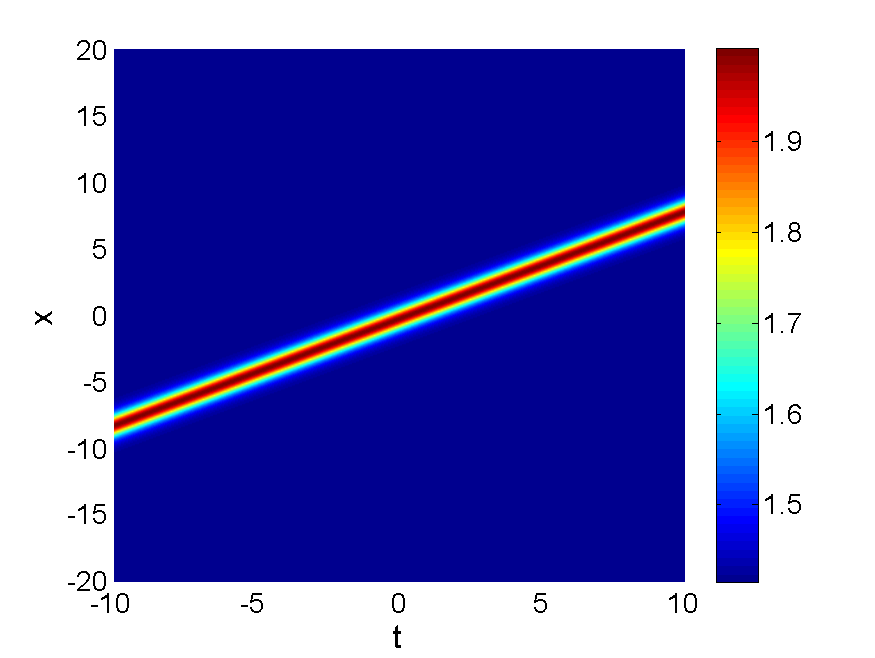}}}\\
$~~~~~~~~~~~~~~~~~~(\textbf{b1})
~~~~~~~~~~~~~~~~~~~~~~~~~~~~~~~~~(\textbf{b2})
~~~~~~~~~~~~~~~~~~~~~~~~~~~~~~(\textbf{b3})$\\

\noindent
{\rotatebox{0}{\includegraphics[width=4.8cm,height=3.2cm,angle=0]{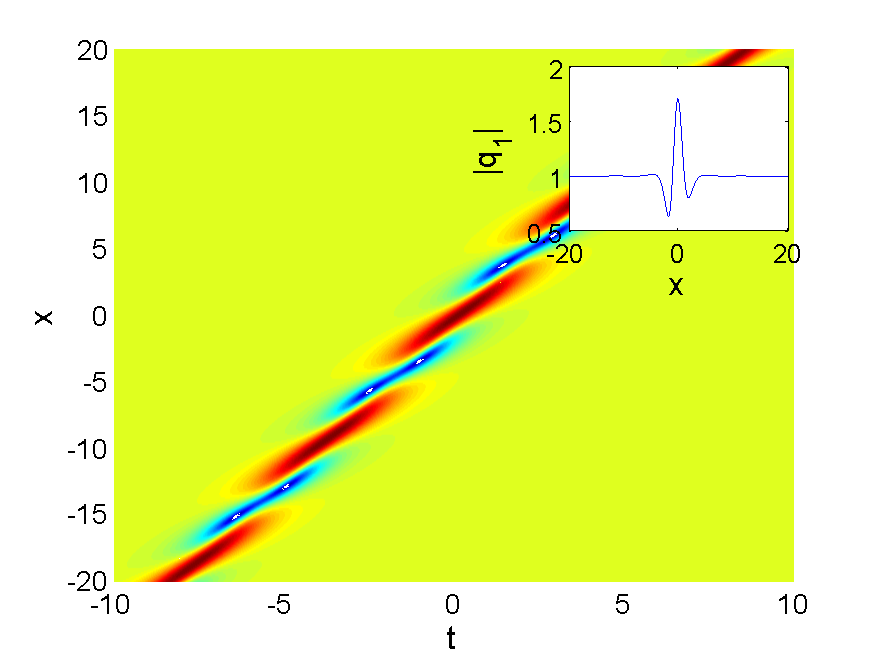}}}
{\rotatebox{0}{\includegraphics[width=4.8cm,height=3.2cm,angle=0]{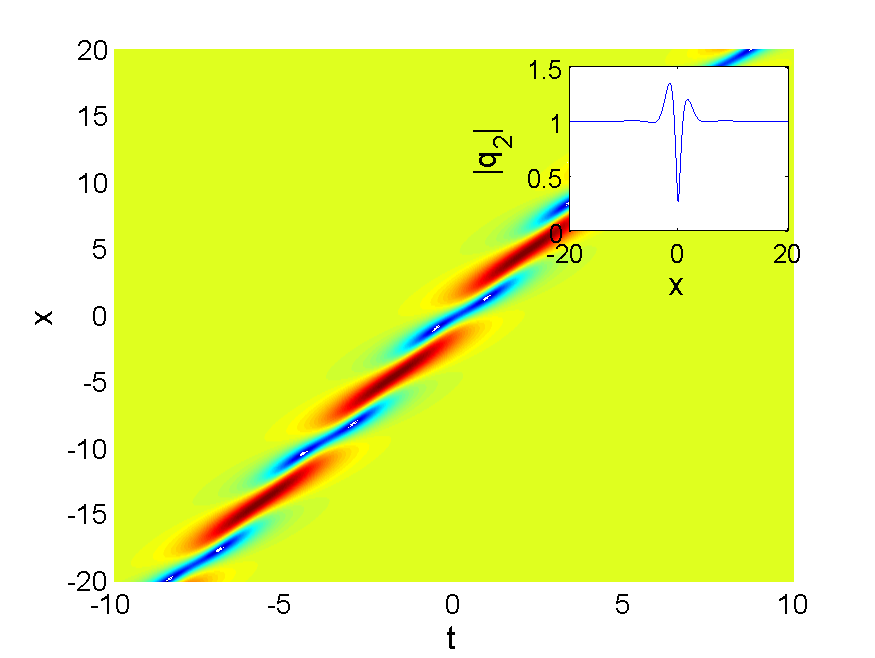}}}
{\rotatebox{0}{\includegraphics[width=4.8cm,height=3.2cm,angle=0]{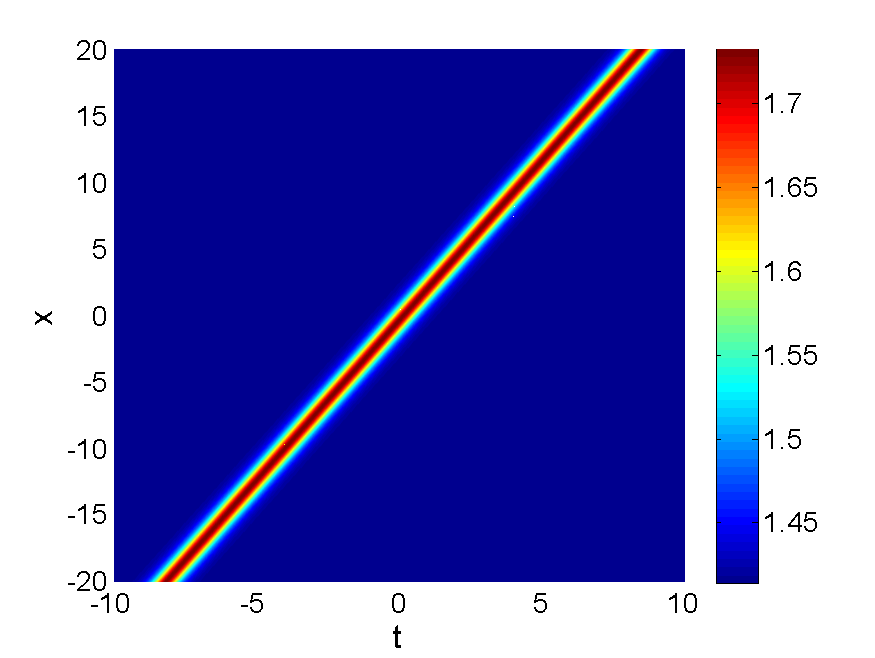}}}\\
$~~~~~~~~~~~~~~~~~~(\textbf{c1})
~~~~~~~~~~~~~~~~~~~~~~~~~~~~~~~~~(\textbf{c2})
~~~~~~~~~~~~~~~~~~~~~~~~~~~~~~(\textbf{c3})$\\

\noindent { \small \textbf{Figure 5.} Amplitude distributions of vector breathers of the type II with spectral parameters defined by Equations \eqref{A},\eqref{B},\eqref{C}. (a) Spectral parameter belongs to the Kuznetsov region, see Eq.\eqref{A}. The corresponding solution is a standing dark-bright oscillating breather. (b) Spectral parameter belongs to the Peregrine region, see Eq.\eqref{B}. (c) Spectral parameter belongs to the Akhmediev region, see Eq,\eqref{C}. From left to right, \(|q_1|\), \(|q_2|\), and \(\sqrt{|q_1|^2+|q_2|^2}\). Other parameters are defined as \(a_1 = 1, ~a_2 = 1, ~\epsilon = 0.1 \).\\}
\\

\noindent

Figure 5 shows a type II breather wave, whose prominent feature is that the two components do not satisfy linear scaling, forming ``dark-bright'' composite structure. Figure 5(a) belongs to the Kuznetsov region, its space-time structure is a standing dark-light breath wave pair. Both components are spatially localized, but their local shapes are different. One component  presents an amplitude depression (dark breather wave), while the other component  presents an amplitude bulge (bright breather wave), and the two coincide perfectly in space. At the same time, it exhibits periodic oscillations in time. Different from type I, this structure cannot be obtained through simple scalar transformation and is a direct product of vector nonlinear coupling. The polarization state on the background is reversed on both sides of the wave packet. Figure 5(b) is the value of type II breather wave at \(  \xi = 0 \), but no mathematical degradation has occurred. The waveform appears as a common, localized wave packet with a certain broadening in space and time. Although it also rises from the background and then falls, its maximum amplitude is far less than 9 times the background intensity of the Peregrine strange wave, and the waveform is asymmetrical. This figure clearly conveys to readers that there is no Peregrine type II strange wave in the vector system. Even if the spectral parameters are taken at the ``strange wave point'', the wave field performance is still ``moderat''. The spatio  temporal structure in Figure 5(c) is a localized wave packet moving to the right (or left) at a non-zero group velocity. The waveform maintains its locality during propagation and does not have the spatial periodicity of the Akhmediev breather. This means that the ``spatial periodic modulation'' in the scalar case completely disappears in the type II vector case, degenerating into an isolated moving wave packet. The tail of the wave packet decays exponentially.

\noindent
{\rotatebox{0}{\includegraphics[width=4.8cm,height=3.2cm,angle=0]{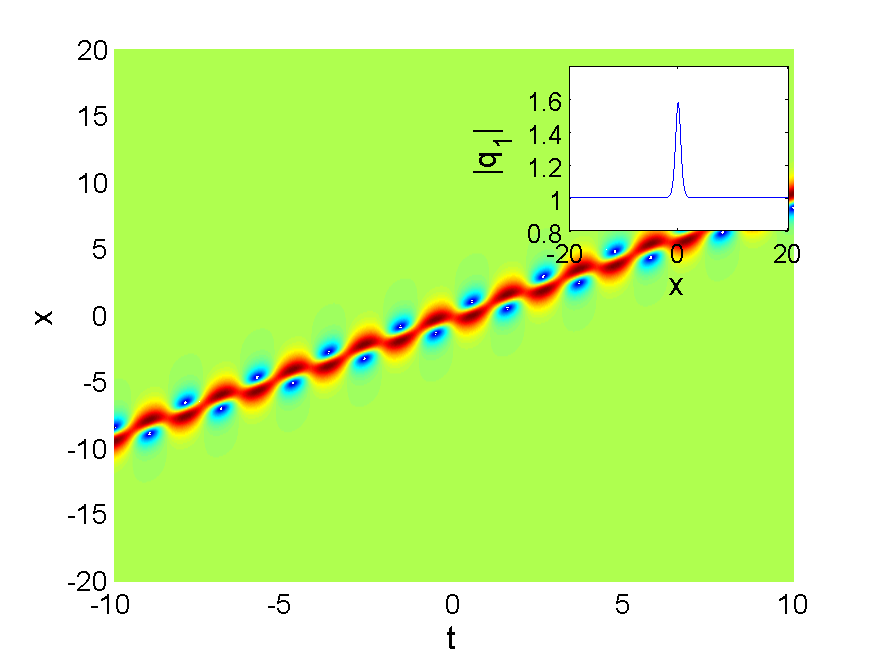}}}
{\rotatebox{0}{\includegraphics[width=4.8cm,height=3.2cm,angle=0]{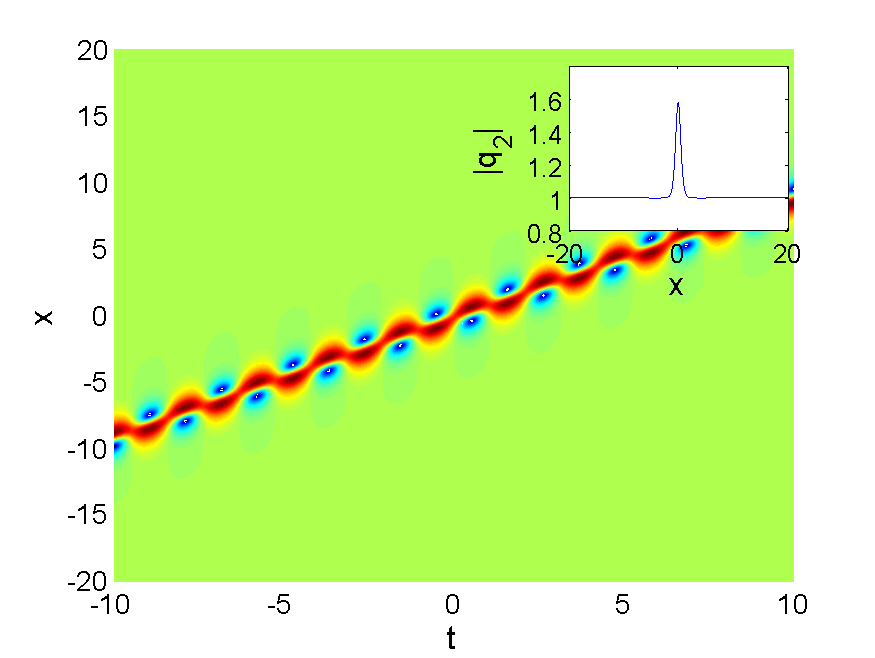}}}
{\rotatebox{0}{\includegraphics[width=4.8cm,height=3.2cm,angle=0]{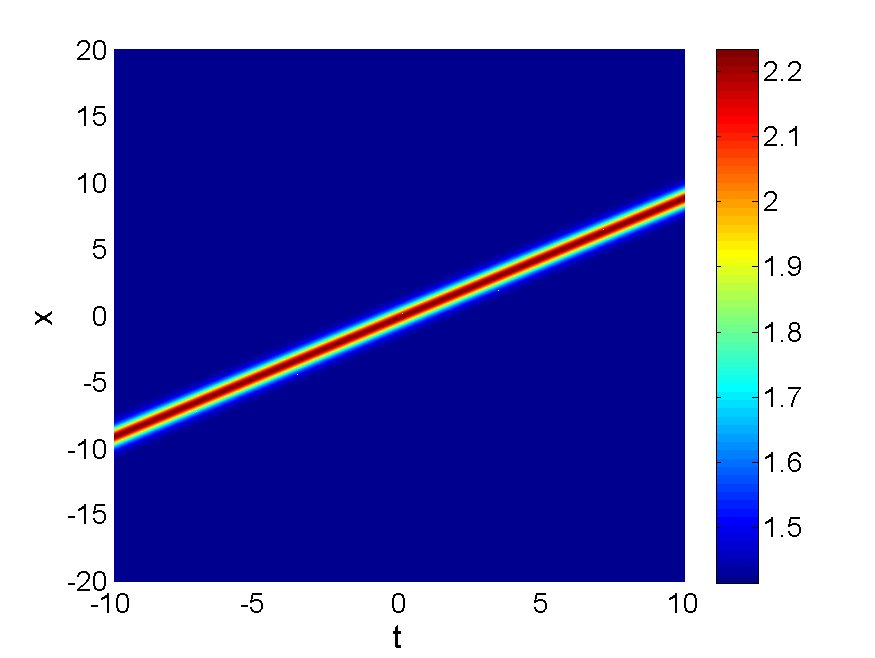}}}\\
$~~~~~~~~~~~~~~~~~~(\textbf{a1})
~~~~~~~~~~~~~~~~~~~~~~~~~~~~~~~~~(\textbf{a2})
~~~~~~~~~~~~~~~~~~~~~~~~~~~~~~(\textbf{a3})$\\

\noindent
{\rotatebox{0}{\includegraphics[width=4.8cm,height=3.2cm,angle=0]{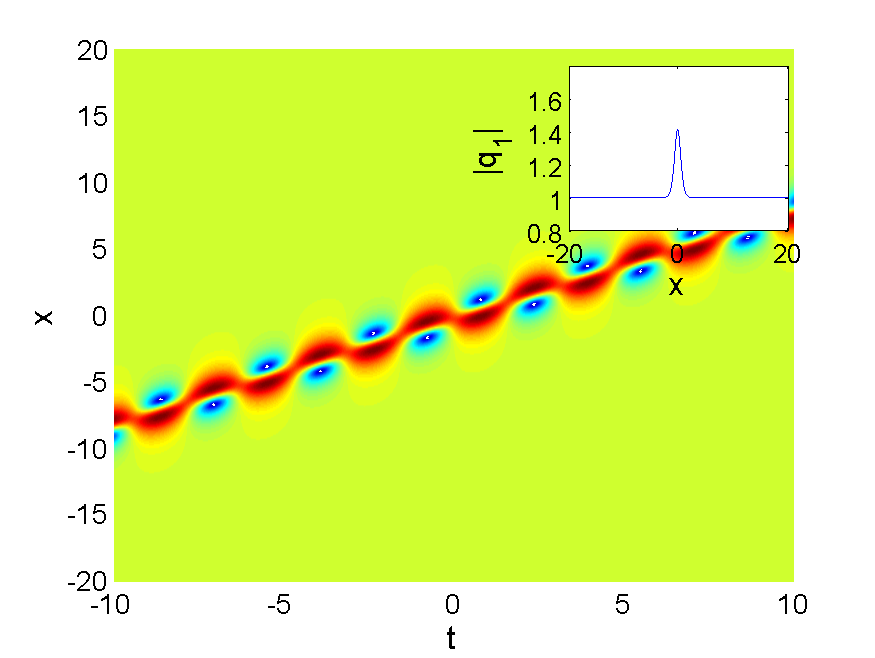}}}
{\rotatebox{0}{\includegraphics[width=4.8cm,height=3.2cm,angle=0]{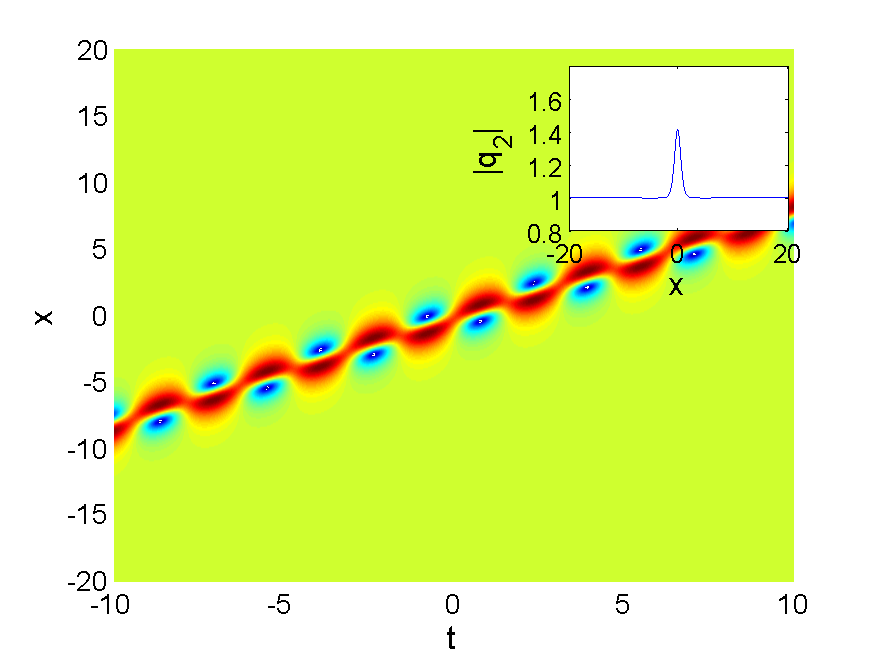}}}
{\rotatebox{0}{\includegraphics[width=4.8cm,height=3.2cm,angle=0]{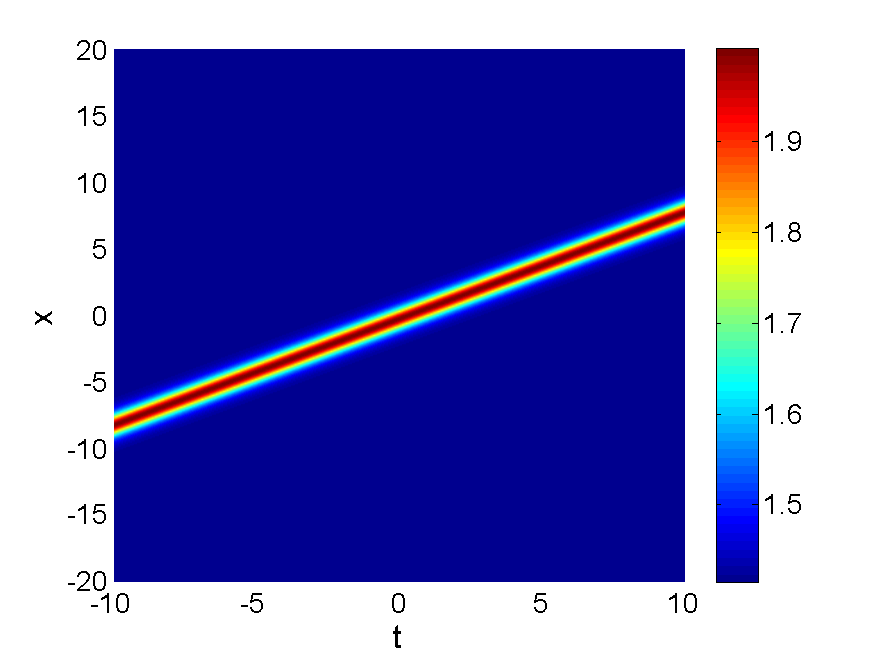}}}\\
$~~~~~~~~~~~~~~~~~~(\textbf{b1})
~~~~~~~~~~~~~~~~~~~~~~~~~~~~~~~~~(\textbf{b2})
~~~~~~~~~~~~~~~~~~~~~~~~~~~~~~(\textbf{b3})$\\

\noindent
{\rotatebox{0}{\includegraphics[width=4.8cm,height=3.2cm,angle=0]{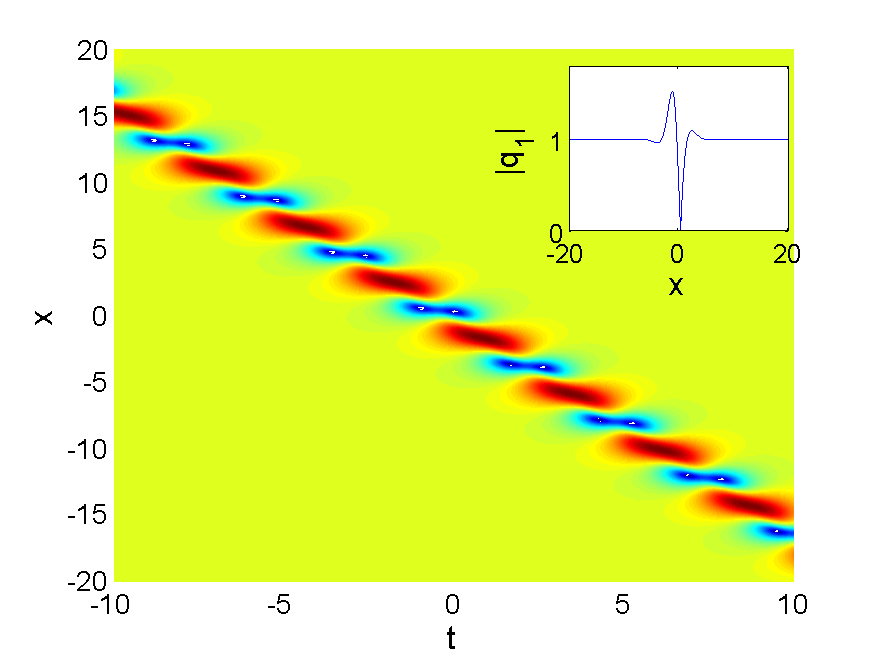}}}
{\rotatebox{0}{\includegraphics[width=4.8cm,height=3.2cm,angle=0]{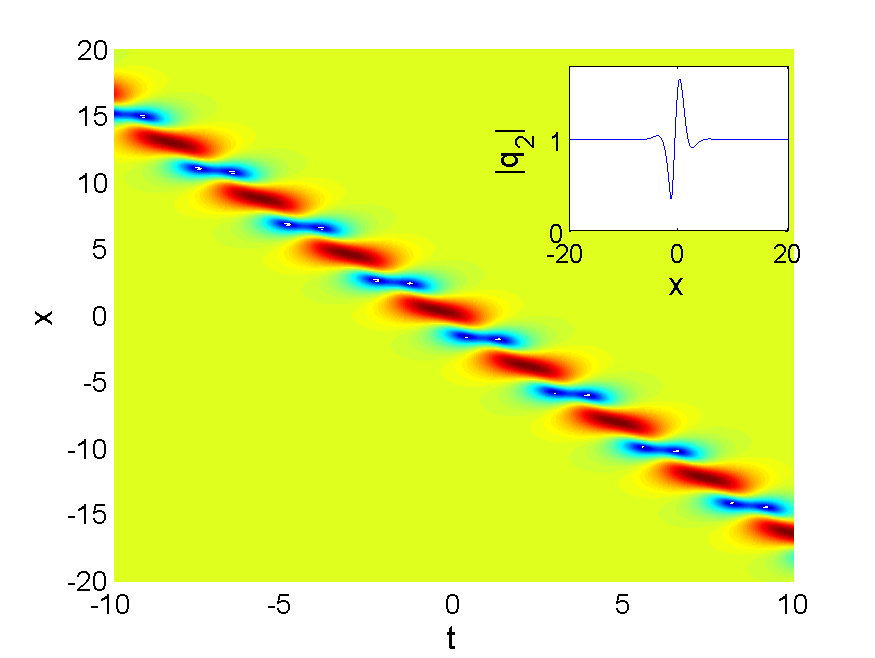}}}
{\rotatebox{0}{\includegraphics[width=4.8cm,height=3.2cm,angle=0]{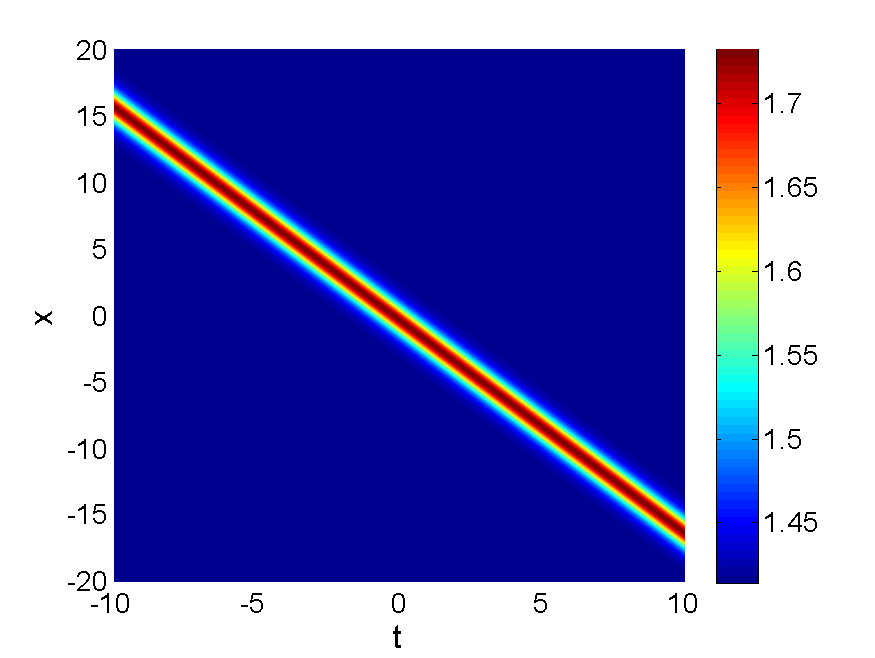}}}\\
$~~~~~~~~~~~~~~~~~~(\textbf{c1})
~~~~~~~~~~~~~~~~~~~~~~~~~~~~~~~~~(\textbf{c2})
~~~~~~~~~~~~~~~~~~~~~~~~~~~~~~(\textbf{c3})$\\

\noindent { \small \textbf{Figure 6.} Amplitude distributions of vector breathers of type III with spectral parameters defined by Eq.\eqref{A},\eqref{B},\eqref{C}. (a) Spectral parameter belongs to the Kuznetsov region, see Eq.\eqref{A}. The corresponding solution is a standing dark-bright oscillating breather. (b) Spectral parameter belongs to the Peregrine region, see Eq.\eqref{B}. (c) Spectral parameter belongs to the Akhmediev region, see Eq.\eqref{C}. From left to right, \(|q_1|\), \(|q_2|\), and \(\sqrt{|q_1|^2+|q_2|^2}\). Other parameters are defined as \(a_1 = 1, ~a_2 = 1, ~\epsilon = 0.1 \).\\}
\\

\noindent

Figure 6 shows the type III breather wave. Under the same spectral parameters, the direction, phase change and characteristic size of type III are completely opposite to those of type II, but their core physical mechanisms are the same. The spatio temporal structure in Figure 6(a) is also a standing dark-bright breather pair, which is consistent with Type II (Figure 5a) in structure, but there are mirror differences in details. For exactly the same \(\alpha\) and \(\xi\), the characteristic length of Type III is smaller than that of Type II, and the group velocity of Type III is different in magnitude and opposite in direction to that of Type II.The spatio temporal structure shown in Figure 6(b) is the same as that of Type II. There is no degeneration here, and there is no rational rogue wave event. It presents as a localized wave packet, and the evolution of the waveform in space time is smooth. The amplitude changes of the two components are gentle, only slightly affected by local perturbations, and there are no extreme wave peaks. The spatio temporal structure shown in Figure 6(c) is a moving localized breather wave packet, which is similar to that in Figure 5(c). However, it is worth noting that when the spectral parameter is located here (i.e., on the branch cut), Type II and Type III solutions merge into the same category. The wave packet propagates along the space at a constant speed, and its waveform is highly symmetric. It does not have any spatial periodicity, nor does it oscillate repeatedly over time. It is simply a stable, moving localized nonlinear excited state.

\section{Resonance interactions of vector breathers}\label{sec4}

In this section we investigate a distinctive class of inelastic interactions in which two vector breathers merge into a single one, or conversely, a single breather splits into two, provided that the characteristic wave numbers and frequencies satisfy exact matching conditions. Such resonant processes, originally predicted for three-wave systems without dispersion, have later been found in integrable multicomponent dispersive models, including the Manakov system. Here we demonstrate that the same phenomenon occurs in the coupled Hirota equation \eqref{HT}, and that the higher-order parameter \(\epsilon\) modifies the propagation velocities and oscillation periods of the participating breathers without destroying the resonance conditions themselves.

 This phenomenon holds significant value for understanding the long-term evolution and energy redistribution mechanisms of nonlinear systems. Zakharov and Manakov \cite{Zakharov1975} first theoretically predicted resonant interactions in the nondispersive three-wave model of nonlinear media. This phenomenon manifests as soliton fusion in three-wave packets within nonlinear media, where two solitons merge to form a new excited state (i.e., a pump soliton), or an initial soliton decays into secondary solitons. The causes of this phenomenon \cite{Kaup1976} were further explored. The physical basis of soliton fusion was initially attributed to the absence of dispersion in the system. Subsequently, research shifted toward multicomponent systems with dispersion. Taking the Manakov system as a milestone, researchers employed methods such as the dressing method and bilinear approach to successfully construct multi-breather wave solutions. They discovered that special configurations of poles (spectral parameters) in the complex plane such as merging could lead to fusion or decay, i.e., resonant interactions. This explicitly challenged the earlier notion that ``resonance requires the absence of dispersion'' Barashenkov et al. identified resonance phenomena between subluminal and superluminal solitons in the sine-Gordon \cite{Barashenkov1988,Barashenkov1993} model. Subsequently, a series of generalized models (e.g., multicomponent NLSE families) were confirmed to exhibit similar inelastic processes.
		
The resonant interaction of three vector breathers is described by the first-order solution \eqref{YJJ2}, when all the integration constants $C_0$, $C_1$, and $C_2$ are nonzero. In the following analysis, we firstly consider the situation when the eigenvalue is of general type, see Eq.\eqref{W}, and after that, we give the particular choices \eqref{A} and \eqref{C}.

Figure 7a illustrates the space-time pictures for the modulus of the components $q_1$ and $q_2$ under the condition $\xi>0, ~\pi/2 > \alpha > 0$. In Figure 7a, we choose a typical parameter set $a_1=1, ~a_2=1,~ \alpha=\pi/6, ~\xi=1/5, ~C_0 = C_1 = C_2 = 1 $. The intermediate region describes the fusion process (\(\mathrm{I} + \mathrm{II} \rightarrow \mathrm{III}\)). The fusion of the breathers then occurs at $x = 0,~ t = 0$. Breather I, II, and III have the respective characteristics $u_{I} = u = \mathrm{const}$, $u_{II} = u_{0} - u = \mathrm{const}$, and $u_{III} = u_{0} + u = \mathrm{const}$. We see that the characteristics $u_{I} = u = \mathrm{const}$ and $u_{II} = u_{0} - u = \mathrm{const}$ terminate at the instant of their intersection and are then replaced with the characteristic $u_{III} = u_{0} + u = \mathrm{const}$. When the spectral parameter lies in the range $\pi > \alpha > \pi/2$, the time arrow reverses, giving the decay process (\( \mathrm{III} \rightarrow \mathrm{I} + \mathrm{II}\)). Thus, the same algebraic solution describes both fusion and decay, depending on the sign of \(\cos\alpha\).

Then, we show the particular choices \eqref{A} and \eqref{C} in Figure 7b, Figure 7c, respectively. In Figure 7b, we choose the parameter values $a_1=1,~ a_2=1,~ \alpha = \pi/2, ~\xi = 1/5, ~C_0 = C_ 1 = C_2 = 1 $. From the figure, we can see that the fusion of the breathers occurs at $x = 0, ~t = 0$.  In Figure 7c, we choose the parameter values $a_1=1, ~a_2=1,~ \alpha = \pi/4, ~\xi = 0, ~C_0 = C_1 = C_2 = 1 $. From the figure, we still see that the fusion of the breathers occurs at $x = 0, ~t = 0$.

A crucial observation is that the expressions \eqref{LX4}, \eqref{LX24}, and \eqref{LX34} that the sum of the wave numbers and frequencies of the type I and II breathers are precisely equal to the wave number and frequency of the type III breather
\begin{align}\label{R43}
&k_{\mathrm{I}} + k_{\mathrm{II}} = k_{\mathrm{III}},\notag\\
&\omega_{\mathrm{I}} + \omega_{\mathrm{II}} = \omega_{\mathrm{III}}.
\end{align}

We conclude that the fusion of breathers I and II into a breather III represents a resonance process, because the wave numbers and frequencies of the type I and II breathers satisfy the standard resonance conditions.

\noindent
{\rotatebox{0}{\includegraphics[width=4.8cm,height=3.2cm,angle=0]{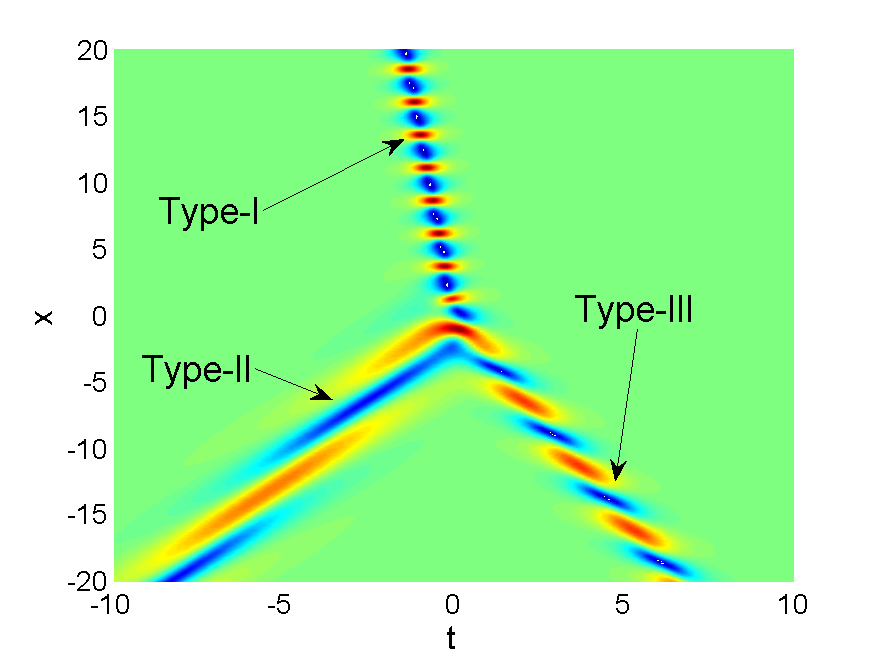}}}
{\rotatebox{0}{\includegraphics[width=4.8cm,height=3.2cm,angle=0]{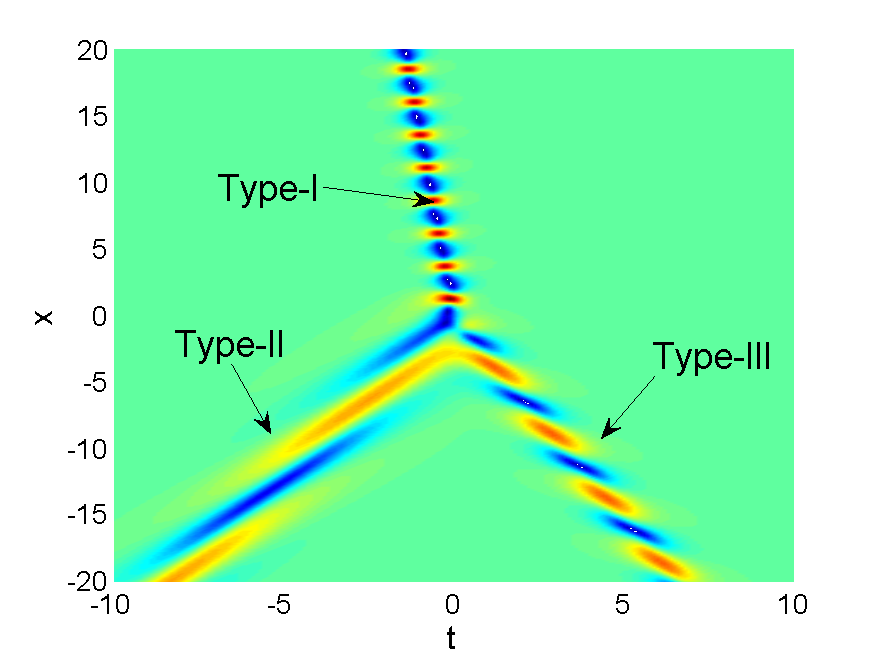}}}
{\rotatebox{0}{\includegraphics[width=4.8cm,height=3.2cm,angle=0]{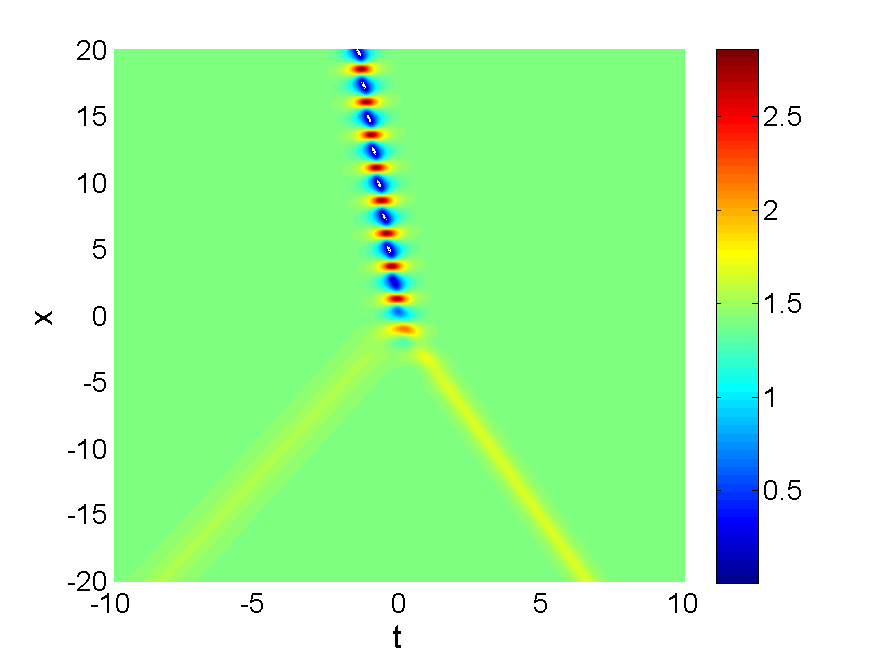}}}\\
$~~~~~~~~~~~~~~~~~~(\textbf{a1})
~~~~~~~~~~~~~~~~~~~~~~~~~~~~~~~~~(\textbf{a2})
~~~~~~~~~~~~~~~~~~~~~~~~~~~~~~(\textbf{a3})$\\

\noindent
{\rotatebox{0}{\includegraphics[width=4.8cm,height=3.2cm,angle=0]{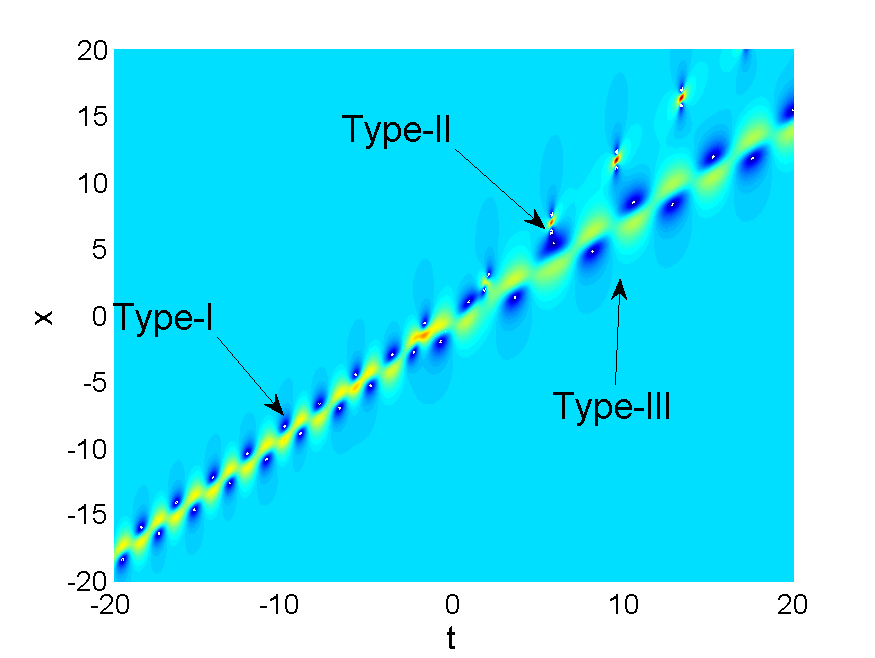}}}
{\rotatebox{0}{\includegraphics[width=4.8cm,height=3.2cm,angle=0]{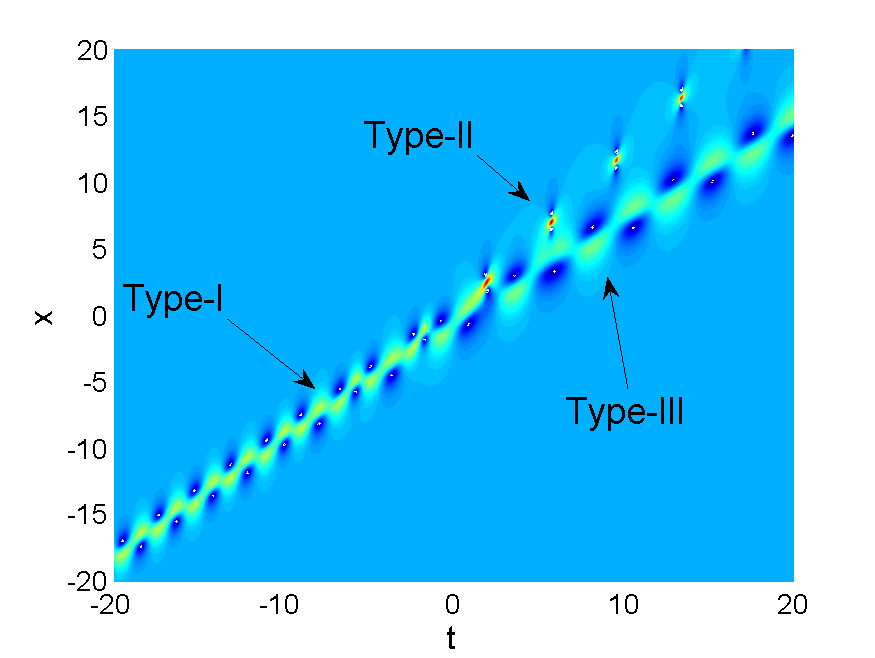}}}
{\rotatebox{0}{\includegraphics[width=4.8cm,height=3.2cm,angle=0]{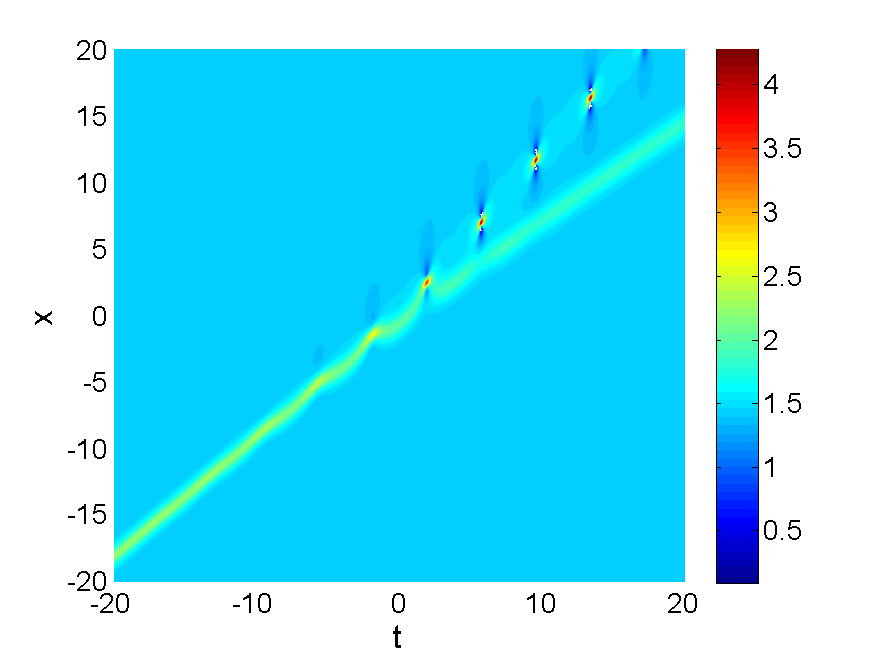}}}\\
$~~~~~~~~~~~~~~~~~~(\textbf{b1})
~~~~~~~~~~~~~~~~~~~~~~~~~~~~~~~~~(\textbf{b2})
~~~~~~~~~~~~~~~~~~~~~~~~~~~~~~(\textbf{b3})$\\

\noindent
{\rotatebox{0}{\includegraphics[width=4.8cm,height=3.2cm,angle=0]{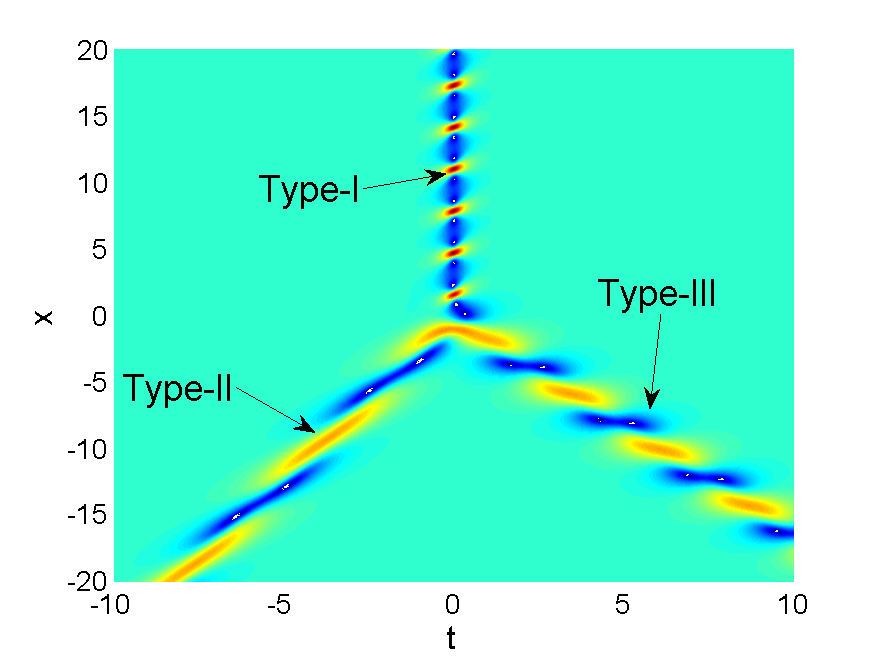}}}
{\rotatebox{0}{\includegraphics[width=4.8cm,height=3.2cm,angle=0]{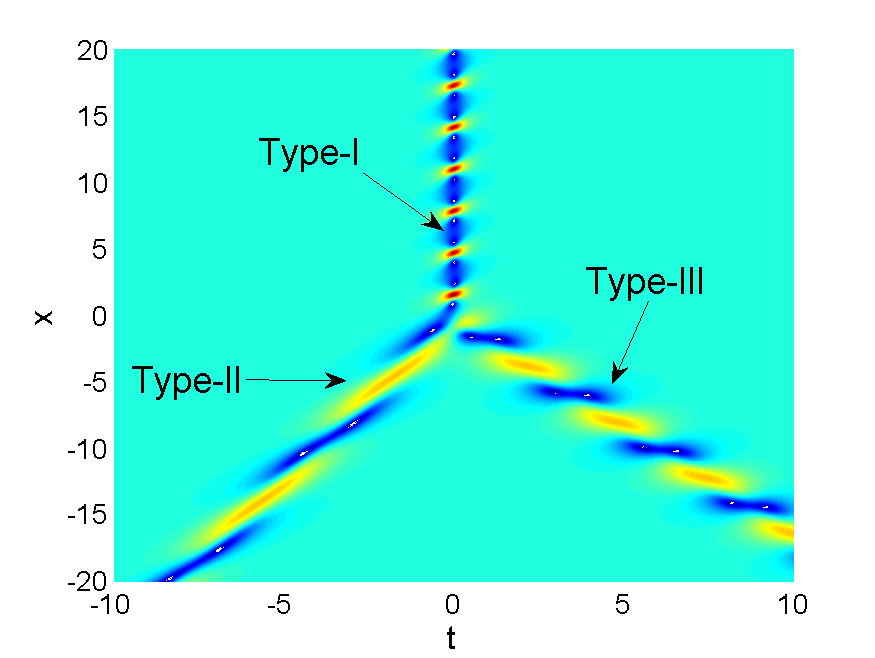}}}
{\rotatebox{0}{\includegraphics[width=4.8cm,height=3.2cm,angle=0]{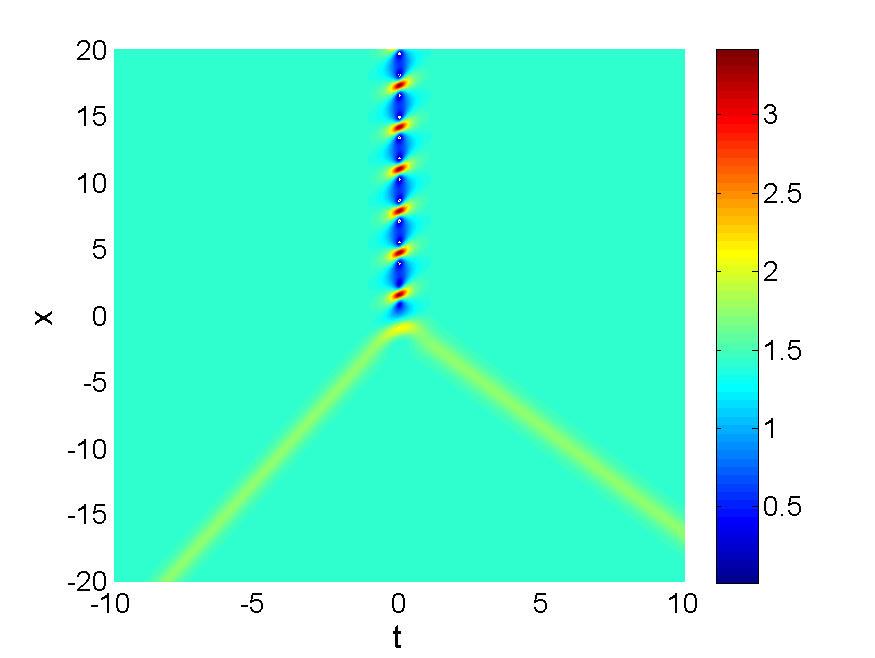}}}\\
$~~~~~~~~~~~~~~~~~~(\textbf{c1})
~~~~~~~~~~~~~~~~~~~~~~~~~~~~~~~~~(\textbf{c2})
~~~~~~~~~~~~~~~~~~~~~~~~~~~~~~(\textbf{c3})$\\

\noindent { \small \textbf{Figure 7.} Resonance interaction (\( \text{I} + \text{II} \rightarrow \text{III} \)) of vector breathers described by single-pole solution \eqref{YJJ2} with \(C_{0} = C_{1} = C_{2} = 1 \). And resonance interaction of vector breathers with spectral parameters defined by Eq.\eqref{A}, \eqref{C}. (a) In the general situation. (b) Spectral parameter belongs to the Kuznetsov region, see Eq.\eqref{A}. (c) Spectral parameter belongs to the Akhmediev region, see Eq.\eqref{C}. From left to right, amplitude distributions of different component, \(|q_1|\), \(|q_2|\), and \(\sqrt{|q_1|^2+|q_2|^2}\). Other parameters are defined as \(a_1 = 1, ~a_2 = 1, ~\epsilon = 0.1 \).\\}
\\

\noindent

Note that these equalities are not a consequence of the linear dispersion laws but rather follow from the nonlinear breather parameters. Moreover, the wave-number relations do not involve \(\epsilon\)), because the spatial phase factors in the dressing method are independent of the higher-order coefficient. The frequency relations, on the other hand, are automatically satisfied due to the algebraic structure of the solution, and they hold for any value of \(\epsilon\)). In practice, \(\epsilon\)) only renormalises the group velocities and the internal oscillation frequencies of the breathers, thereby shifting the spatiotemporal position of the merging point and altering the local dynamics near the collision, but it never prevents the resonance from occurring.

\section{Second-order breather solutions and its dynamics}\label{sec5}

In this section, we now turn to the two-eigenvalue solutions of the coupled Hirota equation \eqref{HT}, obtained from the general formula Eq.\eqref{YJ2} with \(N = 2\). These solutions describe a rich variety of interaction scenarios between vector breathers, including elastic collisions, resonant fusion/decay patterns, and mixed regimes. The explicit form reads	
\begin{align}\label{TPJ1}	
q_{1} =e^{2\textrm{i}A^2 t}\left[ a_{1} + 2\frac{\widetilde{\mathbf{M}}_{12}}{\mathbf{M}} \right],\notag\\
q_{2} =e^{2\textrm{i}A^2 t}\left[ a_{2} + 2\frac{\widetilde{\mathbf{M}}_{13}}{\mathbf{M}}  \right],
\end{align}
where
\begin{equation*}\label{TPJ2}
			\begin{aligned}
&\widetilde{M}_{12} = \big[-m_2 q_{12}^* q_{11} |\mathbf{q}_2|^2 + n_2 q_{12}^* q_{21} (\mathbf{q}_1,\mathbf{q}_2^*)  + n_1 q_{22}^* q_{11} (\mathbf{q}_2,\mathbf{q}_1^*) - m_1 q_{22}^* q_{21} |\mathbf{q}_1|^2\big],\\
&\widetilde{M}_{13} = \big[-m_2 q_{13}^* q_{11} |\mathbf{q}_2|^2 + n_2 q_{13}^* q_{21} (\mathbf{q}_1,\mathbf{q}_2^*)+ n_1 q_{23}^* q_{11} (\mathbf{q}_2,\mathbf{q}_1^*) - m_1 q_{23}^* q_{21} |\mathbf{q}_1|^2\big],\\
&M =\big[m |\mathbf{q}_1|^2 |\mathbf{q}_2|^2 - n (\mathbf{q}_1^*,\mathbf{q}_2) (\mathbf{q}_2^*,\mathbf{q}_1)\big],
			\end{aligned}
\end{equation*}
and
\begin{equation*}\label{TPJ3}
		\begin{aligned}
&{q_{n1} = e^{\phi_n}C_{n1} + \textrm{i}r_n e^{-\phi_n}C_{n2},}\\
&{q_{n2} = \frac{1}{A} [-a_2e^{-\phi_{0n}}C_{n0} + a_1(r_n e^{\phi_n}C_{n1} - \textrm{i} e^{-\phi_n}C_{n2})],}\\
&{q_{n3} = \frac{1}{A} [a_1e^{-\phi_{0n}}C_{n0} + a_2(r_n e^{\phi_n}C_{n1} - \textrm{i} e^{-\phi_n}C_{n2})],}\\
&|\mathbf{q}_n|^2 = |e^{\phi_n}C_{n1} +  \textrm{i}r_n e^{-\phi_n}C_{n2}|^2 +|r_n e^{\phi_n}C_{n1} - \textrm{i} e^{-\phi_n}C_{n2}|^2 +|e^{-\phi_{0n}}C_{n0}|^2.
			\end{aligned}
\end{equation*}

Here, we define \(m_{1,2} \equiv 1 / (\lambda_{1,2} - \lambda_{1,2}^{*})\), \(n_{1,2} \equiv 1 / (\lambda_{1,2} - {\lambda}_{2,1}^{*})\), and \(m = m_{1}m_{2}\), \(n = n_{1}n_{2} \leq 0\). Through the parameterization \eqref{CSH} in the third section, the coefficients take the compact forms provided
\begin{equation*}\label{TPJ4}
			\begin{aligned}
&m_{i} = \frac{2|r_{i}|^{2}}{A(r_{i}^{*} - r_{i})(1 + |r_{i}|^{2})},\\
&n_{i} = \frac{2r_{i}r_{j}^{*}}{A(1 + r_{i}r_{j}^{*})(r_{j}^{*} - r_{i})},~~
	i = 1,2,~~ j = 3 - i;\\
&m = \frac{4|r_{1}r_{2}|^{2}}{A^{2}(1 + |r_{1}|^{2})(1 + |r_{2}|^{2})(r_{1} - r_{1}^{*})(r_{2} - r_{2}^{*})} = -\big[4A^{2}\sin \alpha_{1}\sin \alpha_{2}\cosh \xi_{1}\cosh \xi_{2}\big]^{-1},\\
&n = -\frac{4|r_{1}r_{2}|^{2}}{A^{2}|1 + r_{1}r_{2}^{*}|^{2}|r_{1} - r_{2}^{*}|^{2}}.
			\end{aligned}
\end{equation*}

Based on the front part of the paper and the relevant summary, the second-order solution describes a rich variety of vector breather interactions. This solution is fully determined by two eigenvalues and six integration constants \(C_{ij}\) (with \(i=1,2\); \(j=0,1,2\)). By appropriately choosing which of these constants vanish, solution \eqref{TPJ1} can exhibit various physical scenarios, including elastic collisions between two breathers of arbitrary types (I, II, or III), interactions between a breather and fusion/decay resonance wave patterns, and even combinations of two resonance wave patterns. Taking into account the three fundamental types of breathers (types I, II, and III) present in the system, solution \eqref{TPJ1} can uniformly describe more than ten different breather interaction scenarios, such as collisions, fusion, or decay processes between types I and II, or II and III. This result systematically generalizes the resonance behavior revealed by the one-pole solution and provides a concise and complete framework for studying the dynamics of vector breathers.	

Firstly, we suppose $C_{ij} \neq 0$, in the case, it describes the formation or fusion of the two pairs I+II/III of breathers in the coupled Hirota equation~\eqref{HT}. The first and second pairs depend on the complex numbers $\lambda_1$ and $\lambda_2$. At some point in time, each pair either collapses into a single breather (III or II) or, conversely, arises from it; breathers from different pairs, on the other hand, simply pass through each other, with the sole effects being additional shifts in their phases and centre locations.

\noindent
{\rotatebox{0}{\includegraphics[width=4.8cm,height=3.2cm,angle=0]{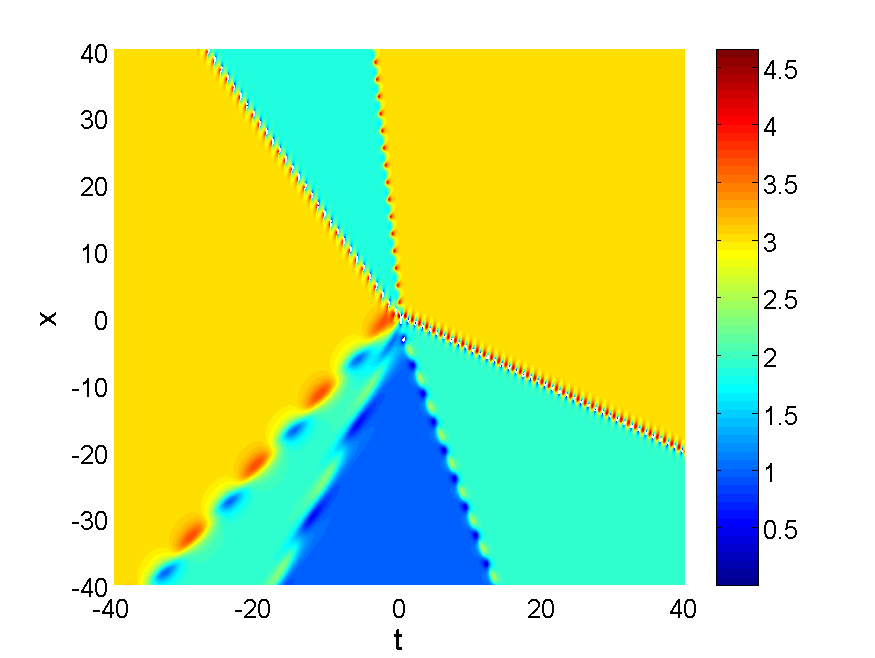}}}
{\rotatebox{0}{\includegraphics[width=4.8cm,height=3.2cm,angle=0]{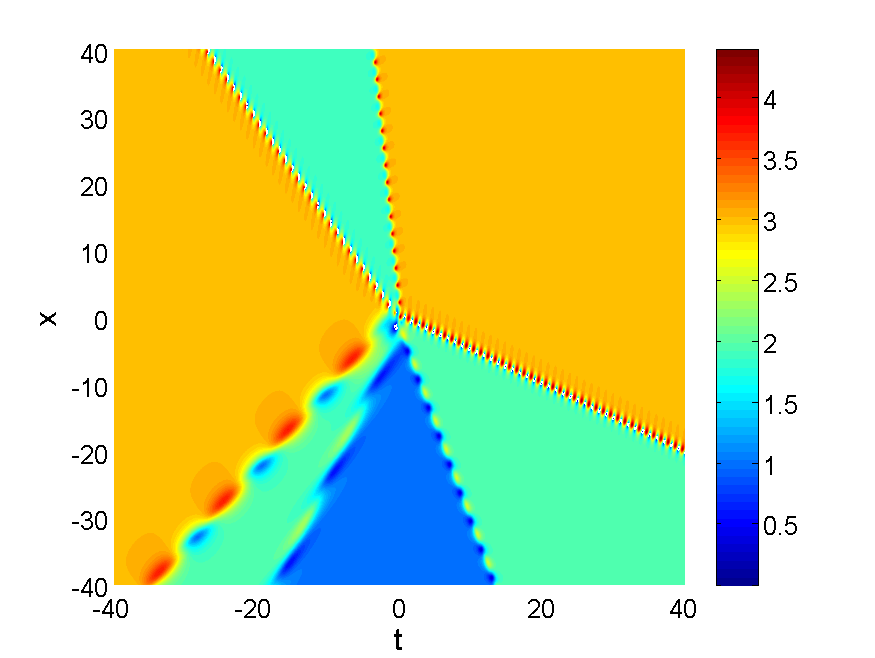}}}
{\rotatebox{0}{\includegraphics[width=4.8cm,height=3.2cm,angle=0]{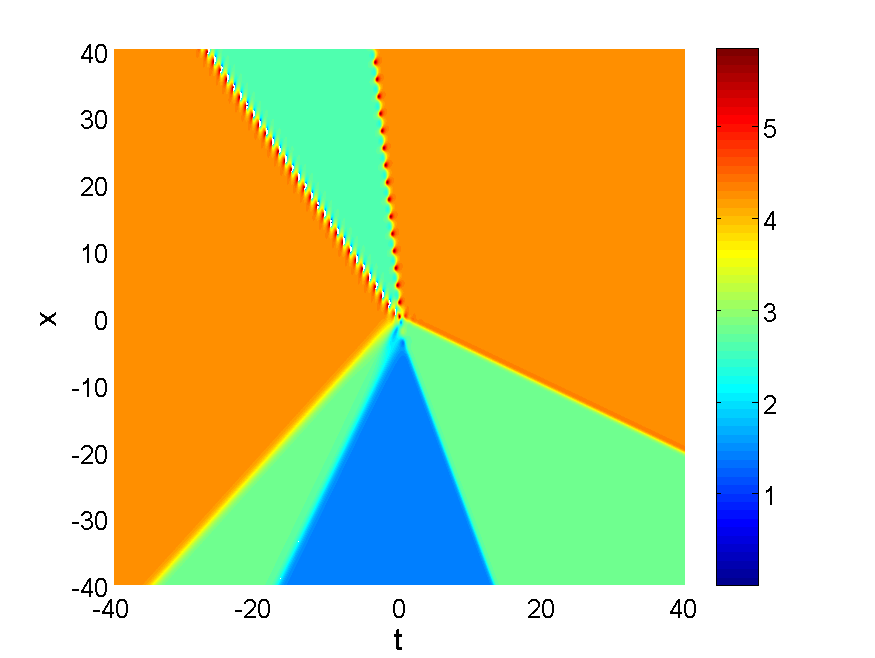}}}\\
$~~~~~~~~~~~~~~~~(\textbf{a1})
~~~~~~~~~~~~~~~~~~~~~~~~~~~~~~~~~(\textbf{a2})
~~~~~~~~~~~~~~~~~~~~~~~~~~~~~~~~(\textbf{a3})$\\

\noindent { \small \textbf{Figure 8.} Amplitude distributions of second-order solution given by Eq.\eqref{TPJ1} with integration constants ( $C_{ij} \neq 0$), From left to right, \(|q_1|\), \(|q_2|\), and \(\sqrt{|q_1|^2+|q_2|^2}\).  Other parameters here are \(a_1 = 1, ~a_2 = 2, ~\xi_1 = 0.25, ~\xi_2 = 0.5, ~\alpha_1 = \pi /5, ~\epsilon=0.1\) and \(\alpha_2 = 2\pi/5\).\\}
\\

\noindent

As shown in Figure 8, we choose the parameters \(a_1 = 1,~ a_2 = 2,~ \xi_1 = 0.25, ~\xi_2 = 0.5, ~\alpha_1 = \pi/5, ~\epsilon=0.1 \), and \(\alpha_2 = 2\pi/5\), then we can get a combination of two resonance wave patterns.

Secondly, if we only choose one of the integration constants is zero (for example, $C_{2k} = 0$), then solution~\eqref{TPJ1} describes interactions between a breather I, II, or III (at $k = 0, 1$ or 2) and fusion/decay resonance wave patterns. Figure 9 illustrates the interaction of the one-pole excitation with a breather I $(C_{20} = 0)$, a breather II $(C_{21} = 0)$ and a breather III $(C_{22} = 0)$ at \( a_1 = 1,~ a_2 = 2,~ \xi_1 = 0.25, ~\xi_2 = 1.0, ~\alpha_1 = \pi/5\), and \(\alpha_2 = \pi/4\), respectively. And all breathers collide at \(x = 0 , t = 0\).

\noindent
{\rotatebox{0}{\includegraphics[width=4.8cm,height=3.2cm,angle=0]{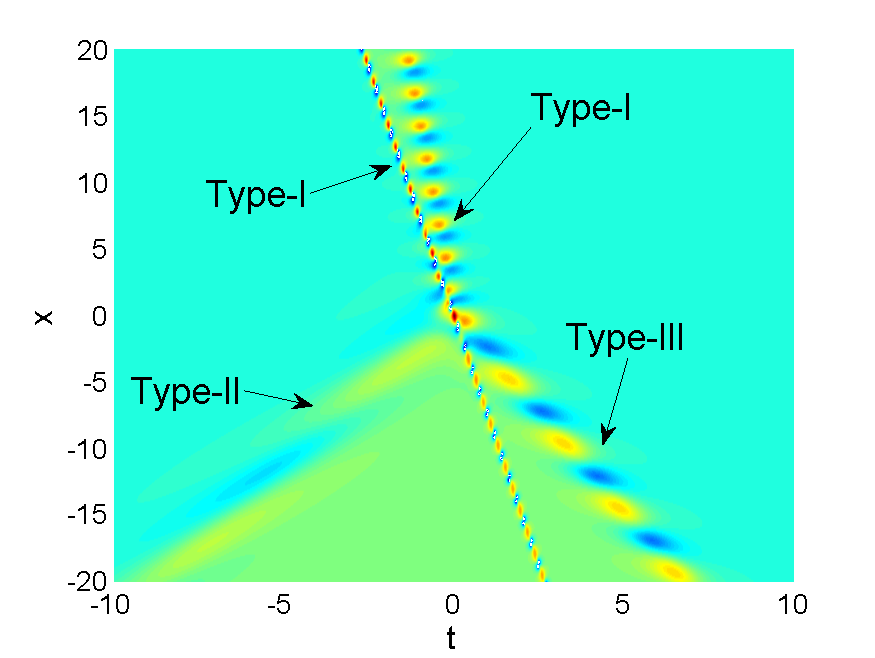}}}
{\rotatebox{0}{\includegraphics[width=4.8cm,height=3.2cm,angle=0]{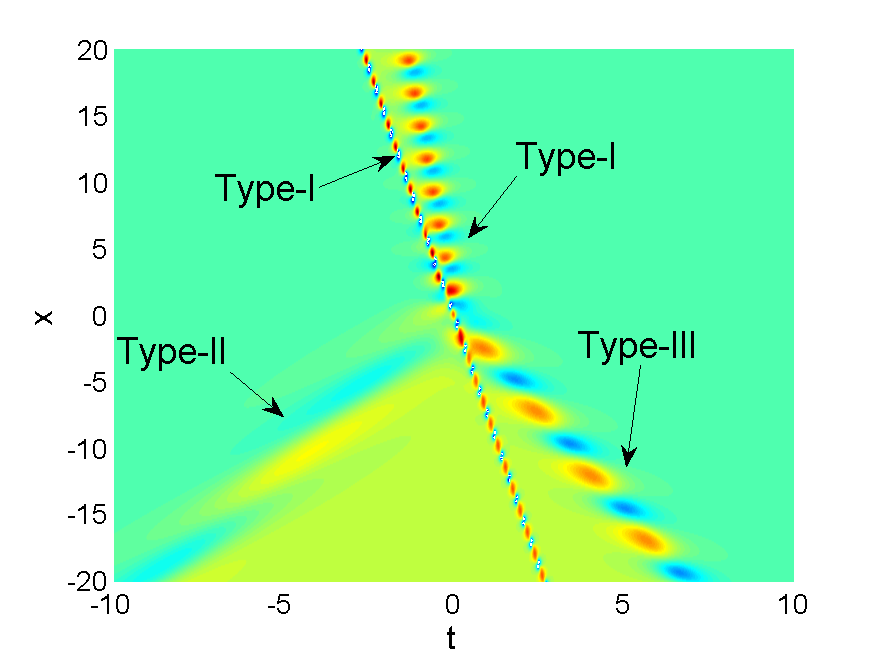}}}
{\rotatebox{0}{\includegraphics[width=4.8cm,height=3.2cm,angle=0]{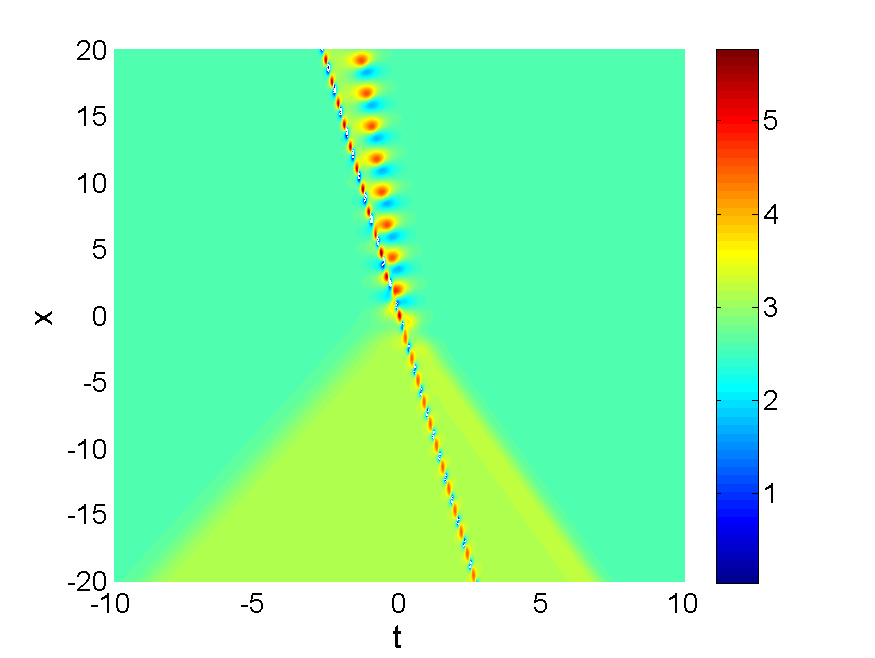}}}\\
$~~~~~~~~~~~~~~~~~~(\textbf{a1})
~~~~~~~~~~~~~~~~~~~~~~~~~~~~~~~~~(\textbf{a2})
~~~~~~~~~~~~~~~~~~~~~~~~~~~~~~(\textbf{a3})$\\

\noindent
{\rotatebox{0}{\includegraphics[width=4.8cm,height=3.2cm,angle=0]{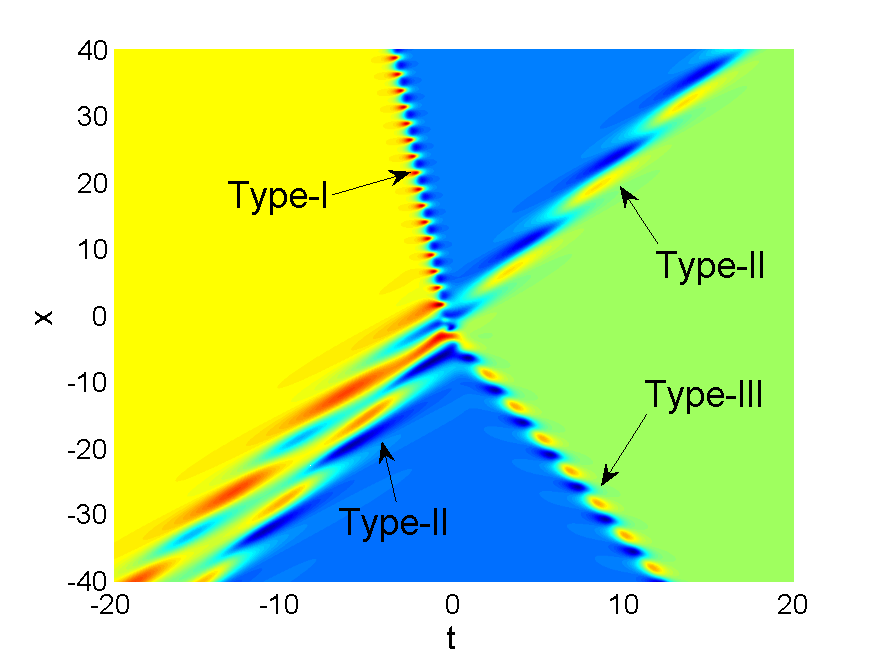}}}
{\rotatebox{0}{\includegraphics[width=4.8cm,height=3.2cm,angle=0]{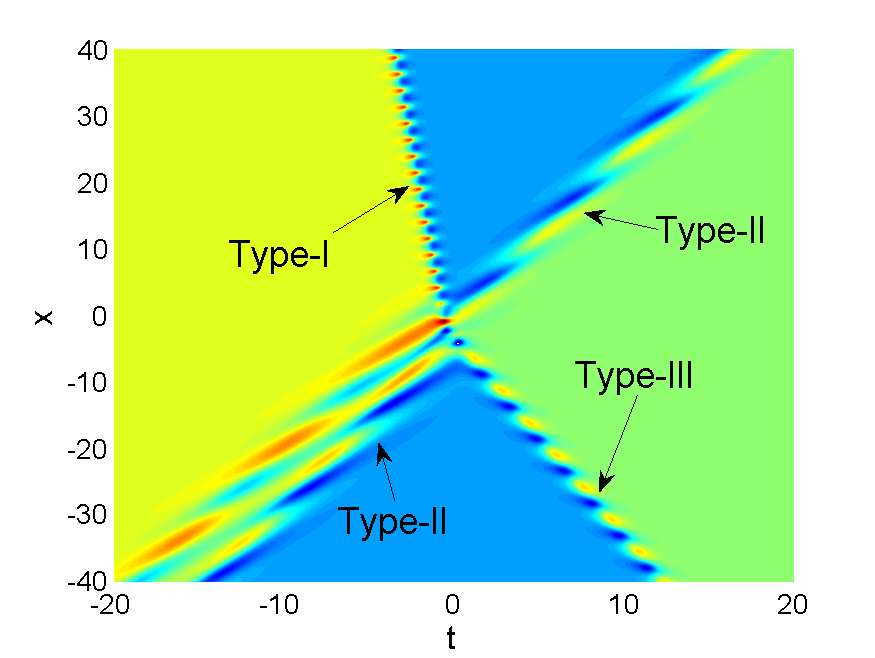}}}
{\rotatebox{0}{\includegraphics[width=4.8cm,height=3.2cm,angle=0]{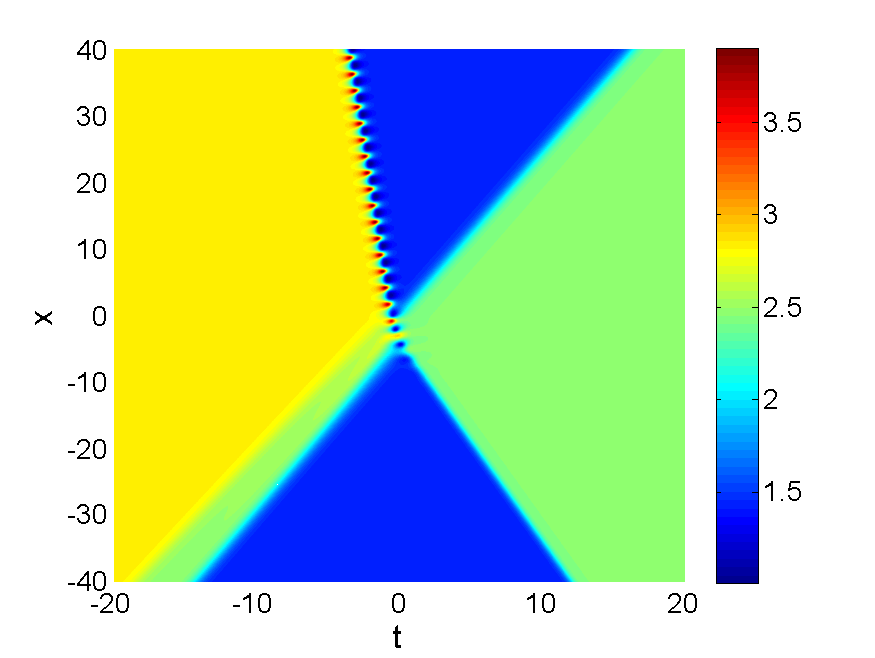}}}\\
$~~~~~~~~~~~~~~~~~~(\textbf{b1})
~~~~~~~~~~~~~~~~~~~~~~~~~~~~~~~~~(\textbf{b2})
~~~~~~~~~~~~~~~~~~~~~~~~~~~~~~(\textbf{b3})$\\

\noindent
{\rotatebox{0}{\includegraphics[width=4.8cm,height=3.2cm,angle=0]{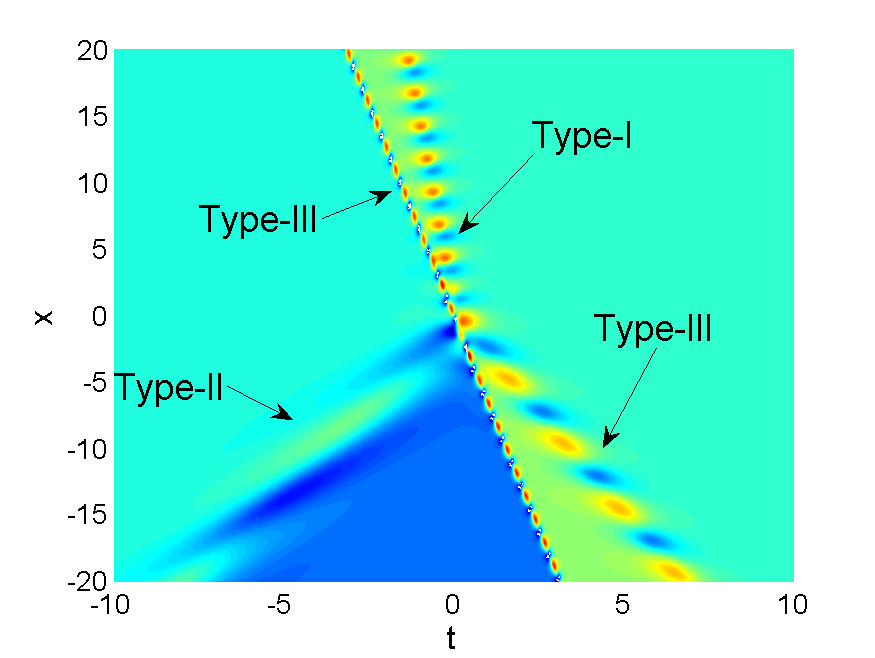}}}
{\rotatebox{0}{\includegraphics[width=4.8cm,height=3.2cm,angle=0]{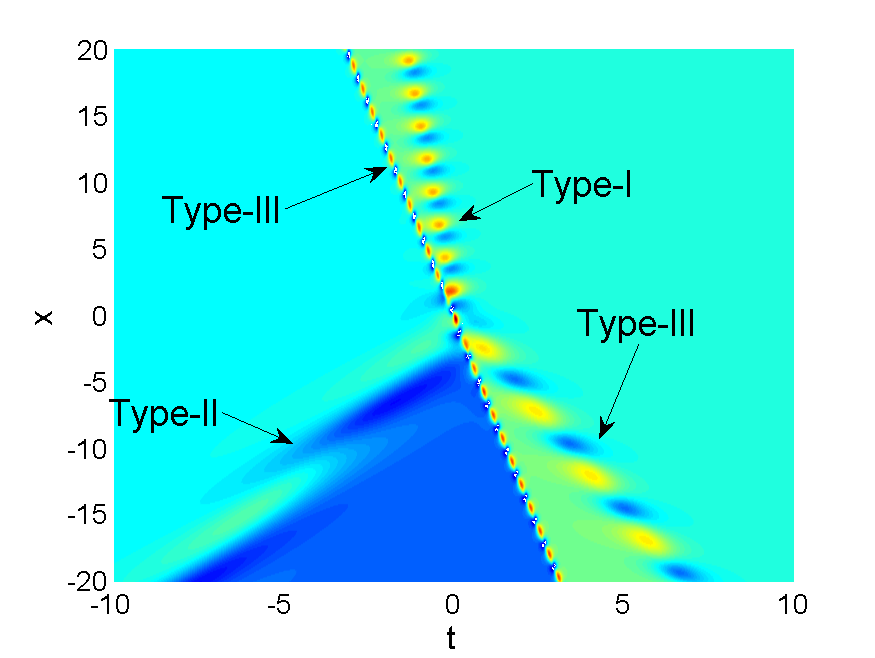}}}
{\rotatebox{0}{\includegraphics[width=4.8cm,height=3.2cm,angle=0]{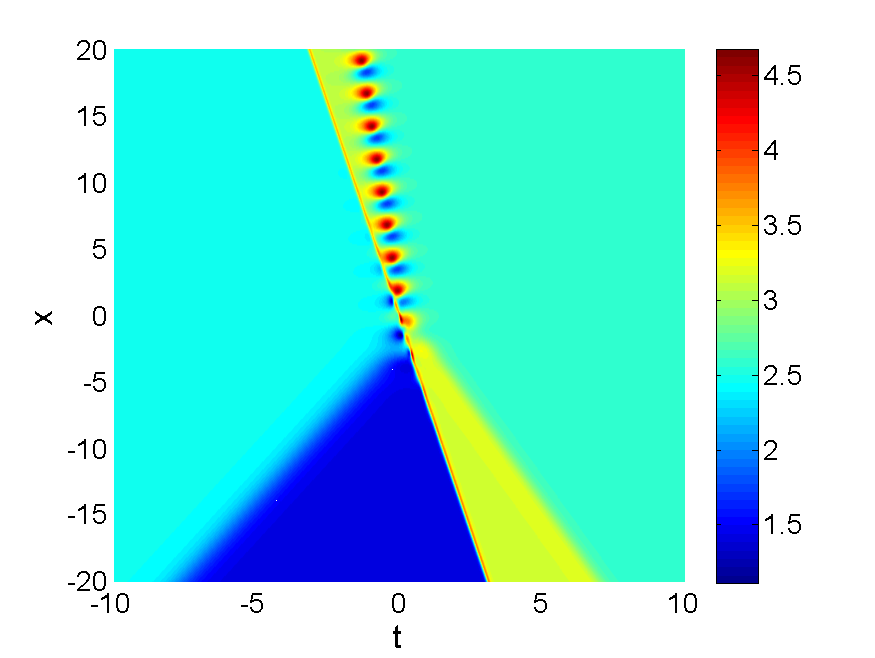}}}\\
$~~~~~~~~~~~~~~~~~~(\textbf{c1})
~~~~~~~~~~~~~~~~~~~~~~~~~~~~~~~~~(\textbf{c2})
~~~~~~~~~~~~~~~~~~~~~~~~~~~~~~(\textbf{c3})$\\

\noindent { \small \textbf{Figure 9.} Second-order solution  with parameters: $a_1 = 1$, $a_2 = 2$, $\xi_1 =0.25$, $\xi_2 = 1.0$, $\alpha_1 = \pi/5$, $\alpha_2 = \pi/4$, $\epsilon = 0.1$, $C_{10} = C_{11} = C_{12} = 1$. (a) $C_{20}=0$; (b) $C_{21} =0$; (c) $C_{22} =0$.\\}
\\

\noindent

When each of the two vectors \( \mathbf{C}_1 \) and \( \mathbf{C}_2 \) has exactly one zero component (so that each eigenvalue generates a single breather of a definite type), the solution \eqref{TPJ1} describes an elastic collision in which two breathers pass through each other, acquiring only phase and position shifts. There are nine possible combinations: I+I, I+II, II+I, II+II, III+III, III+II, I+III, II+III, III+I. We begin with the case of type I breathers collision, that is,  I + I $\rightarrow$ I + I , we choose the parameterization for the vectors \( \mathbf{C}_1 \) and \( \mathbf{C}_2 \): \( C_{10} = C_{20} = 0 \), see the figure 10(a). Furthermore, as an example, figure 11(a) shows the interaction of breathers II and III with the integration constants chosen as \( C_{11} = C_{22} = 0 \).

$~~~~~~~~~~~~~~~~~~~~~~~~~~~~~~~~$
{\rotatebox{0}{\includegraphics[width=4.8cm,height=3.2cm,angle=0]{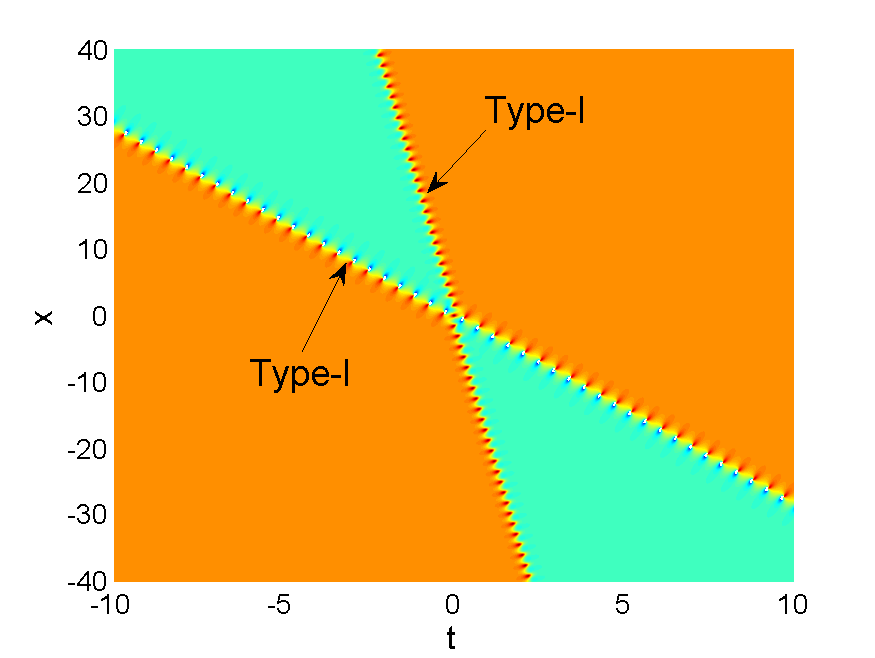}}}
\\
$~~~~~~~~~~~~~~~~~~~~~~~~~~~~~~~~~~~~~~~~~~~~~~~~~~~~~~~~(\textbf{a1})
$\\

\noindent
{\rotatebox{0}{\includegraphics[width=4.8cm,height=3.2cm,angle=0]{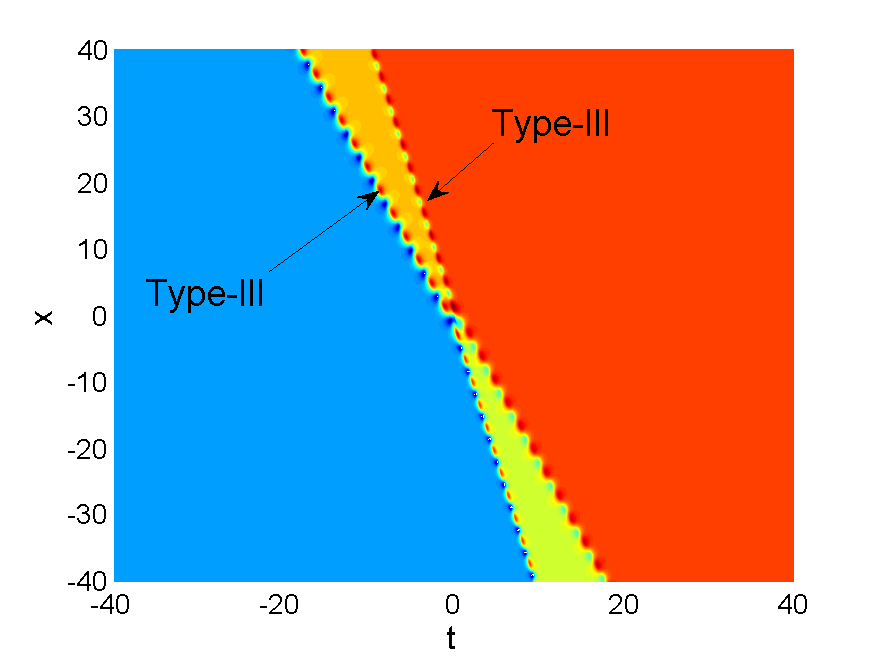}}}
{\rotatebox{0}{\includegraphics[width=4.8cm,height=3.2cm,angle=0]{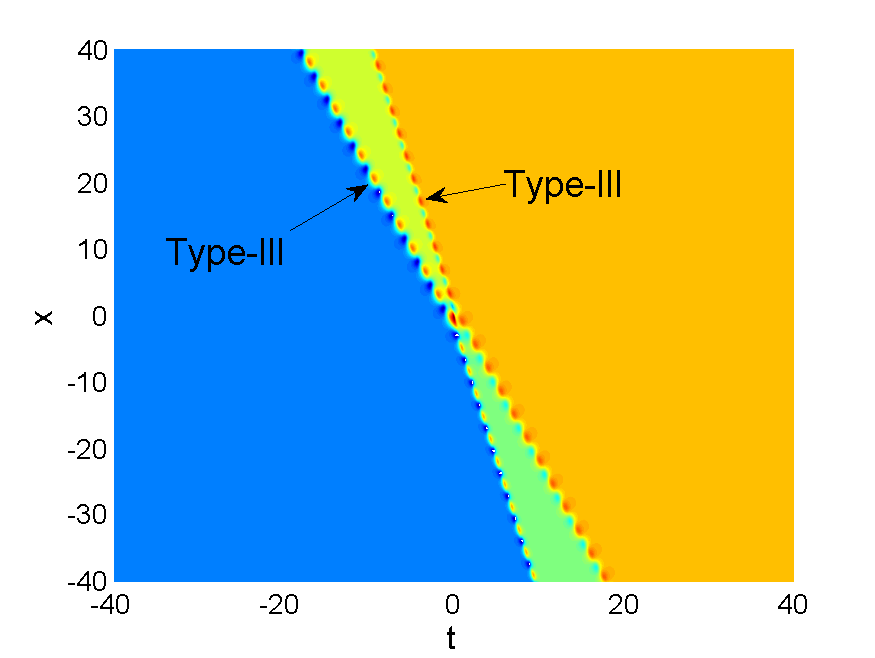}}}
{\rotatebox{0}{\includegraphics[width=4.8cm,height=3.2cm,angle=0]{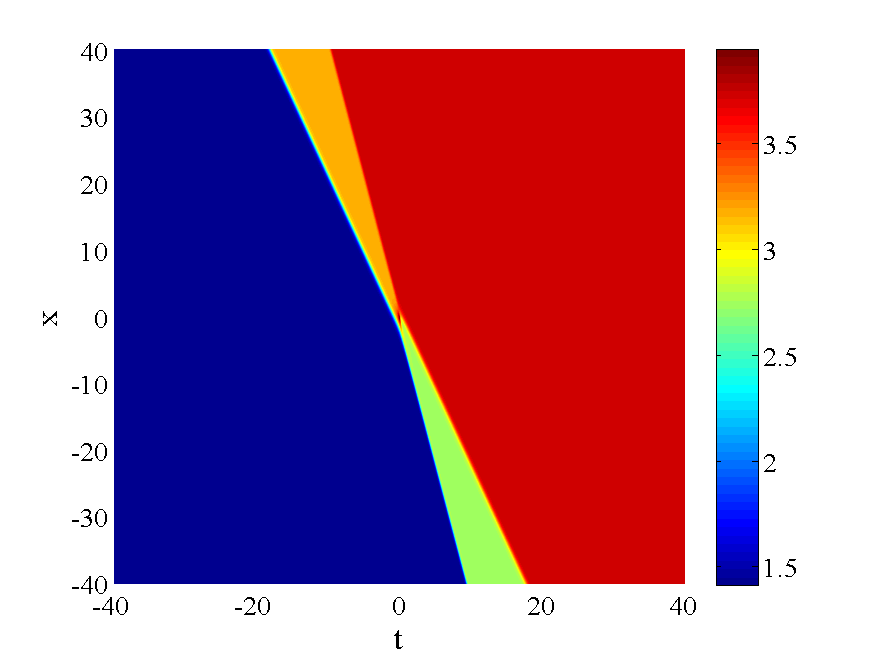}}}\\
$~~~~~~~~~~~~~~~~~~(\textbf{b1})
~~~~~~~~~~~~~~~~~~~~~~~~~~~~~~~~~(\textbf{b2})
~~~~~~~~~~~~~~~~~~~~~~~~~~~~~~(\textbf{b3})$\\

\noindent { \small \textbf{Figure 10.} Amplitude distributions of second-order solution given by Eq.\eqref{TPJ1}. (a) with integration constants ( $C_{10} = C_{20}= 0$), elastic collision I + I of vector breathers. (b) with integration constants ( $C_{11} = C_{21} = 0$), elastic collision III + III of vector breathers.  From left to right, \(|q_1|\), \(|q_2|\), and \(\sqrt{|q_1|^2+|q_2|^2}\).  Other parameters here are: (a) \( a_1 = 1,~ a_2 = 1, ~\xi_1 = 0.25, ~\xi_2 = 0.3,~ \alpha_1 = \pi/5, ~\alpha_2 = 2\pi/5 \)). (b) \(a_1 = 1, ~a_2 = 1, ~\xi_1 = 0.25, ~\xi_2 = 0.5, ~\alpha_1 = \pi /4, ~\epsilon=0.1\) and \(\alpha_2 = \pi/5\).\\}
\\

\noindent

\noindent
{\rotatebox{0}{\includegraphics[width=4.8cm,height=3.2cm,angle=0]{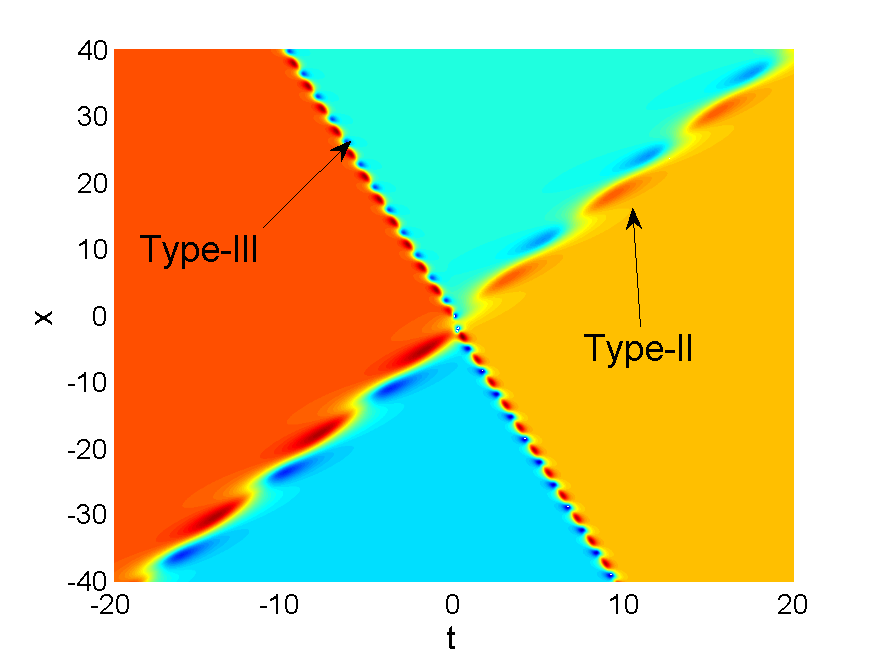}}}
{\rotatebox{0}{\includegraphics[width=4.8cm,height=3.2cm,angle=0]{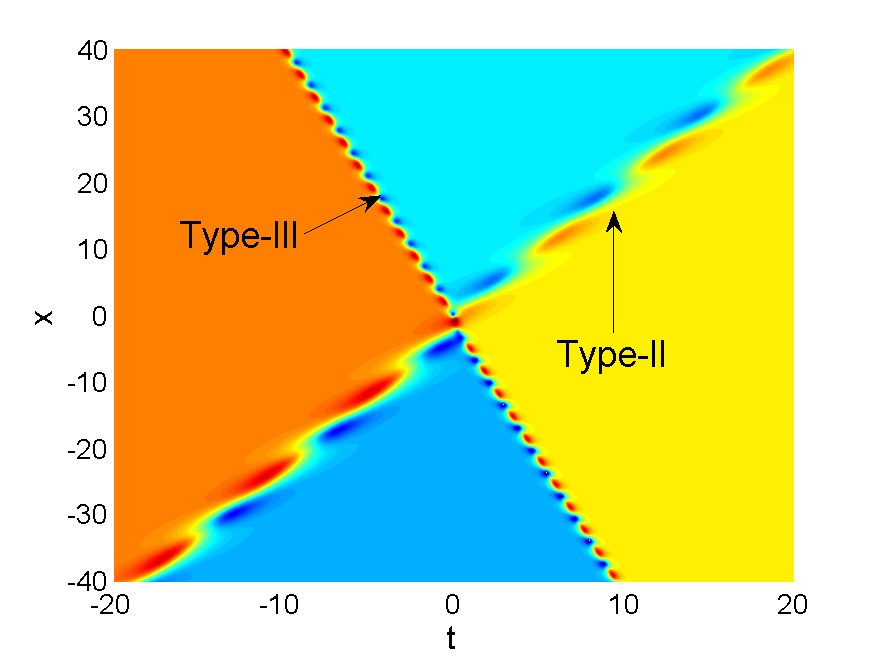}}}
{\rotatebox{0}{\includegraphics[width=4.8cm,height=3.2cm,angle=0]{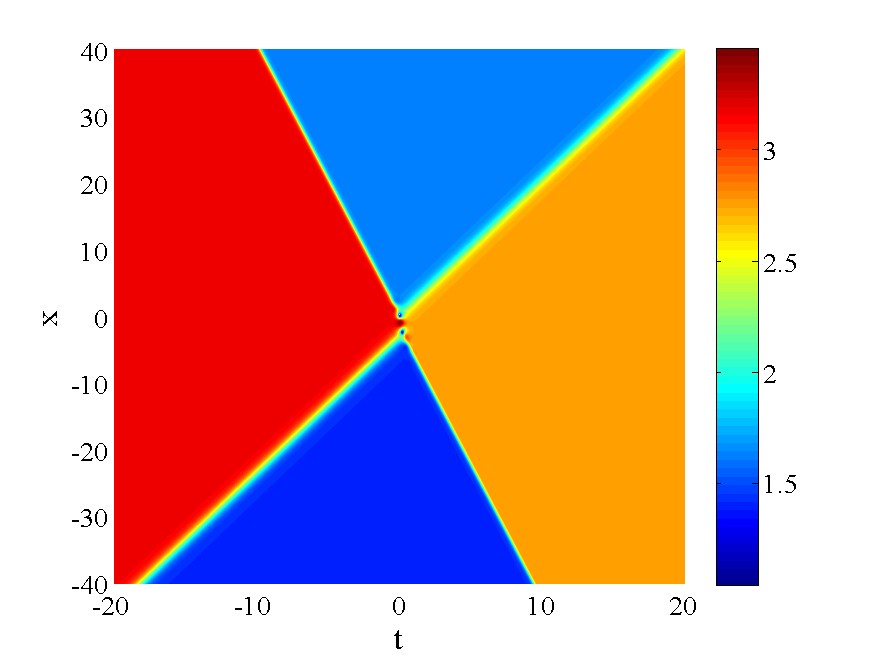}}}\\
$~~~~~~~~~~~~~~~~~~(\textbf{a1})
~~~~~~~~~~~~~~~~~~~~~~~~~~~~~~~~~(\textbf{a2})
~~~~~~~~~~~~~~~~~~~~~~~~~~~~~~(\textbf{a3})$\\

\noindent
{\rotatebox{0}{\includegraphics[width=4.8cm,height=3.2cm,angle=0]{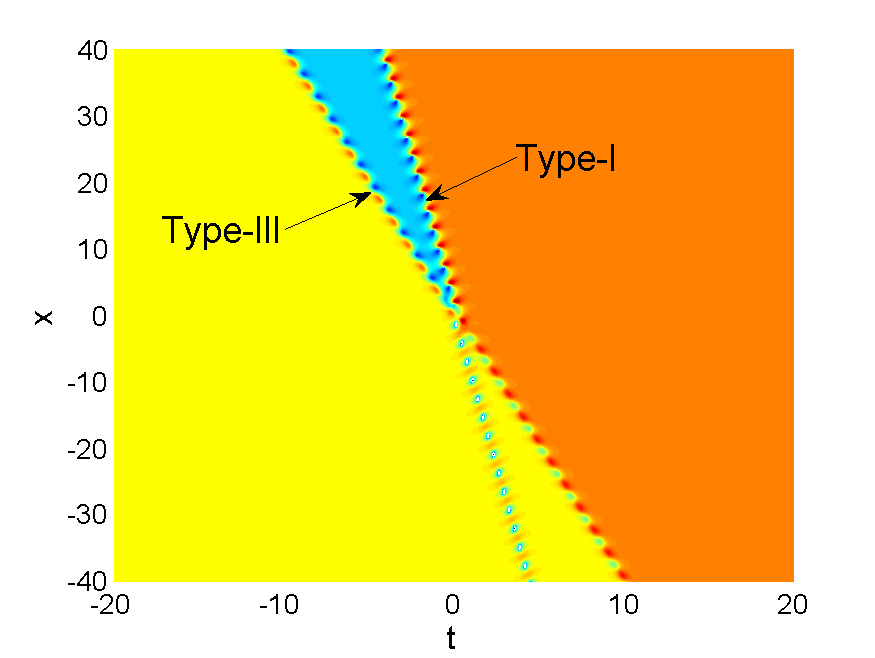}}}
{\rotatebox{0}{\includegraphics[width=4.8cm,height=3.2cm,angle=0]{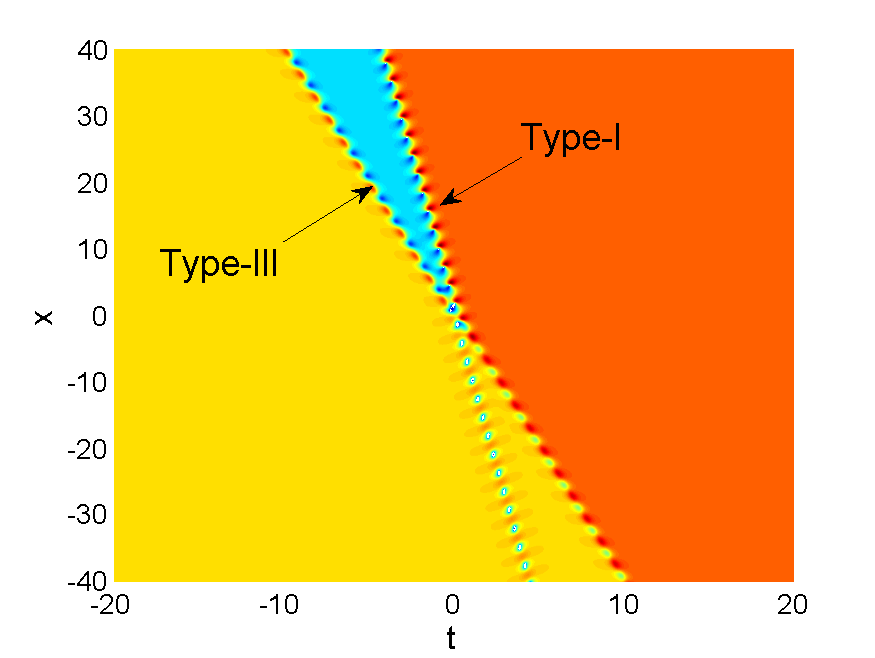}}}
{\rotatebox{0}{\includegraphics[width=4.8cm,height=3.2cm,angle=0]{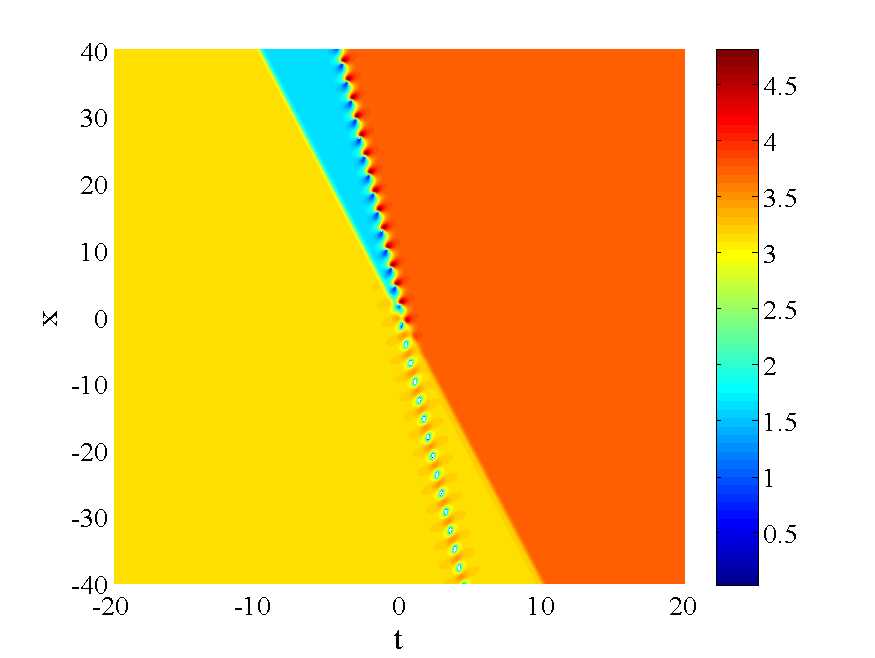}}}\\
$~~~~~~~~~~~~~~~~~~(\textbf{b1})
~~~~~~~~~~~~~~~~~~~~~~~~~~~~~~~~~(\textbf{b2})
~~~~~~~~~~~~~~~~~~~~~~~~~~~~~~(\textbf{b3})$\\

\noindent { \small \textbf{Figure 11.}  Amplitude distributions of second-order solution given by Eq. \eqref{TPJ1}. (a) with integration constants ( $C_{11} = C_{22}= 0$), elastic collision II + III of vector breathers. (b) with integration constants ( $C_{10} = C_{22} = 0$), elastic collision I + III of vector breathers.  From left to right, \(|q_1|\) , \(|q_2|\) , and \(\sqrt{|q_1|^2+|q_2|^2}\).  Other parameters here are \(a_1 = 1, ~a_2 = 1, ~\xi_1 = 0.25, ~\xi_2 = 0.5, ~\alpha_1 = \pi /4, ~\epsilon=0.1\) and \(\alpha_2 = \pi/5\).\\}
\\

\noindent

Then, we discuss a particular case of vector two-breather solution.

In the previous discussion, we discussed two important types of dynamic behaviors of vector breathers respectively. The fourth part reveals a unique inelastic interaction in the Hiorta equation--the resonance process, that is, the fusion (\( \text{I} + \text{II} \rightarrow \text{III} \)) or decay (\( \text{III} \rightarrow \text{I} + \text{II} \)) of the three breather subtypes (I, II, III). From a physical image, this resonance appears as a nonlinear wave complex presented by the single eigenvalue solution \eqref{YJJ2} with three nonzero integration constants, and its asymptotic states correspond to three different types of single breathers.

However, a fundamental theoretical question arises: can this seemingly independent resonance phenomenon, determined only by a single eigenvalue, be derived from the more general two-breather collision framework? In other words, are there deep mathematical homologies between the resonant fusion observed in Section 4 and the elastic collisions systematically studied in precious content?

To answer this question, let's turn our attention to an important assertion in the theory of classical integrable systems. Zakharov and Manakov pointed out in their pioneering work on three-wave resonance that when two discrete eigenvalues of the auxiliary spectral problem tend to coincide, the collision displacement between breathers will tend to infinity, causing two originally independent breathers to degenerate into an inseparable resonant bound state. This idea inspires us that the vector breather resonance in the fourth part is very likely a singular limit of the elastic collision of two breathers at the boundary of the eigenvalue parameter space.

Now, we consider the asymptotic behavior of the double eigenvalue solution when the eigenvalues tend to degeneracy. That is, assume \(\lambda_1 \to \lambda_2\), for the processes (\( \text{I} + \text{II} \rightarrow \text{I} + \text{II} \)) or (\( \text{I} + \text{III} \rightarrow \text{I} + \text{III} \)). To get a feeling of this limit, we plot the spatio-temporal diagrams for the breather collisions corresponding to a set of small differences \(\epsilon_0\) between the breather eigenvalues, defined as

\begin{equation}\label{JX}
\alpha_1=\alpha_2=\pi /5,~~\xi_1=\xi=0.25,~~\xi_2=\xi_1+\epsilon_0.
\end{equation}

As the figures can be also seen in Figure 12, where we plot three collision portraits (a), (b), and (c) with \(\epsilon_0=10^{-2}\), \(\epsilon_0=10^{-4}\), and \(\epsilon_0=10^{-6}\) correspondingly. At small \(\epsilon_0\), the point where the breathers collide transforms into an increasing straight junction, which begins at the point of the breathers association and later ends at the point of the breathers separation.

When \(\epsilon_0=10^{-6}\), since the length of the junction region is close to infinity, the outgoing type I and type II trajectories have not broken free within the visible time range in the figure. Physically, they are pushed to the infinitely distant future, which perfectly explains the core conclusion that resonance is the limit case of elastic collision where infinite spatial displacement occurs when the eigenvalues merge. At this time, the junction region itself asymptotically transforms into a pure type III vector breather wave in the limit of \(\epsilon_0 \to 0\). This mathematically confirms that when the eigenvalues are completely merged \(\epsilon_0=0\), the junction region formed by the collision of two wave packets is the resonance product, the type III breather wave.

It intuitively demonstrates Zakharov-Manakov's idea that resonance stems from the merging of eigenvalues through three consecutive values of \(\epsilon_0\). When the eigenvalue spacing between two colliding breather waves approaches zero, the interaction between them is no longer an elastic scattering with finite displacement, but evolves into an infinitely long junction. In the limit, this junction is exactly equal to the single -pole resonance solution (type III breather wave). The visualization of this viscous effect provides a profound mathematical and physical picture for understanding the inelastic resonant fusion and decay of vector breather waves.

\noindent
{\rotatebox{0}{\includegraphics[width=4.8cm,height=3.2cm,angle=0]{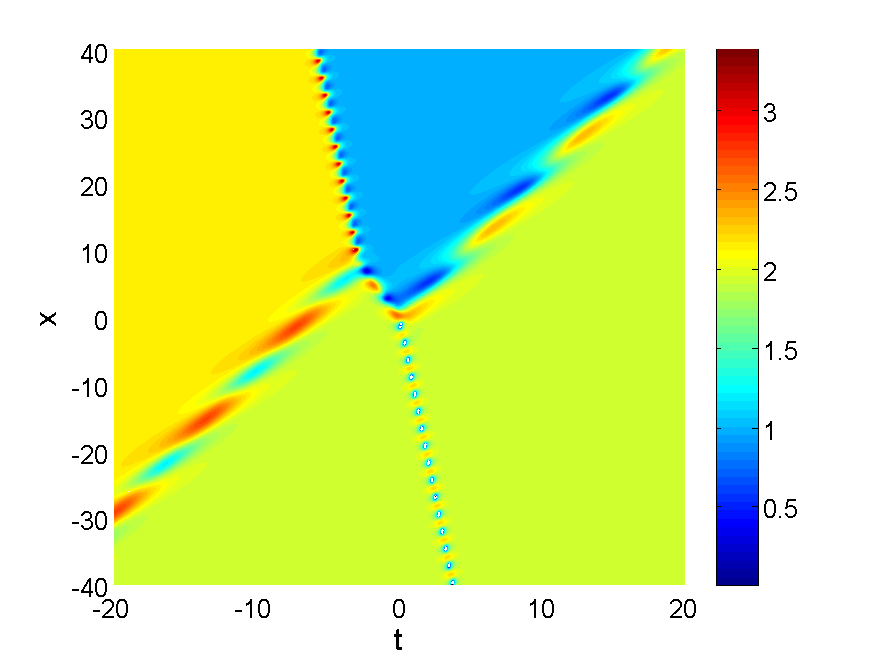}}}
{\rotatebox{0}{\includegraphics[width=4.8cm,height=3.2cm,angle=0]{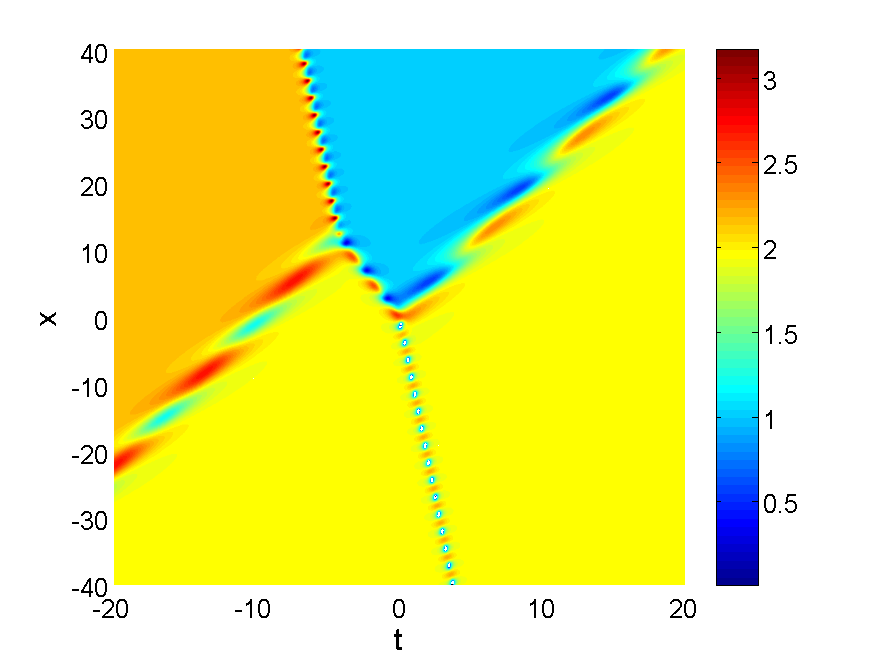}}}
{\rotatebox{0}{\includegraphics[width=4.8cm,height=3.2cm,angle=0]{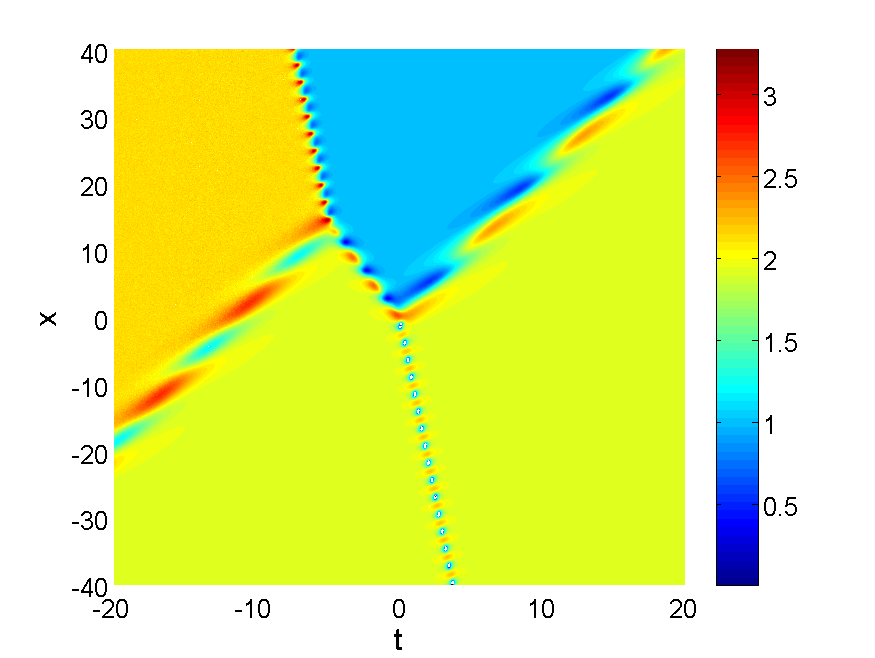}}}\\
$~~~~~~~~~~~~~~~~(\textbf{a})
~~~~~~~~~~~~~~~~~~~~~~~~~~~~~~~~~~(\textbf{b})
~~~~~~~~~~~~~~~~~~~~~~~~~~~~~~~~~~~(\textbf{c})$\\

\noindent { \small \textbf{Figure 12.} Elastic collision (\( \text{I} + \text{II} \rightarrow \text{I} + \text{II} \)) of vector breathers with close eigenvalues chosen according to \eqref{JX}, and (a) \(\epsilon_0=10^{-2}\), (b) \(\epsilon_0=10^{-4}\), (c) \(\epsilon_0=10^{-6}\). The spectral parameters \(\xi\) and \(\alpha\) taken from Eq.\eqref{JX}. Characters I, II, and III in panel (c) indicate the two colliding breathers and the collision junction. The latter transforms to the type III asymptotically at \(\epsilon_0=0\).\\}
\\

\noindent

\section{Numerical simulations}\label{sec6}
To verify the correctness of the analytical solutions obtained via the dressing method and to investigate the propagation robustness of vector breathers under perturbations, this section takes the Type-I, II and III vector breather as the prototypical example and performs numerical simulations of the coupled Hirota equation using the split-step Fourier method \cite{Liu-2021,Chen-2021}.

Figures 13(a) display the spatio temporal evolutions of the two components in the plane. It is clearly observed that the breather exhibits a localized wave-packet structure and propagates along the spatial direction with a constant group velocity. During the entire evolution, the amplitude and shape of the wave packet remain highly consistent, without noticeable dispersion or decay, which visually confirms that the analytical solution is indeed an exact solution of the equation. Subsequently, we compare the cross-sectional profiles of the numerical and exact solutions at a fixed time; the two curves almost completely overlap over the entire spatial domain, demonstrating excellent quantitative agreement.

In realistic physical experiments, such as those in optical-fiber communications, the background wave is inevitably accompanied by noise perturbations. To assess the viability of this vector breather as an information carrier, we add Gaussian white noise with a relative amplitude of \(0.02\) to the exact initial field. The results show that, although minor jagged fluctuations are superimposed on the waveform, the overall envelope, the position of the local peak, and the oscillation period of the breather remain essentially unchanged. This indicates that the vector breather possesses a natural ``immunity'' to background noise, the noise energy is only partially superimposed on the continuous spectrum and does not effectively transfer to the bound state of the breather. This robustness provides a solid theoretical foundation for experimental observation and for potential applications in non-ideal transmission environments.

{\rotatebox{0}{\includegraphics[width=5.4cm,height=4.2cm,angle=0]{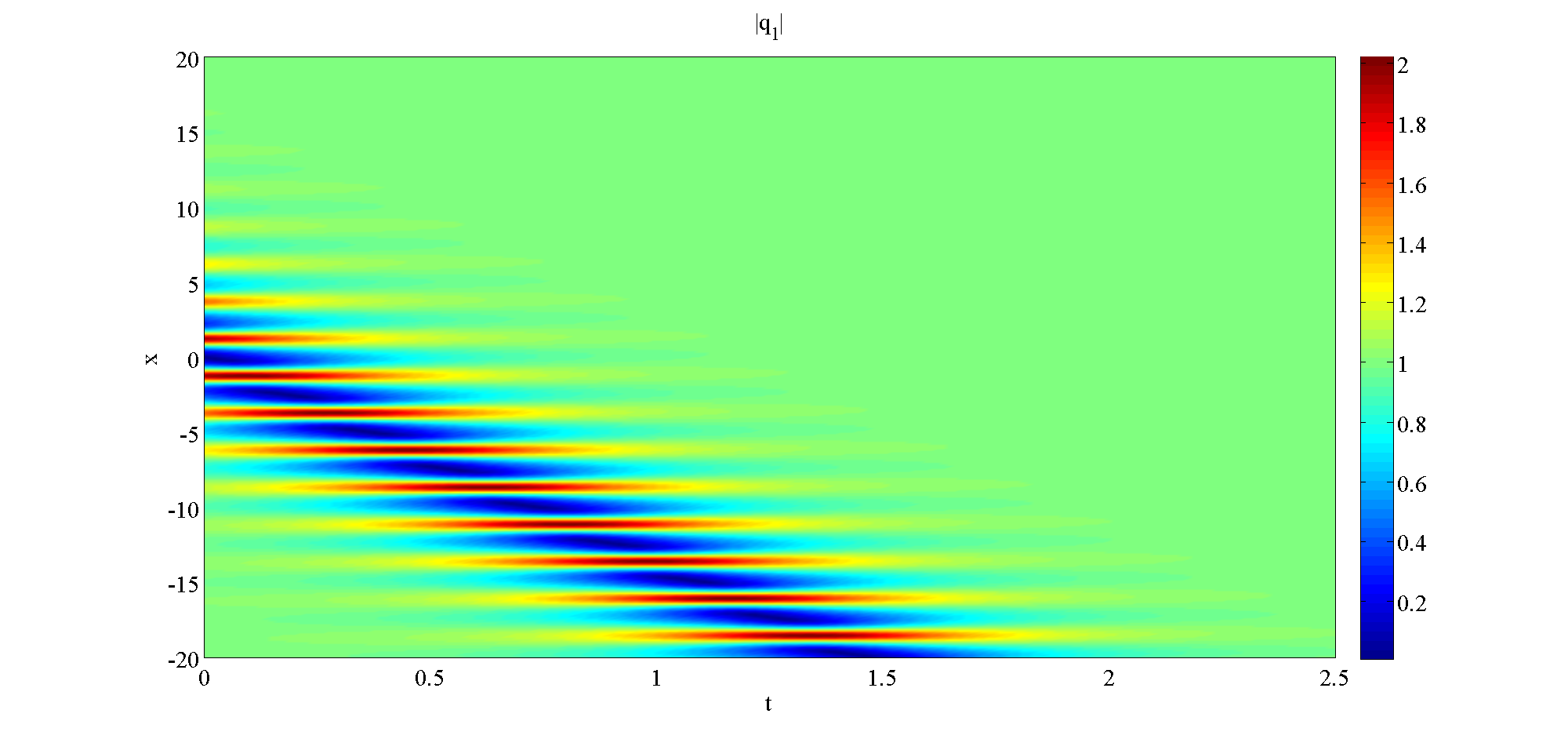}}}
{\rotatebox{0}{\includegraphics[width=5.4cm,height=4.2cm,angle=0]{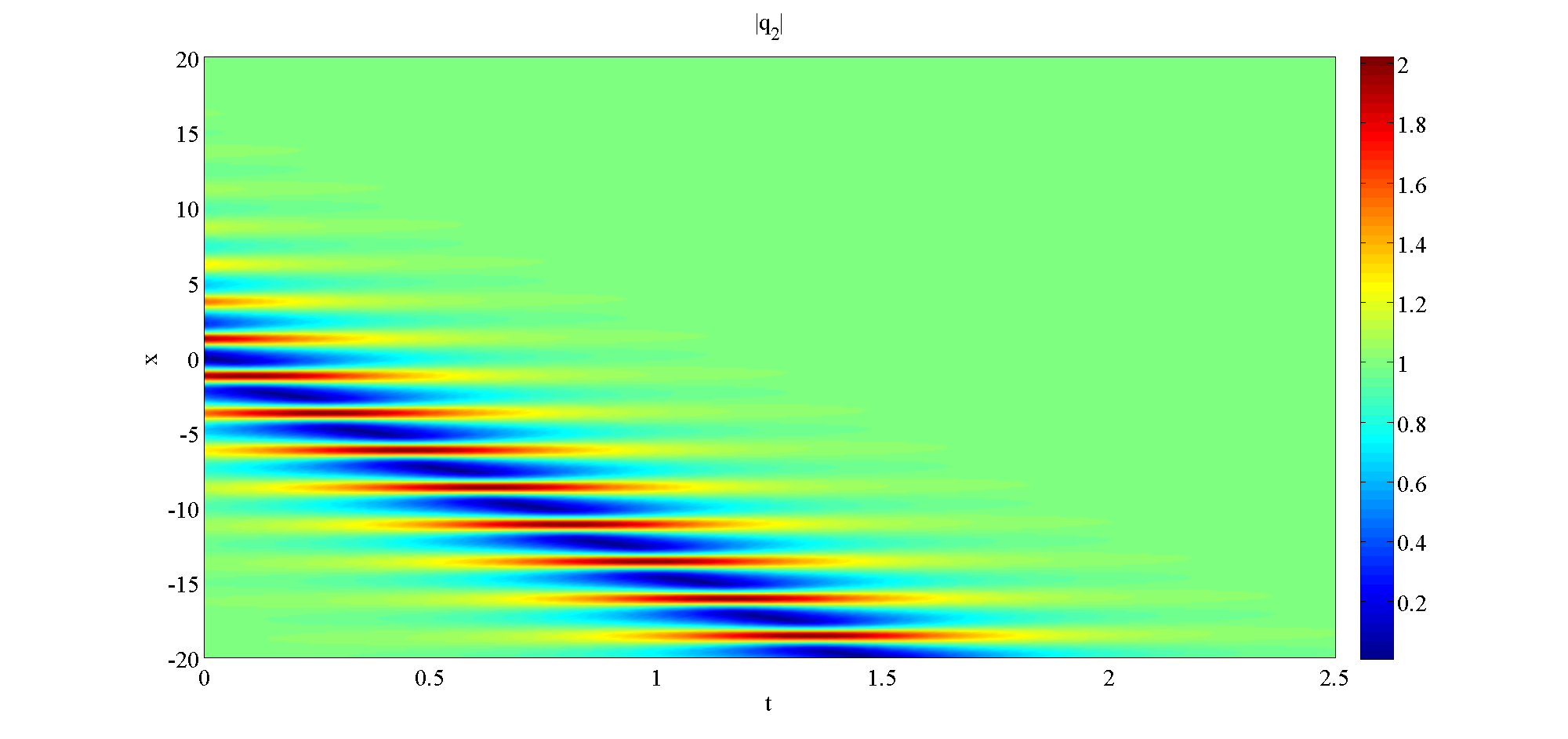}}}

$\qquad\qquad\qquad(\textbf{a1})
\qquad\qquad\qquad\qquad\qquad\qquad(\textbf{a2})
$\\

{\rotatebox{0}{\includegraphics[width=5.4cm,height=4.2cm,angle=0]{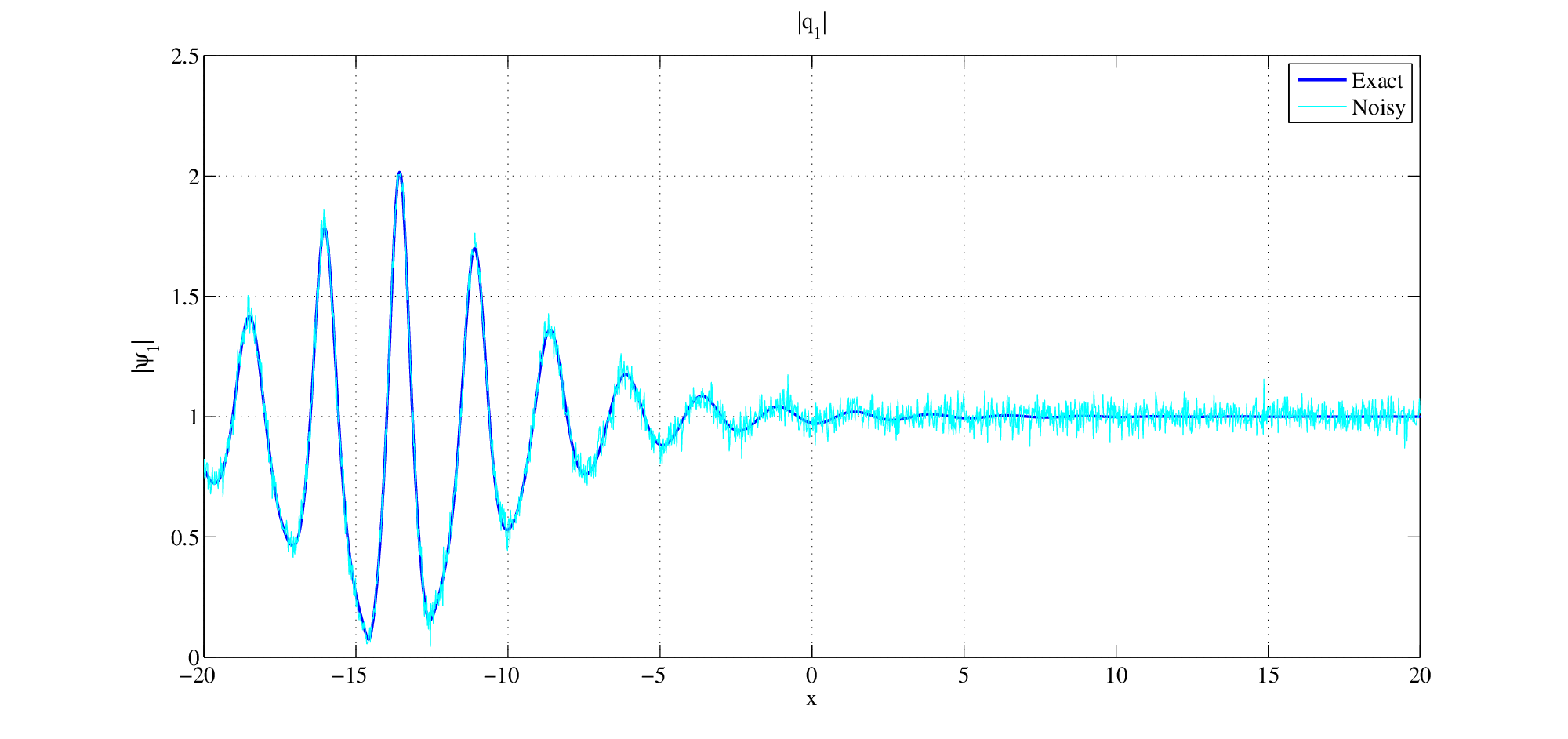}}}
{\rotatebox{0}{\includegraphics[width=5.4cm,height=4.2cm,angle=0]{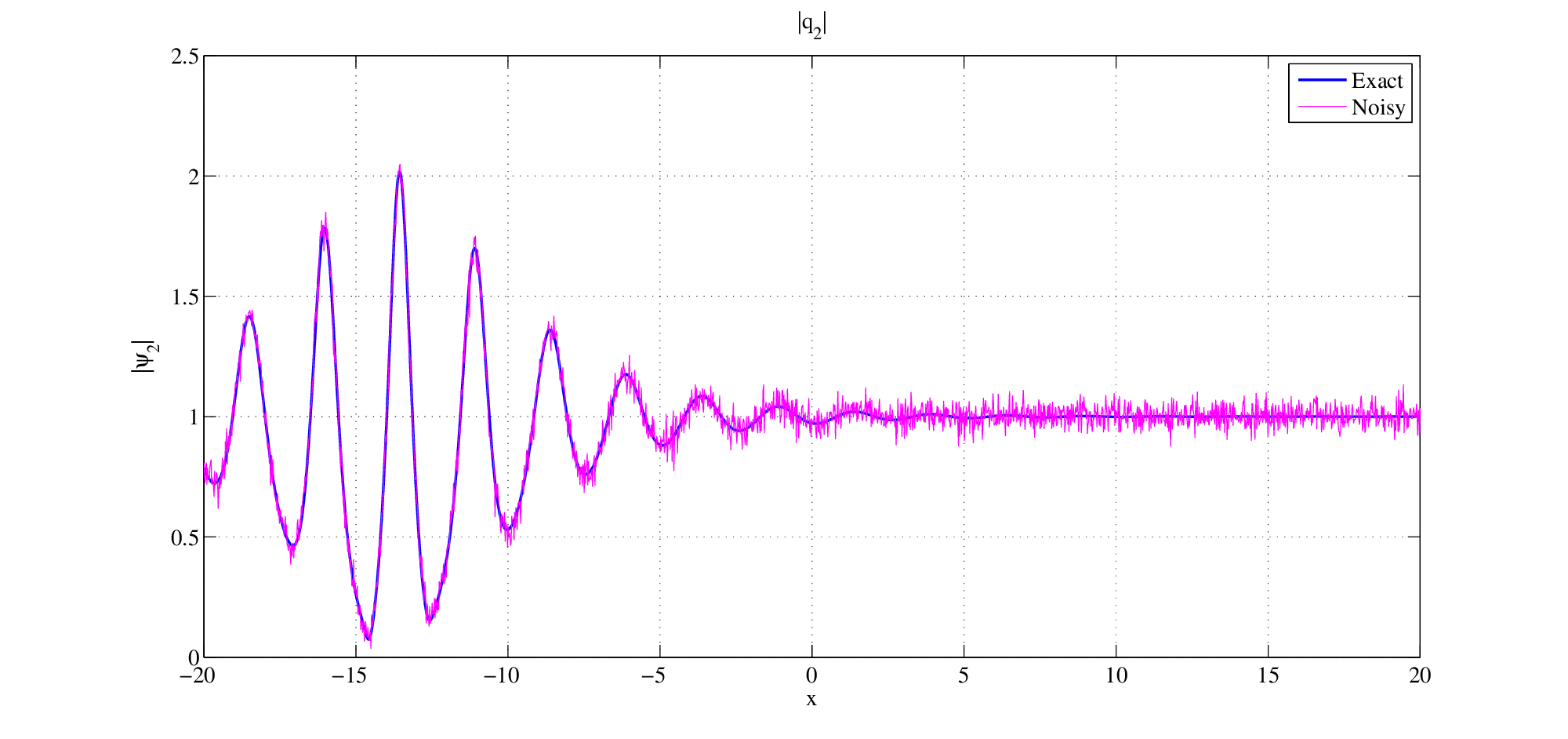}}}

$\qquad\qquad\qquad(\textbf{b1})
\qquad\qquad\qquad\qquad\qquad\qquad(\textbf{b2})
$\\

\noindent { \small \textbf{Figure 13.} (a) Spatiotemporal evolution of the Type-I vector breather obtained from the exact solution. (a1) Amplitude distribution of component \(q_1\); (a2) amplitude distribution of component \(q_2\). The two components exhibit identical localized wave patterns with a constant amplitude ratio. (b) Robustness test of the Type-I breather against initial white-noise perturbation with a relative amplitude of \(0.02\). (b1) Perturbed profile of \(q_1\); (b2) perturbed profile of \(q_2\). The main localized structure remains intact, indicating strong stability against small-amplitude background fluctuations.\\}
\\

\noindent

{\rotatebox{0}{\includegraphics[width=5.2cm,height=4.2cm,angle=0]{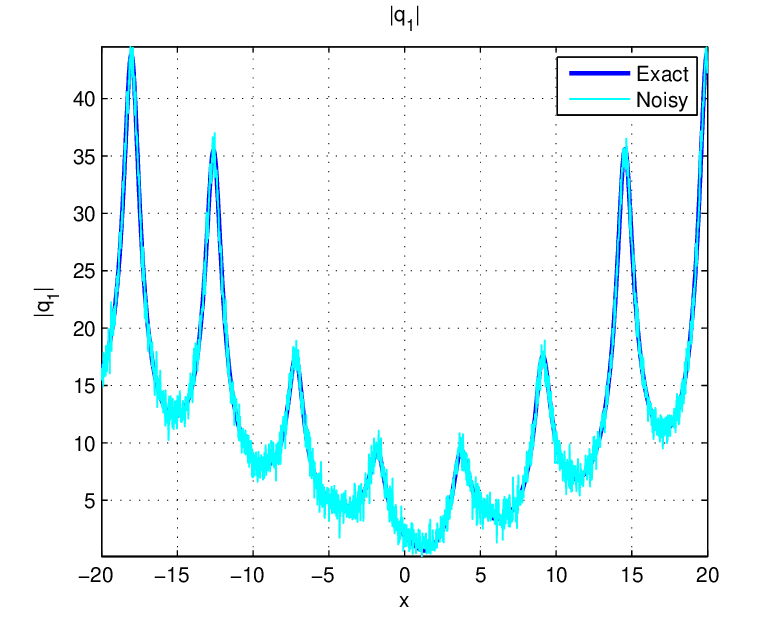}}}
{\rotatebox{0}{\includegraphics[width=5.2cm,height=4.2cm,angle=0]{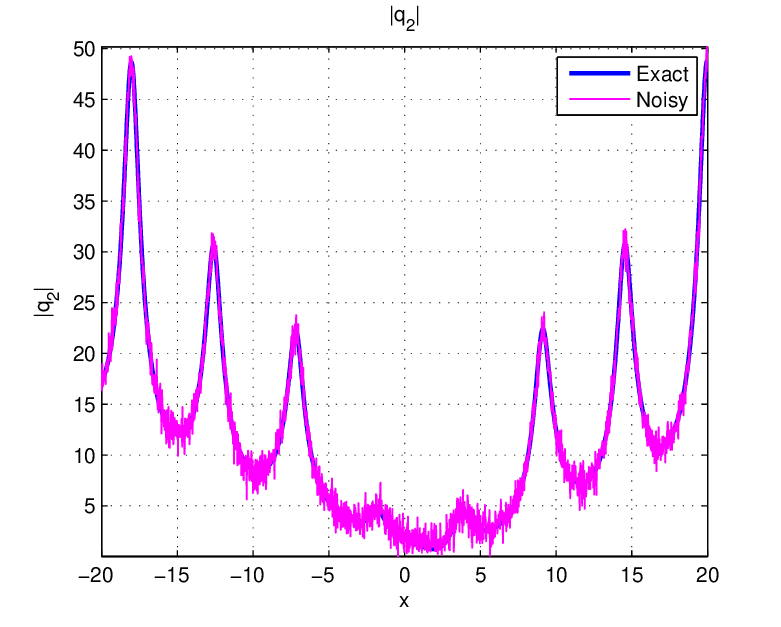}}}

$\qquad\qquad\qquad(\textbf{a1})
\qquad\qquad\qquad\qquad\qquad\qquad(\textbf{a2})
$\\

{\rotatebox{0}{\includegraphics[width=5.2cm,height=4.2cm,angle=0]{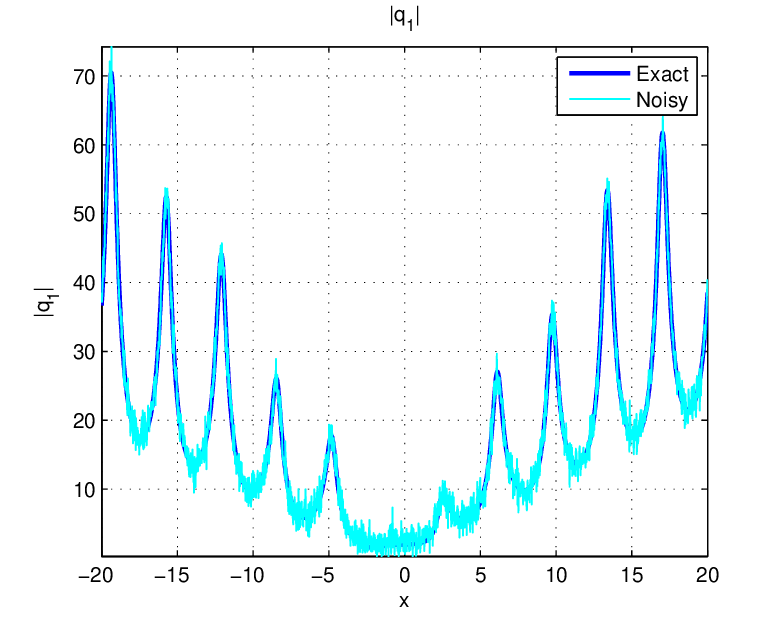}}}
{\rotatebox{0}{\includegraphics[width=5.2cm,height=4.2cm,angle=0]{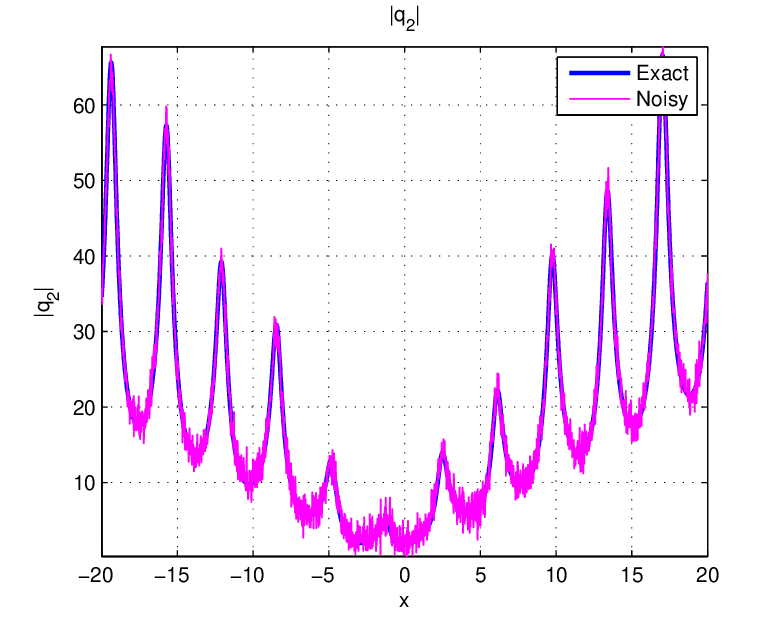}}}

$\qquad\qquad\qquad(\textbf{b1})
\qquad\qquad\qquad\qquad\qquad\qquad(\textbf{b2})
$\\

\noindent { \small \textbf{Figure 14.} (a) Robustness test of the Type-II breather against initial white-noise perturbation with a relative amplitude of \(0.02\). (a1) \(q_1\); (a2) \(q_2\). (b) Robustness test of the Type-III breather against initial white-noise perturbation with a relative amplitude of \(0.02\). (b1) \(q_1\); (b2) \(q_2\).\\}
\\

\noindent

\section{Conclusion}\label{sec7}

This paper presents a systematic and in-depth theoretical investigation of vector breathers and their nonlinear interactions, particularly the modulatory role of the higher-order effect parameter \(\epsilon\), in the focusing-focusing coupled Hirota equation \eqref{HT}, which constitutes a vector generalization of the scalar NLSE. On a non-zero plane wave background, this model describes the simultaneous propagation of ultrashort optical pulses in optical fibers, where the parameter \(\epsilon\) characterizes non-negligible higher-order linear and nonlinear effects such as third-order dispersion and self-steepening. By employing the vector dressing method developed in the study of the Manakov system, we successfully construct multi-breather solutions of this model on a nontrivial background. Using these solutions as a tool, we reveal the rich typology, collision dynamics, and profound connections with higher-order effects of vector breathers.
		
First, by performing asymptotic analysis of the single-breather solution, we rigorously prove the existence of three fundamental types of vector breathers in equation \eqref{HT}: Type I, Type II, and Type III. Type I breathers, which are served as a direct generalization of scalar breathers (such as the Kuznetsov, Akhmediev, and Peregrine solutions); their two field components satisfy a simple proportionality relation. Type II and Type III breathers, which is stable, and exhibit dynamical behaviors unique to the vector case that are fundamentally different from scalar breathers; their wave-field components cannot be expressed in terms of a common scalar field. Subsequently, we analyze in detail how the higher-order effect parameter \(\epsilon\) modifies the eigenvalue distribution, spatial localization scale, group velocity, and asymptotic phase shift of the breathers.
Next, we report for the first time in the coupled Hirota system, the inelastic resonant interaction between vector breathers, namely the fusion of a Type I and a Type II breather into a Type III breather (\( \text{I} + \text{II} \rightarrow \text{III} \)), or the reverse decay process of a Type III breather into Type I and Type II breathers (\( \text{III} \rightarrow \text{I} + \text{II} \)). This process satisfies the resonance matching condition modified by \(\epsilon\).

Then, we systematically analyze the rich interactions and dynamical behaviors among the second-order solutions of the coupled Hirota equation, including resonant interactions and elastic collisions, and point out a special case where the resonance interaction occurs when the eigenvalues coalesce. Finally, numerical simulations using the split-step Fourier method are performed to verify the correctness of the solutions.

\section*{Acknowledgements}
\hspace{0.3cm}
We express our sincere thanks to all the persons who have provided valuable suggestions to this
paper. This work is supported by the National Natural Science Foundation of China under Grant No. 12201622
and the Project Funded by China Postdoctoral Science Foundation under Grant No. 2024M753506.

%\end{CJK*}
%\end{document}

% Converted from Microsoft Word to LaTeX
% by Chikrii SoftLab Word2TeX converter (version 2.4)
% Copyright (C) 1999-2001 Kirill A. Chikrii, Anna V. Chikrii
% Copyright (C) 1999-2001 Chikrii SoftLab.
% All rights reserved.
% http://www.word2tex.com/
% mailto: info@word2tex.com, support@word2tex.com

% Warning: You are using UNREGISTERED Chikrii SoftLab Word2TeX!
%          In UNREGISTERED mode some restrictions will apply.
%          For more information please visit http://www.word2tex.com/
% YOU CAN USE THIS FILE WITH THE SOLE PURPOSE OF EVALUATING Word2TeX.

%\documentclass [12pt]{article}

%\begin{document}

\end{document}